\documentclass[11pt, oneside]{article}
\usepackage{graphicx} %
\usepackage{geometry}
\usepackage{float}
\usepackage{xcolor}         %
\usepackage{subcaption}
\usepackage{amsmath}
\usepackage{amssymb}
\usepackage{CormorantGaramond}
\usepackage[font={it}]{caption}
\usepackage[parfill]{parskip}
\usepackage{hyperref}
\usepackage{footnote}
\usepackage{xcolor}
\usepackage{colortbl}
\usepackage{array}
\usepackage{longtable}
\usepackage{textcomp}
\usepackage{bbm}
\usepackage{enumitem}

\usepackage[T1]{fontenc}
\usepackage[utf8]{inputenc}
\usepackage{natbib}
\usepackage{comment}
\usepackage{framed}
\usepackage{todonotes}
\usepackage{wrapfig}
\usepackage{url}
\usepackage{pdfpages}
\makeatletter
\g@addto@macro{\UrlBreaks}{\do\/\do\-}
\makeatother

\usepackage{setspace}

\usepackage{sectsty}
\sectionfont{\sffamily\Large}
\subsectionfont{\sffamily\large}
\subsubsectionfont{\normalfont\large\itshape}

\usepackage{fancyhdr}

\hypersetup{
   colorlinks=true,
   linkcolor=[RGB]{30, 30, 180},
   citecolor=[RGB]{30, 30, 180},
   urlcolor=black,
   pdfborder=0 0 0,
   pdftitle={Safe and Certifiable AI Systems: Concepts, Challenges, and Lessons Learned},
   pdfsubject={}, 
   pdfkeywords={Certification, AI System Audit, Regulation, Functional Trustworthiness, Statistical testing, Robustness},
   pdfauthor={},%
   pdfstartview=FitH,
   breaklinks=true
}

\def\blankpage{
      \clearpage
      \thispagestyle{empty}
      \null
      \clearpage}

\renewcommand{\blankpage}{}

\begin{document}

\pagenumbering{gobble}

\includepdf[pages=1,fitpaper=true]{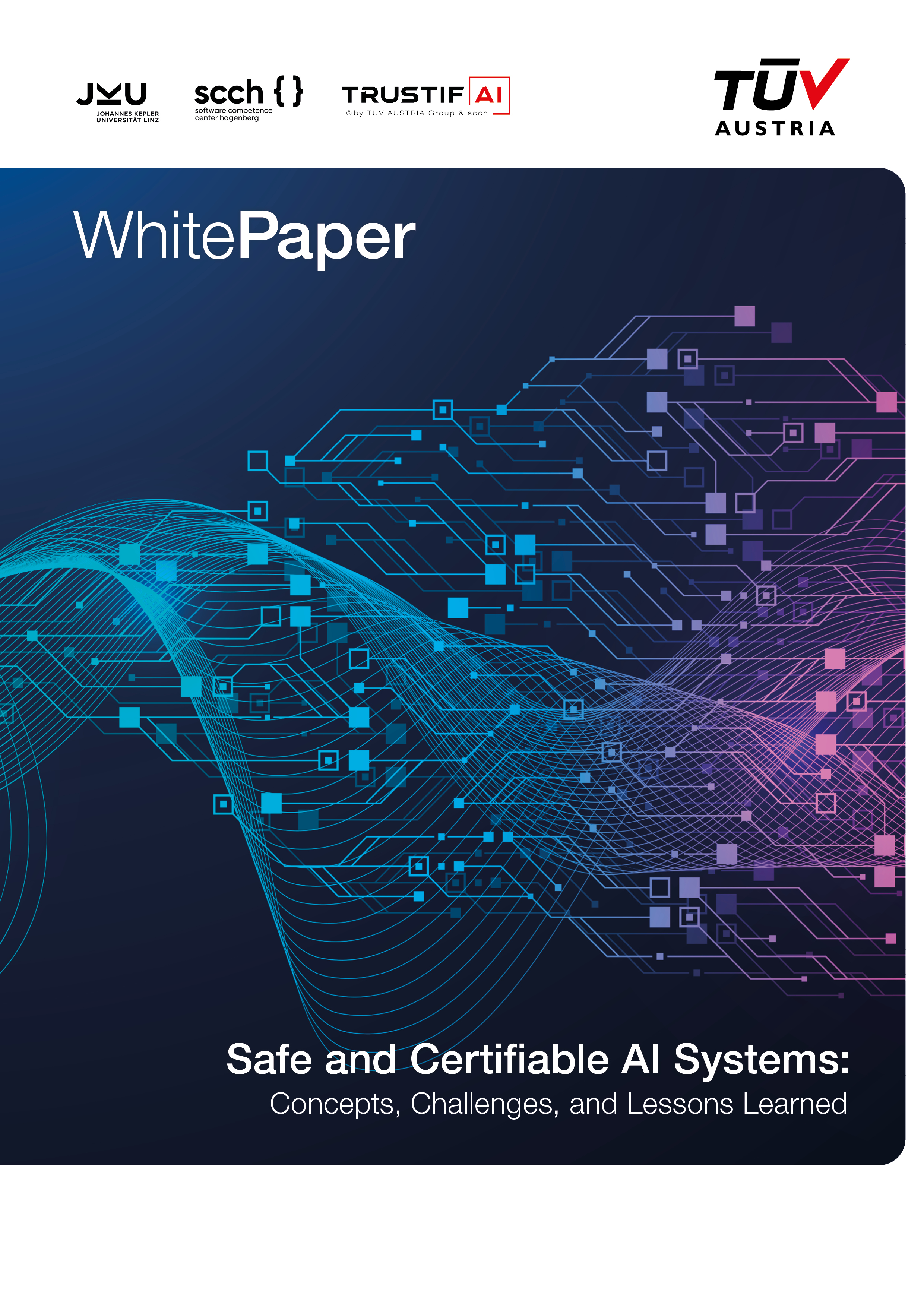}

\begin{titlepage}
\begin{center}
~~~~ 

\vfill

{\Huge Safe and Certifiable AI Systems:\\Concepts, Challenges, and Lessons Learned}

\vfill

\begin{centering}
\textit{Vienna, September 2025} \\[2em]

\textit{
Kajetan Schweighofer$^{4,*}$ \\
Barbara Brune$^{2,*}$ \\ 
Lukas Gruber$^4$ \\
Simon Schmid$^{3,4}$ \\
Alexander Aufreiter$^3$ \\
Andreas Gruber$^1$ \\
Thomas Doms$^{1,2}$ \\
Sebastian Eder$^2$ \\
Florian Mayer$^2$ \\
Xaver-Paul Stadlbauer$^2$ \\
Christoph Schwald$^2$ \\
Werner Zellinger$^4$\\
Bernhard Nessler$^3$ \\
Sepp Hochreiter$^4$ 
}
\end{centering}

\vspace{2cm}

\textbf{
$^1$ TRUSTIFAI GMBH\\ 
$^2$ TÜV AUSTRIA HOLDING AG \\ 
$^3$ Software Competence Center Hagenberg \\ 
$^4$ Johannes Kepler University Linz - Institute for Machine Learning \\
$^*$ Equal Contribution
}
\end{center}
\vfill

\textbf{Imprint:} \\
{\Large\textcopyright}~ TÜV AUSTRIA HOLDING AG, TÜV AUSTRIA-Platz 1, 2345 Brunn am Gebirge, Austria \\
\vspace{0.1cm}

Figures \\
{\Large\textcopyright}~ Figure~\ref{fig:ai_ml_dl}  elenabsl | Shutterstock.

\end{titlepage}

\newpage
\blankpage
~
\vfill

\begin{center}
\begin{minipage}{0.8\textwidth}
\begin{center}   
{\sffamily\huge \textbf{Abstract}}
\end{center}
\vspace{0.5cm}

There is an increasing adoption of artificial intelligence in safety-critical applications, yet practical schemes for certifying that AI systems are safe, lawful and socially acceptable remain scarce. 

\quad This white paper presents the \textit{TÜV AUSTRIA Trusted AI} framework an end-to-end audit catalog and methodology for assessing and certifying machine learning systems. 
The audit catalog has been in continuous development since 2019 in an ongoing collaboration with the Institute for Machine Learning at Johannes Kepler University Linz, which was further extended with the Software Competence Center Hagenberg and the resulting joint-venture TRUSTIFAI.

\quad Building on three pillars – Secure Software Development, Functional Requirements, and Ethics \& Data Privacy – the catalog translates the high-level obligations of the EU AI Act into specific, testable criteria. 
Its core concept of functional trustworthiness couples a statistically defined application domain with risk-based minimum performance requirements and statistical testing on independently sampled data, providing transparent and reproducible evidence of model quality in real-world settings. 
We provide an overview of the functional requirements that we assess, which are oriented on the lifecycle of an AI system.
In addition, we share some lessons learned from the practical application of the audit catalog, highlighting common pitfalls we encountered, such as data leakage scenarios, inadequate domain definitions, neglect of biases, or a lack of distribution drift controls. 

\quad We further discuss key aspects of certifying AI systems, such as robustness, algorithmic fairness, or post-certification requirements, outlining both our current conclusions and a roadmap for future research.
In general, by aligning technical best practices with emerging European standards, the approach offers regulators, providers, and users a practical roadmap for legally compliant, functionally trustworthy, and certifiable AI systems.

\end{minipage}
\end{center}
~
\vfill

\newpage 

\blankpage

\tableofcontents

\newpage
\blankpage 

\pagestyle{fancy}
\renewcommand{\headrulewidth}{0pt}
\fancyhead[LO,LE]{} 
\fancyhead[RO,RE]{}
\fancyhead[CO, CE]{\textit{Safe and Certifiable AI Systems: Concepts, Challenges, and Lessons Learned}}

\pagenumbering{arabic}

\section{Introduction}

Artificial Intelligence (AI) has become an integral part of our daily lives, powering everything from personalized recommendations to automated decision-making in healthcare, finance, and beyond. 
As AI applications become more sophisticated and ubiquitous, the question arises as to how we can ensure that they do not cause any harm?
In 2019, TÜV AUSTRIA launched an initiative to answer this question in collaboration with the world-renowned Institute for Machine Learning at Johannes Kepler University Linz\footnote{\href{https://www.jku.at/en/institute-for-machine-learning/}{https://www.jku.at/en/institute-for-machine-learning/}} headed by Sepp Hochreiter. 
This joint effort resulted in the publication of one of the world's first audit catalogs for trustworthy artificial intelligence \citep[see also][]{winterTrustedArtificialIntelligence2021}. 
Several audits of AI systems in production have been carried out by a team of auditors at TÜV AUSTRIA based on the developed criteria. In turn, those criteria were then further extended and improved on the basis of those practical auditing experiences. 
In 2022, the Hagenberg Software Competence Center joined this research and development effort as an additional partner with its applied research experience, and in 2023 the joint venture TRUSTIFAI was founded in Linz to further support the availability and distribution of the auditing competence of TÜV AUSTRIA in the market.

In response to the increasing adoption of AI systems in the European market, the EU is establishing a legal regulatory framework on AI.
In April 2021, the Commission proposed a draft of the EU AI Act that was revised following intensive negotiations between the European Parliament and the Council. 
The EU AI Act was adopted in 2024 \citep{EU_AI_Act_2024}, came into force on August~1st that same year, and will be implemented gradually with staggered transition periods. 
We have commented on, and sometimes criticized, this development process in our publications \citep{Nessler:23, Wendehorst:24, eli:2024guidelines-aisystem}. 
To further specify and clarify the provisions of the AI Act, additional interpretative guidance is being developed both at the national and EU level, including Commission guidelines on the definition of AI systems \citep{eucomm:2024guidelines-aisystem} and prohibited practices \citep{eucomm:2024guidelines-prohibited}, as well as the Living Repository of the Commission for AI literacy practices \citep{eucomm:2024literacy}. 

The EU AI Act is supplemented by a concerted standardization process conducted by CEN/CENELEC as requested and mandated by the EU \citep{eucomm:2023standardisationrequest} and is accompanied by a shared process in ISO. %
The AI Act provides only broad guidelines, and the standardization process that aims to further specify detailed requirements is very complex and still ongoing. Consequently, there is no detailed conformity assessment scheme yet.
We are monitoring the latest developments closely and aim to actively contribute to the EU AI Act's implementation. 
Our goal is to enrich the standardization process and the ongoing regulatory efforts with our theoretical background knowledge and practical expertise, and with the insights gained from our audit catalog as presented in this white paper.

By evaluating and certifying AI applications today, we aim to ensure that they are functional, safe, and robust in real-world settings, as well as compliant with emerging regulations.
The core of our approach to ensuring that the AI system fulfills its intended purpose is the assessment of its functional trustworthiness \citep{Nessler:23}. In order to achieve a statistically valid assessment of the performance of the AI system in its intended domain, we require
\begin{itemize}[noitemsep]
    \item A precise statistical definition of the application domain of the AI system
    \item Risk-based minimum performance requirements, including all quantitatively measurable functional properties of the model
    \item A statistical test of said performance requirements on independent samples from the application domain.
\end{itemize}
Building on this basis, we assess further properties, from the competence of the development team, the data gathering and modeling process, right through to the monitoring environment for the AI system, in order to detect potential issues and deficiencies. 

The contents of this white paper are as follows.
In Section~\ref{sec:background}, we provide a brief overview and describe important concepts in the field of AI.
In Section~\ref{sec:regulatory-background}, we briefly summarize the current state of the growing regulatory frameworks.
Section~\ref{sec:description-audit-catalog} introduces the three pillars on which our audit catalog is based: Secure Software Development, Functional Requirements, and Ethics \& Data Privacy. 
This is where we outline the structure of the audit catalog and introduce the central concepts to ensure the functional trustworthiness and safety of AI applications. 
Section~\ref{sec:details} then provides deeper insights into the essential aspects of the proposed certification scheme, including robustness, fairness, and representativeness of the test data. 
Finally, Section~\ref{sec:maintaining} addresses topics in monitoring and continuous updating of AI systems and Section~\ref{sec:conclusions} contains our conclusions.

\newpage

\section{Technical Background} \label{sec:background}

This section provides a brief introduction to the most important concepts in AI, focusing on Machine Learning and Deep Learning.
It includes an overview of current topics in Deep Learning, along with recent advances that are crucial to understanding the challenges involved in the certification of safety-critical AI applications.

\begin{figure}[b!]
    \centering
    \includegraphics[width=\linewidth]{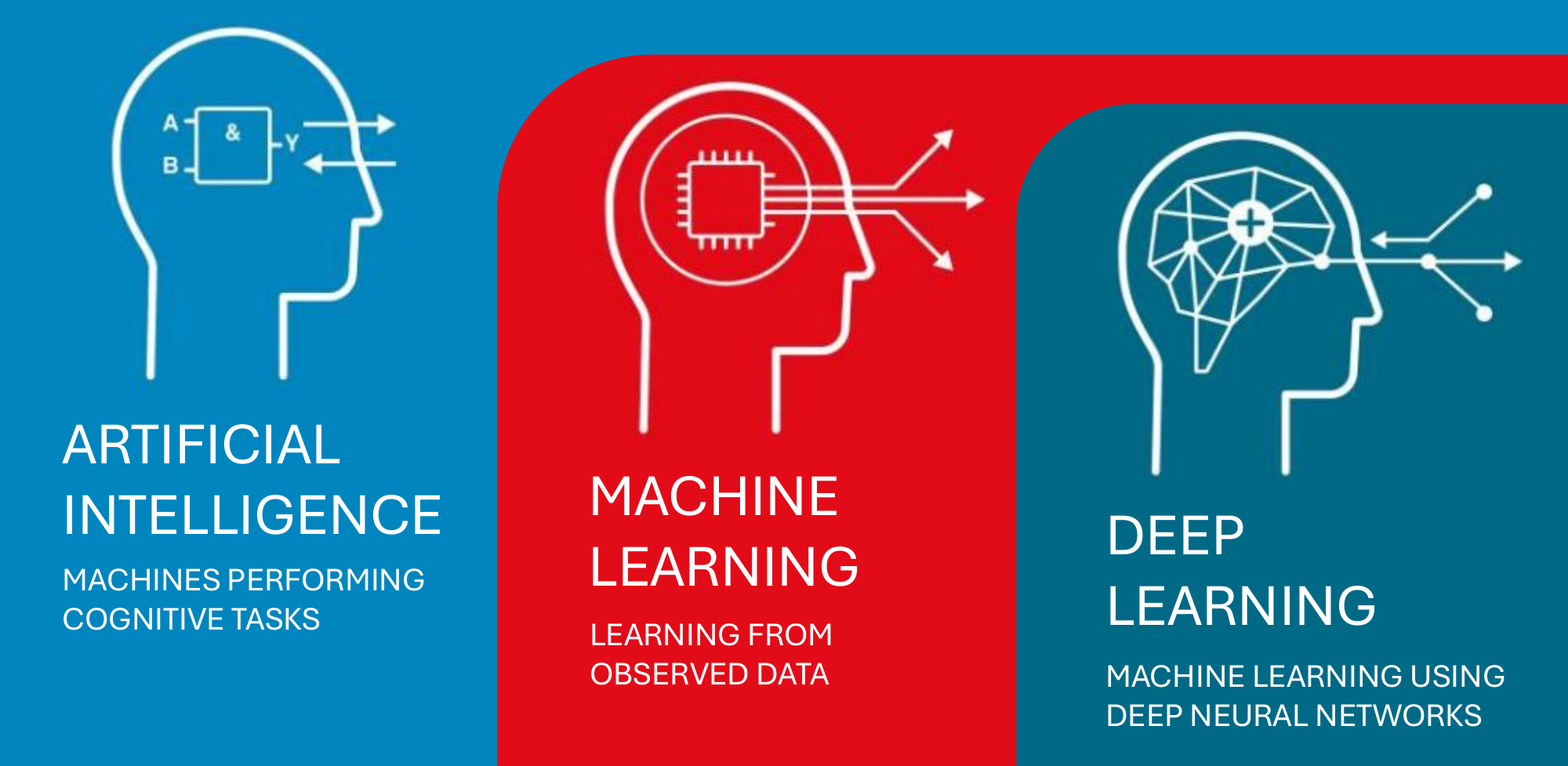}
    \caption{Deep Learning is a subfield of Machine Learning, which in turn is a subfield of Artificial Intelligence.}
    \label{fig:ai_ml_dl}
\end{figure}

\subsection{Artificial Intelligence with a Focus on Deep Learning}

Artificial Intelligence (AI) encompasses a broad range of techniques and systems designed to perform cognitive tasks commonly associated with human intelligence.
We focus on AI systems that are based on Machine Learning and its subfield, Deep Learning.
Figure~\ref{fig:ai_ml_dl} depicts the relationship between AI, Machine Learning, and Deep Learning. 
Machine Learning is often used synonymously for subsymbolic AI, which relies on rules, representations, and models that are learned from data. 
In contrast, symbolic AI comprises knowledge representations (expert systems and knowledge graphs), search methods, logic systems, formal methods, and similar.

We present a brief overview of AI in the following, focusing on Machine Learning (subsymbolic AI) and particularly on Deep Learning. This has been the technology behind many recent advancements in AI, including chatbots and image or video generation tools, used by billions of users nowadays.

\paragraph{Artificial Intelligence (AI):}

Research on intelligent machines began after the second world war, with pioneers such as Alan Turing and his test on the intelligence of a machine \citep{Turing:50}.
Later in 1956, the term ``Artificial Intelligence'' (AI) was coined by John McCarthy in 
the Dartmouth Summer Project on Artificial Intelligence \citep{McCarthy:55,McCarthy:07}.
A definition of AI was provided in \citet{Barr:81}: 
\begin{quote} 
“Artificial Intelligence (AI) is the part of computer science 
concerned with designing intelligent computer systems, that is, 
systems that exhibit characteristics we associate with intelligence in human behavior – 
understanding language, learning, reasoning, solving problems, and so on.”
\end{quote}

In contemporary discourse, AI refers to systems that can perform cognitive tasks typically attributed to humans, such as learning, planning, reasoning, problem solving, and skill acquisition. 
Research typically focuses on one or a few of these tasks.

\paragraph{Machine Learning (ML):}

Machine Learning (ML) refers to learning a model from available data (past experiences) that is then applied to future unknown data.
The term \emph{Machine Learning} was introduced in \citet{Samuel:59}.
Contrary to deductive reasoning, where specific conclusions are derived from general principles, ML relies on inductive reasoning, where general conclusions are derived from specific observations, i.e., learning from observed data.
A definition of ML was given by \citet{Mitchell:97}:

\begin{quote}
    ``A computer program is said to learn from experience E with respect to some class of tasks T, and performance measure P, if its performance at tasks in T, as measured by P, improves with experience E.''
\end{quote}

The performance on future data is called ``generalization'', in other words, how well a model generalizes from past data to future data.
The quality of the learned model usually depends on the complexity of the problem and the amount of data available.
The explosive growth in the amount of data available in many domains is thus a huge factor to the success of ML, alongside methodological advances.

\begin{figure}[h]
    \centering
    \includegraphics[width=\linewidth]{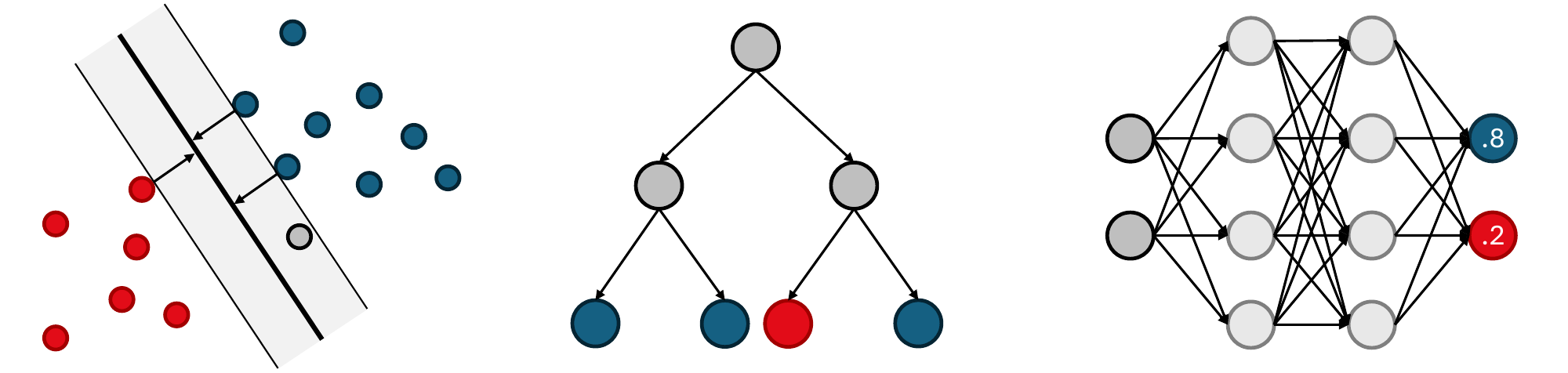}
    \caption{Common ML models: Support vector machines, decision trees and neural networks.}
    \label{fig:ml_algorithms}
\end{figure}

ML approaches have historically been categorized into three types of learning settings: supervised, unsupervised, and reinforcement learning, depending on the available information.
We will introduce these in the next subsection.
At present, these areas are no longer as disjoint from each other, and many subareas such as weakly supervised, semi-supervised, self-supervised, or active learning have been established.
Moreover, ML techniques from different subareas are often combined for practical applications.
There are many methodologies that can be applied in these learning scenarios.
Over the years, methods such as decision trees, support vector machines, or Gaussian processes have been introduced, see Figure~\ref{fig:ml_algorithms}.
However, the main methodology driving the current AI boom are multi-layered (deep) artificial neural networks, which are referred to as Deep Learning \citep{LeCun:15, Schmidhuber:15} models.

\paragraph{Deep Learning (DL):}

The term \emph{Deep Learning} (DL) was introduced in \citet{Aizenberg:00}, while DL became
popular with layer-wise pretraining \citet{Hinton:06, Bengio:06, Ranzato:06}.
DL is a subcategory of ML that uses multi-layered artificial neural networks.
A more general definition was given by \citet{LeCun:15}:

\begin{quote}
    ``Deep-learning methods are representation-learning methods with multiple levels of representation, obtained by composing simple but non-linear modules that each transform the representation at one level (starting with the raw input) into a representation at a higher, slightly more abstract level.''
\end{quote}

\begin{figure}[h]
    \centering
    \includegraphics[width=\linewidth]{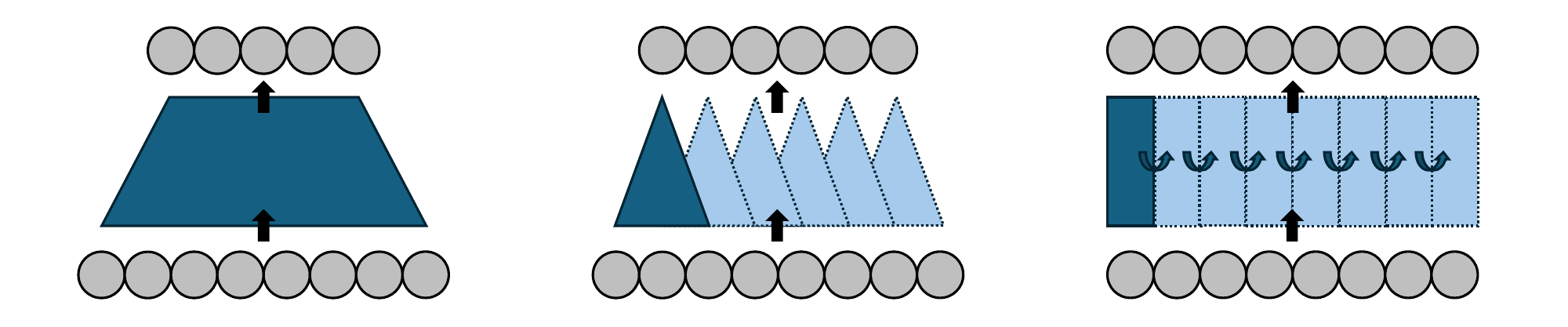}
    \caption{FCNNs (left), CNNs (middle) and RNNs (right). The kernel of CNNs (dark blue) is shared and applied to the whole input sequence. Input and output weights of an RNN are also shared over the whole sequence (dark blue) and additionally a memory that is updated when iterating over an input sequence.}
    \label{fig:neural_net}
\end{figure}

Multi-layered artificial neural networks (ANNs) are the typical framework for implementing deep learning (DL) models. 
These models are composed of layers of interconnected neurons and can be categorized into fully-connected neural networks (FCNNs) \citep{Ivakhnenko:65}, convolutional neural networks (CNNs) \citep{Fukushima:80, LeCun:89}, or recurrent neural networks (RNNs) \citep{Jordan:86, Elman:90}, see Figure~\ref{fig:neural_net}. 
These architectures encode different inductive biases, model assumptions that guide learning, and have been designed for different data modalities, such as spatial or sequential inputs.

Each ANN consists of an input layer, one or more hidden layers, and an output layer. 
Non-linear activation functions are applied after each layer to increase the representational capacity of the network. 
Neurons in one layer receive inputs only from the previous layer and send outputs only to the next (see Figure~\ref{fig:ml_algorithms} right). 
The connections between neurons have associated weights and the function computed by the network is determined by the values of these weights. 
Learning in DL models typically uses variants of stochastic gradient descent \citep{Robbins:51,Kiefer:52,Rumelhart:86} to iteratively update these weights based on the prediction error over many training passes (epochs).
Typically, early layers tend to learn simple features, such as edges and textures, while deeper layers learn increasingly abstract representations, such as object parts, the eye of a dog, or the wheel of a car \citep{Zeiler:14}. 

Advanced architectural innovations further improve performance.
CNNs exploit local connectivity and weight sharing, making them particularly efficient for structured inputs such as images. 
RNNs, which include loops in their architecture, are capable of modeling temporal dependencies by compressing the history into a hidden state. 
Variants such as Long Short-Term Memory networks (LSTMs) \citep{Hochreiter:91,Hochreiter:97} address issues such as the vanishing gradient problem \citep{Hochreiter:91}, enabling the learning of long-range dependencies in sequential data.
In addition, design principles such as residual connections \citep{He:16}, which add shortcut paths between layers, help mitigate training difficulties in very deep networks by facilitating gradient flow. 
The attention mechanism \citep{Bahdanau:15} allows models to dynamically focus on the most relevant parts of the input during prediction. 
Those principles led to the transformer architecture \citep{Vaswani:17} which is the basis of current successes in language generation \citep{Brown:20}.

\subsection{Types of Learning Settings}

ML approaches, and thus also DL approaches as their subset, can be categorized into three different groups by the nature of the available learning signal.

\paragraph{Supervised learning:}

\begin{figure}[t]
    \centering
    \includegraphics[width=\linewidth]{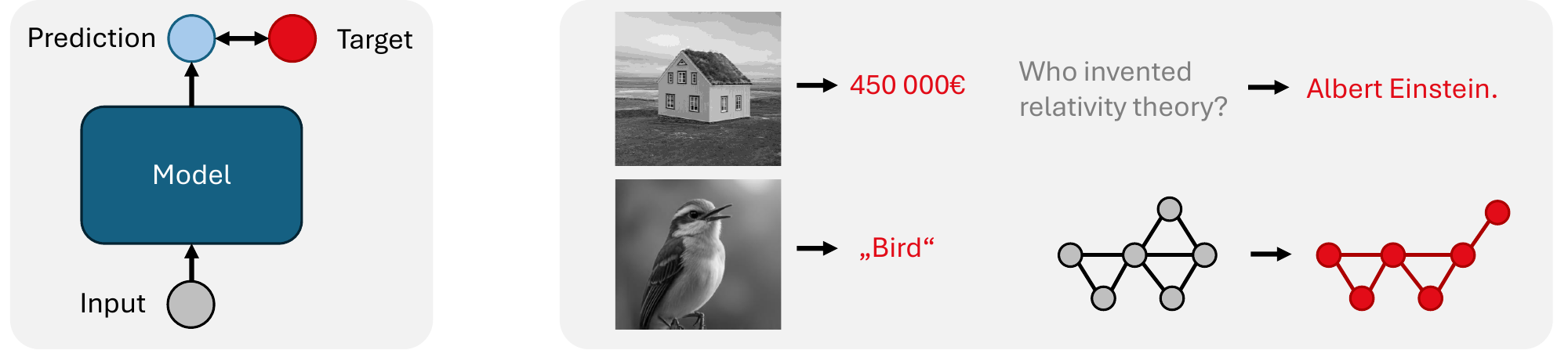}
    \caption{Supervised learning: each input has a target which the model aims to predict (left). Different supervised learning tasks differ by the nature of the target value (right). Input (gray) and output (red) can have different structures, for example the input is an image and the target is a price or class label. However, they can also have the same structure, for example both input and target are text or graphs.}
    \label{fig:sl}
\end{figure}

Supervised learning is the ML task of learning a function (the model) that maps an input to an output based on a given training set of inputs and their targets, also called desired outputs or labels \citep{Russell:09}, see Figure~\ref{fig:sl} left panel.
The name ``supervised'' refers to the fact that when the model processes an input, the quality of its output can be measured against the target value. Thus, there is direct supervision for each prediction.
Formally, the objective of supervised learning is to minimize the risk $R$ of predicting with a model $f: \mathcal{X} \to \mathcal{Y}$ that maps the input space $\mathcal{X}$ to the output space $\mathcal{Y}$.
The risk is defined as the expected loss over the data distribution:
\begin{align} \label{eq:risk}
    R(f) \ &= \ \int_{\mathcal{X}} p(x) \int_{\mathcal{Y}} L(f(x), y) \ p(y \mid x) \ \mathrm{d}y \ \mathrm{d}x \ .
\end{align}
Here, $L$ is a real-valued loss function that compares the output of the model with the target, and $p(x,y) = p(y \mid x) p(x)$ is the data distribution.

Supervised learning tasks can be categorized according to the nature of their targets (see Figure~\ref{fig:sl} right).
These can be unstructured outputs such as single discrete values (classification), single continuous values (regression), or sets (e.g. point clouds).
They can also be structured outputs \citep{Bakir:07}, for example, sequences (e.g., text and time series), graphs (e.g., molecules and social networks), trees (e.g., syntactic parse trees), images or videos.

Given a training set, supervised learning consists of selecting a specific model from a model class (also called hypothesis class), which may be parametric or non-parametric.
A typical example is an artificial neural network with a specific architecture, where the learnable parameters are the connection weights.
The training process involves selecting the best parameters for this model class using the training data, with the goal of achieving a good performance on future unseen data (good generalization).
Model selection may be based on direct analytical solutions (e.g., in naive Bayes classifiers or linear regression), or more complex procedures such as convex optimization (e.g., support vector machines) and gradient-based stochastic methods (e.g., deep learning).
Supervised learning methods typically assume that the data used for training and future prediction are drawn from the same underlying distribution.
Most methods rely on the i.i.d.\ assumption, which means that all data samples are independent and identically distributed, in other words, sampled independently from the same (static) real-world process.
This assumption allows to estimate the expected performance on future data using test data that was completely held out and unseen during model development.

Variations of supervised learning include weakly-supervised learning\footnote{\href{https://ai.stanford.edu/blog/weak-supervision/}{https://ai.stanford.edu/blog/weak-supervision/}} and semi-supervised learning \citep{Chapelle:10}. 
They are used if noisy, partial, or imprecise labels are given or when only a small portion of the data is labeled due to costly and time-consuming labeling processes.
A very important variant of supervised learning is {\em self-supervised learning},\footnote{\href{https://ai.meta.com/blog/self-supervised-learning-the-dark-matter-of-intelligence/}{https://ai.meta.com/blog/self-supervised-learning-the-dark-matter-of-intelligence/}} which is at the heart of the recent successes of large-scale ML models.
Here, the target is constructed from the input itself, bypassing the need for costly labeling.
This is easily possible in time series data, where the target can be automatically created using future timepoints.
The same holds true for language, where the next word (token) is predicted, given the text so far.
Similarly, predicting masked parts of the input is also a widely used self-supervised learning objective \citep{Devlin:19}.
For language, this means predicting a word given the surrounding sentence, e.g. ``The \rule{0.5cm}{0.4pt} sleeps on the couch'', where ``cat'' needs to be predicted.
Both can also be applied to image data, i.e. for next pixel prediction \citep{vandenOord:16a, vandenOord:16b} or to predict masked parts of the image \citep{He:22}.
Although models trained in such a fashion learn very rich features of the training data, they often cannot be used for a specific application.
However, these foundation models (see Section~\ref{subsec:foundation_models}) can be fine-tuned on a smaller, task-specific training dataset in the usual supervised learning setting.

\paragraph{Unsupervised learning:}

Unlike supervised learning, which relies on labeled data, unsupervised learning works with unlabeled data. 
Unsupervised learning methods are designed to uncover structure within the data, create more compact or useful representations, or model the underlying data-generating process. 
In contrast to supervised tasks, where model quality is evaluated on the basis of individual predictions, unsupervised models are typically assessed based on their performance across the entire dataset or subsets of it. 
This results in more complex objective functions, making it harder to determine how individual data points contribute to the overall objective and how their processing should be improved.
While supervised methods are used to predict on future data, unsupervised methods allow to explore, visualize, compress, or find structure in the data. 
Thus, unsupervised methods can help to understand the data and generate new knowledge. 
Unsupervised methods can be grouped into recoding methods
and generative methods.%

\begin{figure}[h]
    \centering
    \includegraphics[width=\linewidth]{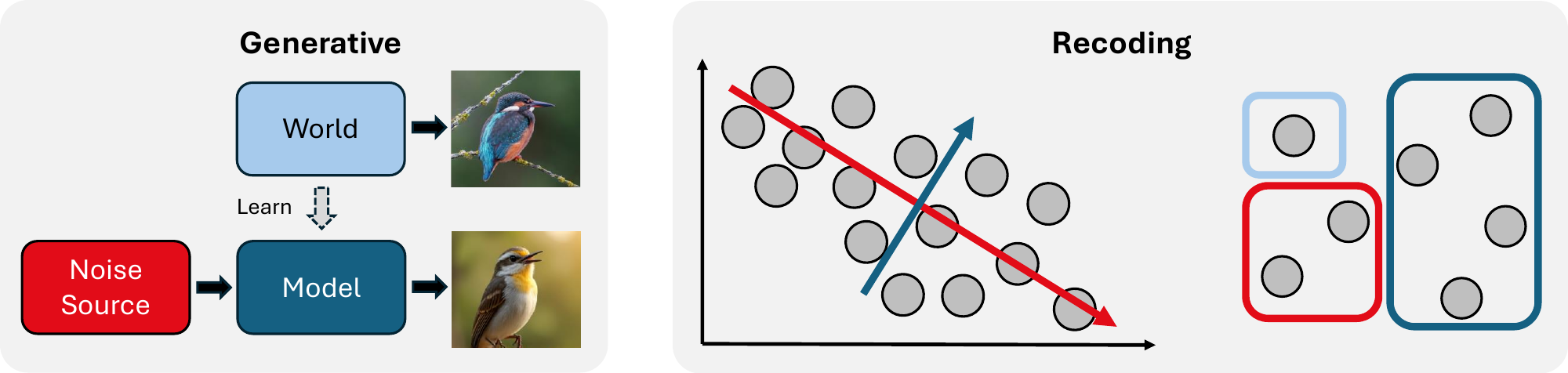}
    \caption{Unsupervised learning: Generative methods replicate the data-generating sampling process (left); recoding methods include dimensionality reduction (e.g., PCA, middle) and clustering (right).}
    \label{fig:usl}
\end{figure}

Recoding methods create new representations of objects based on their original feature vector representations. 
They typically reduce or compress these vectors into a lower-dimensional space to eliminate redundancy and irrelevant components. 
Common projection techniques include principal component analysis \citep{Hotelling:33,Watanabe:65,Oja1982,Jolliffe:86,Jackson:91}, independent component analysis \citep{jutten1991blind,Comon:94,Bell:97,Bell:97nips}, factor analysis \citep{Watanabe:65,Joreskog:67,Rubin:82,harman1976modern}, and projection pursuit \citep{kruskal1969toward,Friedman:74,Huber:85,Jones:87,Friedman:81}.
Cluster analysis (see Figure~\ref{fig:usl}, middle) is a key area within recoding methods, primarily used as an analytical tool in exploratory data analysis. 
It can be defined as follows:\footnote{\href{https://en.wikipedia.org/w/index.php?title=Cluster_analysis}{https://en.wikipedia.org/w/index.php?title=Cluster\_analysis}}

\begin{quote}
    ``Cluster analysis or clustering is the task of grouping a set of objects in such a way that objects in the same group (called a cluster) are more similar or more closely connected (in some sense) to each other than to those in other groups (clusters).''
\end{quote}

Since there is no universal definition of similarity or connectedness, the specific objective of clustering must be tailored to each application.
Examples of clustering methods \citep{Hartigan:72,Hartigan:75,Anderberg:73,Forgy:65,Buhmann:95} include k-means \citep{Hartigan:79,lloyd1982least, macqueen1967some}, hierarchical clustering \citep{ward1963hierarchical,gower1969minimum,Cormack:71,Anderberg:73,Hartigan:75,gowda1978agglomerative, milligan1979ultrametric,Gordon:87,nielsen2016hierarchical}, mixture models \citep{Hasselblad:66,McLachlan:88,marin2005bayesian}, and self-organizing maps \citep{Kohonen:82,Kohonen:95book,Kohonen:2007}, where the latter combine
down-projection with clustering.

Generative models aim to replicate the data-generating process, i.e., they attempt to model the real world, at least within a specific application domain. 
Mathematically, the real world is viewed as a stochastic process that produces data samples according to an underlying probability density distribution. 
Generative unsupervised learning seeks to construct a probabilistic model that closely approximates this distribution, enabling the generation of new synthetic data with similar statistical properties.
The goal is to create a model whose generated data points have a density that matches the observed data. 
The data generation process may also involve input variables or random components, which can be incorporated into the model. 
A key aspect of the generative approach is embedding as much prior knowledge or desired structural properties into the model as possible, thereby constraining the set of plausible models that could explain the observed data.
Examples of generative unsupervised learning methods include density estimation techniques (e.g., kernel density estimation \citep{Rosenblatt:56, Parzen:62}, Gaussian mixtures \citep{Hasselblad:66,marin2005bayesian}), hidden Markov models \citep{Baum:66}, and belief networks \citep{pearl1985bayesian}, which are often subsumed under Markov networks or Markov random fields. 
More advanced approaches include restricted Boltzmann machines \citep{smolensky1986information, Hinton:2007}, (modern) Hopfield networks \citep{ramsauer2021hopfield}, neural network auto-associators \citep{Oja:91,Kramer:91,Hochreiter:99iscas,Hochreiter:99nc}, variational autoencoders \citep{kingma2014autoencoding}, generative adversarial networks \citep{Goodfellow:14}, and normalizing flows \citep{rezende15}.
Recently, generative models have greatly improved at generating high-quality images \citep{Rombach:22, Esser:24} using diffusion \citep{Song:21} and flow matching \citep{Lipman:23}.
Video generation is one of the current frontiers, where many big IT companies have introduced such models.\footnote{\href{https://openai.com/sora/}{https://openai.com/sora/}}\footnote{\href{https://deepmind.google/models/veo/}{https://deepmind.google/models/veo/}}

\paragraph{Reinforcement learning:}

The reinforcement learning (RL) setting is different from both classical supervised and unsupervised learning tasks.
\citet{Sutton:18} give the following definition of RL:

\vspace{0.2cm}
\begin{quote}
    ``Reinforcement learning is learning what to do --- how to map situations to actions --- so as to maximize a numerical reward signal. The learner is not told which actions to take, but instead must discover which actions yield the most reward by trying them. In the most interesting and challenging cases, actions may affect not only the immediate reward but also the next situation and, through that, all subsequent rewards. These two characteristics --- trial-and-error search and delayed reward --- are the two most important distinguishing features of reinforcement learning.''
\end{quote}
\vspace{0.2cm}

\begin{wrapfigure}{r}{0.32\textwidth} 
   \vspace{-0.8cm}
    \includegraphics[width=\linewidth]{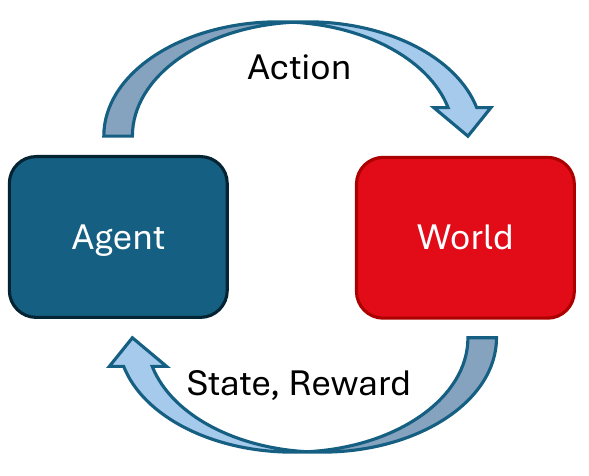}
    \caption{The agent-environment interaction loop in RL.}
    \label{fig:rl}
\vspace{-0.5cm}
\end{wrapfigure}

In RL, an agent selects actions based on the current state, receives a reward from the environment, and transitions to a new state, see Figure~\ref{fig:rl}. 
The agent's performance is typically measured by the return, the cumulative reward collected over an episode. 
This setting is challenging due to the often sparse and delayed nature of the reward signal. 
For example, watering a garden daily may only yield a harvest months later, making the connection between actions and outcomes difficult to learn.
Additionally, training an RL agent from scratch in a large state-action space can be computationally expensive and inefficient. 
In the garden example, random exploration is unlikely to succeed under extreme conditions such as a heat wave.

A common RL approach is to learn a policy, a mapping from states to actions that directly maximizes expected return. 
Alternatively, many methods focus on learning a world model that predicts future states or rewards. 
This often happens implicitly by learning a value function that estimates the expected future return from that state onwards. 
The agent can then act greedily with respect to this function.

The big success of RL has been to surpass human performance in challenging games such as Atari \citep{mnih2015human}, StarCraft \citep{vinyals2019grandmaster}, Chess or Go \citep{schrittwieser2020mastering}.
Furthermore, it has been used to control data center cooling, reduce energy consumption\footnote{\href{https://deepmind.google/discover/blog/deepmind-ai-reduces-google-data-centre-cooling-bill-by-40/}{https://deepmind.google/discover/blog/deepmind-ai-reduces-google-data-centre-cooling-bill-by-40/}} or to control the plasma shapes in a Tokamak fusion reactor \citep{degrave2022magnetic}.
It has also been applied to learn robotic control in various tasks \citep{OpenAI:19}.
It is important to note that a good and fast simulator is crucial for RL applications, as training in the real-world is often risky, slow, and expensive.
If the final RL policy is not applied in the simulator, sim-to-real transfer is the dominant approach, where the policy is trained in a simulation and fine-tuned using real-world data \citep{Tobin:17}.
RL techniques have also recently been applied to large language models (LLMs).
Here, the state space are the possible output sequences and the action of the next selected token.
Possible reward signals currently considered are (1) human preferences approximated by another neural network \citep{Ouyang:22}, and (2) verifiable rewards such as whether the generated code compiles and passes all unit tests or whether the answer to a math question is correct \citep{deepseekr12025, Wang:25}.

\subsection{On the Shoulders of Giants: Foundation Models} \label{subsec:foundation_models}

Most classical ML approaches consider a dataset and a learning algorithm.
The ML model is then learned from an arbitrary starting condition, often an educated guess or randomly drawn according to a distribution.
Another approach is to not train DL models from scratch, but fine-tune them with task-specific data starting from a model that was pre-trained on a more general task.
For instance, when building a model to distinguish two breeds of dogs, one can start with a network that is pre-trained to distinguish different animals and therefore provides a foundation of general visual features, such as recognizing shapes, textures, and patterns.
This pre-trained network can then be specialized to identify the nuances between specific dog breeds.

\begin{figure}
    \centering
    \includegraphics[width=\linewidth]{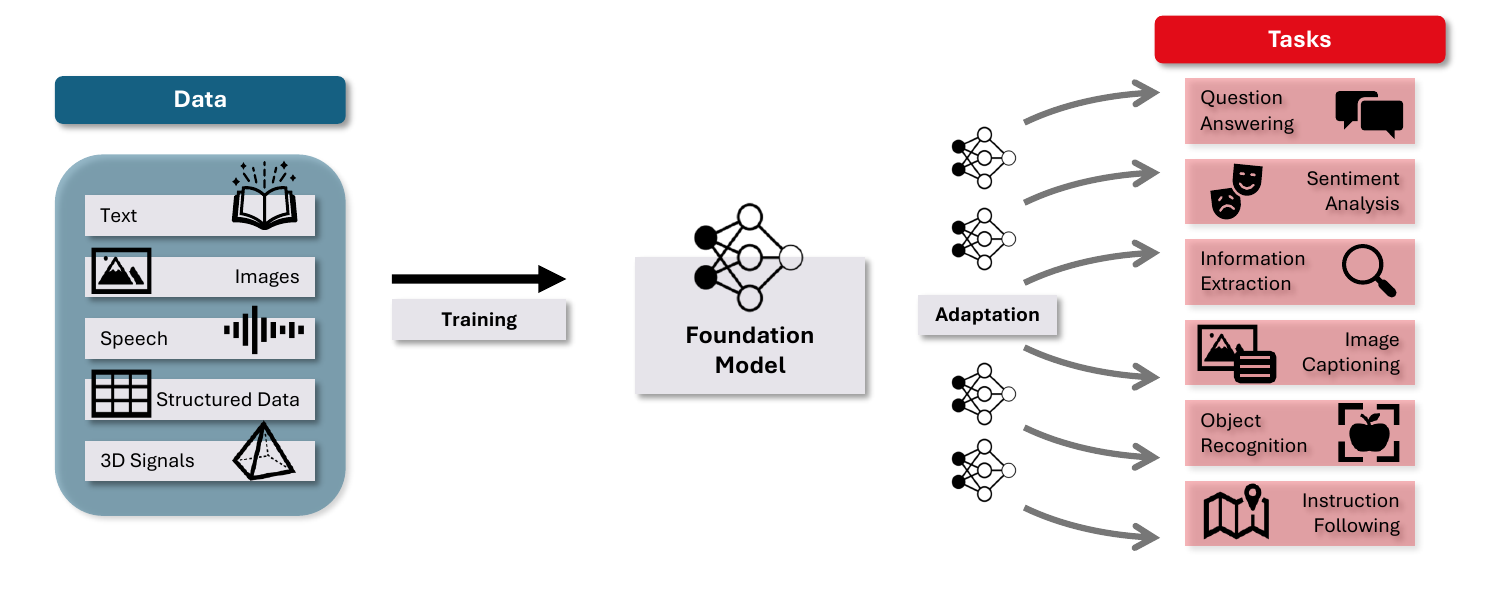}
    \caption{A foundation model is trained on large-scale data, centralizing general knowledge of a domain. The foundation model can then be adapted to various specific tasks. Figure recreated after \citet{bommasani2021opportunities}.}
    \label{fig:foundation_model}
\end{figure}

This trend of building on general models has recently evolved further with the emergence of foundation models. These are highly versatile models that can be adapted to a wide range of specific tasks (see Figure~\ref{fig:foundation_model}). Foundation models can be easily fine-tuned for downstream applications, enabling their efficient reuse across different tasks. Moreover, many prominent foundation models support in-context learning, where adaptation occurs by modifying parts of the input.
For example, in LLMs, exemplary question-answer pairs or task-specific information are provided \citep{Brown:20}, often also referred to as prompting.
Moreover, retrieval augmentation enables the dynamic enrichment of foundation models with task-specific information, without the need for explicit fine-tuning. In the following, we introduce the concept of foundation models, explore key techniques for fine-tuning, and discuss the capabilities offered by in-context learning and retrieval augmentation.

\paragraph{Foundation model:}

The term foundation model was coined by the Stanford Center for Research on Foundation Models \citep{bommasani2021opportunities}.
Their definition of a foundation model is as follows:

\begin{quote}
    ``A foundation model is any model that is trained on broad data (generally using self-supervision at scale) that can be adapted (e.g., fine-tuned) to a wide range of downstream tasks''
\end{quote}

Foundation models are typically large and complex, trained using self-supervised learning on vast amounts of data. Training such models is computationally expensive and resource intensive. As a result, training them from scratch is generally infeasible for most academic research groups.
Big IT companies tend to have a monopoly on foundation models as they are the only ones who can afford to make them.
However, the adaptation or fine-tuning of a foundation model to specific tasks can be carried out by academic groups and smaller companies.
Currently, the best known foundation models are LLMs \citep{Devlin:19, Brown:20}, but vision \citep{Alayrac:22}, time series \citep{Ansari:24, Auer:25}, or audio \citep{Radford:22} foundation models have also been constructed.

In the following, we give some examples of foundation models.
{\em Large language foundation models} are, for example,
Megatron-LM (NVIDIA) \citep{Shoeybi:19},
GPT3 (OpenAI) \citep{Brown:20,Brown:20arxiv},
Gopher (DeepMind) \citep{Rae:21},
GLaM (Google) \citep{Du:21glam},
Chinese M6 (Alibaba) \citep{Lin:21},
mutilingual AlexaTM 20B (Amazon Alexa AI) \citep{Soltan:22},
the more efficient OPT (Meta AI) \citep{Zhang:22opt},
Chinchilla (DeepMind) \citep{Hoffmann:22},
LaMDA (Google) \citep{Thoppilan:22}, or
PaLM (Google) \citep{Chowdhery:22}.
LaMDA attracted media attention after a tester claimed that it might be conscious. 
PaLM can explain jokes and perform inference chaining, in other words,
it explains step by step why something is funny and
can perform deduction step by step.
The large language model that arguably had the most impact 
on the research community and also on the public is ChatGPT \citep{Schulman:22}.
The Llama model series (Meta AI) \citep{Touvron:23, Touvron:23b, Dubey:24} has revolutionized industry adoption of language models, due to its open weights, permissive license, and strong performance.
{\em Large vision foundation models} are very prominent as well. 
For example,
Dino (Facebook/Meta) \citep{Caron:21} is a
semantic segmentation model
that can discover and segment objects in an image or a video
with absolutely no supervision and without being given a target segmentation.
Another foundation model for computer vision is Florence (Microsoft) \citep{Yuan:21}.
CLIP (OpenAI) \citep{Radford:21,Radford:21arxiv} is
a visual model that is trained with natural language supervision.
Flamingo (DeepMind) \citep{Alayrac:22} is 
a vision language model that can handle
multiple tasks depending on the input. 
DALL-E 2 (OpenAI) \citep{Ramesh:22} became the new standard
for text-guided image generation, also garnering much public attention.
Gato (DeepMind) \citep{Reed:22} is a
large multi-task (e.g., robotics, language, vision) model serving as a
generalist policy that can play Atari, caption images, chat, stack blocks with
a real robot arm and much more.

\begin{figure}[b]
    \vspace{0.5cm}
    \centering
    \includegraphics[width=\linewidth]{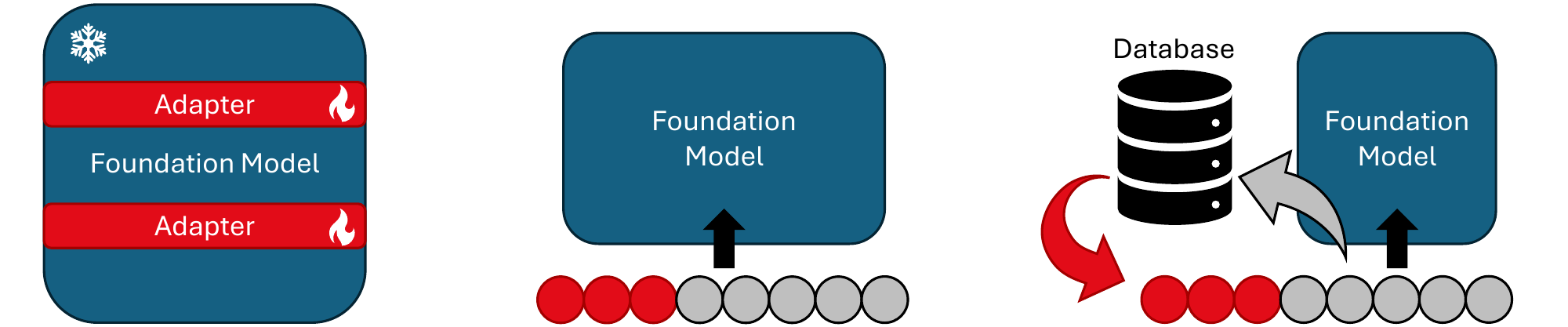}
    \caption{Utilizing a foundation model for specific tasks through fine-tuning, in-context learning or retrieval augmentation.}
    \label{fig:ft_icl_rag}
\end{figure}

\paragraph{Fine-tuning:}
Fine-tuning in DL refers to the practice of utilizing knowledge incorporated in a pre-trained neural network and adapting it to a specific task.
Often, only lower layers are kept, which correspond to more general purpose low-level features, while higher layers are adapted to or even trained from scratch for a specific task.
Fine-tuning is performed on a smaller dataset than pre-training,
thus, fine-tuning should be parameter-efficient to avoid overfitting.
Consequently, the so-called adapters \citep{Houlsby:19} or their low-rank version \citep{Hu:22} were very successful in fine-tuning.
These methods leave the original network unchanged and introduce additional parameters in a multiplicative or additive way, which are the only parameters adjusted during fine-tuning.
Parameter-efficient fine-tuning also allows the efficient storage of a multitude of task-specific models, where only a single large foundation model and a set of small adapters per task have to be stored.

RL was recently introduced into the process of fine-tuning LLMs to make them more useful \citep{Ouyang:22}.
While the foundation model is pre-trained on a large corpus of text, its outputs often do not correspond to human preferences, such as relatedness of the answer to the input question, politeness or refusal to respond to certain input questions.
In this case, proxy models for human preferences are learned from a large amount of labeled conversation data and turned into a reward model.
Standard RL algorithms are then used to optimize the LLM responses such that the reward is maximized.

\paragraph{In-context learning:}

In-context learning refers to a model's ability to adapt and perform tasks based solely on the context provided in its input, without additional fine-tuning \citep{Brown:20}.
The ``context'' can be related information, constraints, knowledge, settings, but also examples or demonstrations of a new task.
Widespread techniques include providing a verbal description of a task or providing some examples of how the task is solved for specific problem instances.
This approach leverages the model's pre-trained knowledge to generalize across diverse tasks, making it flexible and efficient. 
Although widely regarded in LLMs, other non-language foundation models are also amenable to in-context learning, e.g. in time-series prediction \citep{Ansari:24, Auer:25}.
In-context learning is closely related to ``learning to learn'' or meta-learning, which
allows a model to solve new tasks without changing its parameters. 
If it is infeasible to train each task from scratch, one model should 
learn how to learn new tasks from a few given examples, 
rather than learning each new task separately.

\paragraph{Retrieval augmentation:}

Retrieval augmentation \citep{Lewis:20} enhances a model’s capabilities by integrating external information from a database or knowledge source during inference. 
Instead of relying solely on pre-trained knowledge stored in its parameters, 
the model retrieves relevant data from external sources to improve its responses.
Retrieval augmentation enhances accuracy and adaptability, 
especially for tasks that need to be up-to-date or have detailed information. 
Furthermore, it makes the use of foundation models much more practical.
For example, instead of costly fine-tuning an LLM based on company internal documentation, a search index is created based on the text embedding to the respective document, and documents relevant to a specific input to the LLM are retrieved based on the similarity of the embedded text
and used to provide a meaningful output.
This allows new knowledge to be made available for the foundation model and quickly adapt it to specific tasks.

\subsection{Research Frontiers}

Finally, we would like to provide a brief overview of current research trends, outlining what we believe could become the next generation of systems that need to be certified.

\begin{figure}[t]
    \centering
    \includegraphics[width=\linewidth]{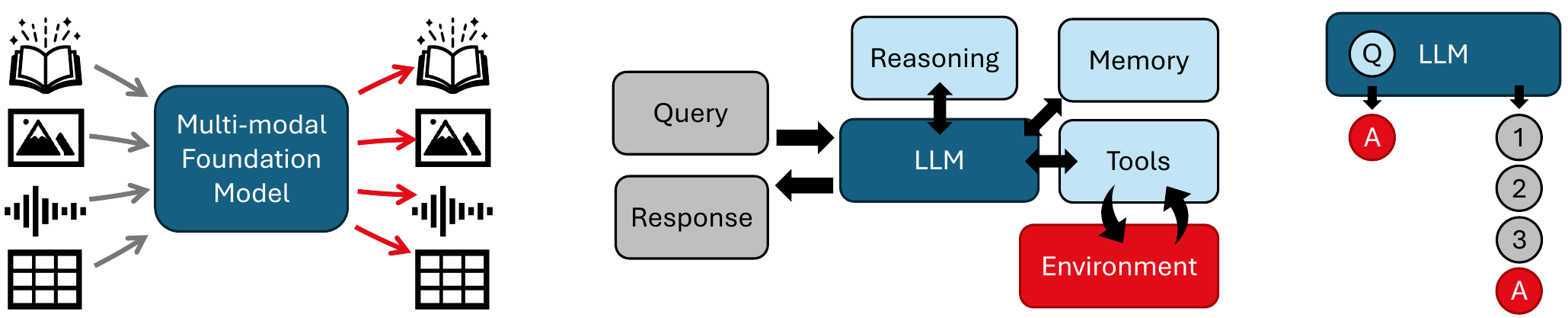}
    \caption{Multi-modal foundation models can process one or multiple input modalities and output them in one or multiple output modalities (left). High-level structure of agentic LLM systems. For certain input queries, they can call tools such as a search engine or a python interpreter, have access to a memory system such as a vector database and can call reasoning subroutines in order to obtain a response (middle). The reasoning LLM system generates intermediate steps before providing an answer to a question. The quality of the final answer usually improves the more computing power is used to obtain such intermediate steps (right).}
    \label{fig:research_frontiers}
\end{figure}

Firstly, there is a trend towards multi-modality of foundation models, meaning that the model can operate on more than one data type, e.g. images, speech and text, similar to how humans can process and operate on different sensory inputs.
Secondly, agentic AI systems are emerging that are designed to perform complex tasks autonomously by breaking them into sub-tasks.
Agentic AI systems exhibit autonomous decision-making, goal-directed behavior, and adaptability, allowing them to perform complex tasks.
They typically combine LLMs with agents that can act on different platforms and systems. 
Thirdly, there are attempts to train LLMs to generate longer intermediate outputs before providing a final answer for certain tasks, such as mathematical reasoning.
These reasoning systems are tuned to break down a problem into intermediate steps tackled one after another and perform self-correction if necessary to finally yield correct answers.
Figure~\ref{fig:research_frontiers} shows these three types of systems.

\paragraph{Multi-modal foundation models:}

Multi-modal foundation models process and integrate data from multiple input types, such as text, images, audio, or video. 
These can be used to obtain a description of the data in a different modality, e.g. a caption for a given image or an image for a given caption.
One of the first widespread successes of multi-modal models was CLIP \citep{Radford:21}, a vision-language model.
Similarly, Gato \citep{Reed:22} is a generalist agent trained to work with many different input modalities such as speech, images, joint moments, or button presses and to output in a suitable modality.
These models are very useful as search engines, for example, allowing a private person using CLIP to search for beach pictures on their mobile phone, or in industry using CLOOME \citep{Sanchez-Fernandez:23} to search for bioimages with chemical structures.
By early 2025, multimodal models were widely used by many innovative AI companies.
Speech, image, and text can be used as input to generate an output text, image or video, potentially adding more capabilities in the future.

\paragraph{Agentic AI systems:}

Agentic AI systems are designed to autonomously solve complex tasks by breaking them into easier sub-tasks.
Such systems use iterative steps to plan, execute, and evaluate actions, often guided by user-defined goals.
They can integrate tools, retrieve information, and adapt dynamically to changing requirements \citep{Dinu:24}.
The open-source AutoGPT project\footnote{\href{https://github.com/Significant-Gravitas/AutoGPT}{https://github.com/Significant-Gravitas/AutoGPT}} is a prominent example of such a system, which operates on the basis of LLMs.
Similarly, the proprietary Claude 3.5 system from Anthropic allows agentic AI systems to directly control the computer,\footnote{\href{https://www.anthropic.com/news/3-5-models-and-computer-use}{https://www.anthropic.com/news/3-5-models-and-computer-use}} instead of having only web access.
Furthermore, agentic systems intended to act as scientists for open-ended scientific discovery have been proposed \citep{Lu:24}.
Another emerging direction is the use of multi-agent agentic AI systems,\footnote{\href{https://www.anthropic.com/engineering/built-multi-agent-research-system}{https://www.anthropic.com/engineering/built-multi-agent-research-system}} where multiple LLMs autonomously use tools in a loop, possibly in parallel and orchestrated by a lead agentic AI system.

\paragraph{Reasoning LLM systems:}

Reasoning LLM systems aim to enhance the reasoning capabilities 
of large language models (LLMs) by enabling them to perform structured, multi-step computations with substeps, subgoals, and intermediate results.
One major direction of research explores chain-of-thought prompting and optimization, where the model is encouraged to explicitly walk through reasoning steps and possibly self-correct before arriving at a conclusion. The intermediate reasoning steps are often presented to the user; this improves the explainability of answers and the detection of errors.
Since the introduction of the o1 model by OpenAI,\footnote{\href{https://openai.com/o1/}{https://openai.com/o1/}} many attempts have been made to replicate its reasoning capabilities.
Notable publicly accessible variants are DeepSeek-R1 \citep{deepseekr12025} or the S1 replication effort \citep{muennighoff2025s1simpletesttimescaling}.
Those reasoning LLM systems scale with inference computing power: the more they can reason, the better the final output.
Generating long answers is not well suited to transformers due to the quadratic complexity of attention with respect to the sequence length.
We therefore expect alternatives such as state-space models \citep{gu2024mamba} or xLSTM \citep{beck2024xlstm} with linear complexity in the sequence length to become prominent.
Regarding certification, the question arises as to whether only the performance of outputs or also the reasoning sub-steps must be assessed, depending on what information is provided to a potential user.
This will become an important question if such systems are used more and more, e.g., as assistants in technical or scientific fields.

\newpage
\blankpage 

\section{Regulatory Background} \label{sec:regulatory-background}

In this section, we point out the intersections and possible contributions of our AI assessment scheme to the fulfillment of regulatory obligations and evolving trustworthiness standards. We give a short introduction to the regulatory background of this paper. We also provide a brief overview of the EU AI Act and place our activities in the context of the existing infrastructure for conformity assessments and certification.

\subsection{AI Act}

The EU AI Act is the world’s first legal framework to regulate the development, deployment, and use of artificial intelligence. It lays out obligations for developers, providers, and users of AI systems, with the overarching goal to make sure AI technologies are safe, transparent, and aligned with fundamental rights as well as democratic values.

At the heart of the AI Act lies a \textbf{risk-based approach}. AI systems are classified according to their risk level, with unacceptable risk and high risk being the most prominent categories. \textbf{Unacceptable risk} systems such as social scoring and manipulative or deceptive AI systems are strictly prohibited. \textbf{High-risk} systems can be broadly divided into two types. First, AI systems that are intended for specific use cases as listed in ANNEX III of the AI Act - e.g., critical infrastructure, employment, biometric identification, and education. Second, AI systems that are either regulated products themselves or act as safety components of such products. These systems have to meet strict requirements, e.g., related to transparency, data quality, human oversight, and robustness throughout their lifecycle. \textbf{Limited-risk} systems are only subject to transparency requirements (e.g., disclosing AI interaction), while \textbf{minimal-risk} systems remain mostly unregulated.

The AI Act also introduces a separate regime for \textbf{General-Purpose AI} (GPAI) defined as \citep{EU_AI_Act_2024}:
\begin{quote}
    ``\textit{General-Purpose AI model} means an AI model, including where such an AI model is trained with a large amount of data using self-supervision at scale, that displays significant generality and is capable of competently performing a wide range of distinct tasks regardless of the way the model is placed on the market and that can be integrated into a variety of downstream systems or applications, except AI models that are used for research, development, or prototyping activities before they are placed on the market.''
\end{quote} 
GPAI models are typically foundation models (see Section~\ref{subsec:foundation_models}). However, even though the legal definition of GPAI models in the AI Act and the definition and perception of foundation models by the AI community largely overlap, they are not necessarily the same.
The special treatment of GPAI in the EU AI Act reflects the growing concern about the unique challenges associated with foundation models, especially if they are large-scale, very general, and potentially impactful on society. 
The AI Act also introduces the \textbf{systemic risk} that is specific to GPAI models and is defined as \citep{EU_AI_Act_2024}:
\begin{quote}
    ``\textit{Systemic risk} means a risk that is specific to the high-impact capabilities of general-purpose AI models, having a significant impact on the Union market due to their reach, or due to actual or reasonably foreseeable negative effects on public health, safety, public security, fundamental rights, or the society as a whole, that can be propagated at scale across the value chain.''
\end{quote}
Developers of GPAI models with systemic risk must implement additional safeguards, conduct systemic risk evaluations, and maintain detailed technical documentation to support regulatory oversight.

One of the core objectives of the AI Act is the harmonization of AI regulation across the EU to reduce legal fragmentation and uncertainty. This reflects the EU’s broader ``New Approach'' to regulation, which entrusts the development of \textbf{harmonized technical standards} to the independent standardization bodies CEN and CENELEC. These standards are meant to operationalize the high-level requirements of the AI Act such as safety, transparency, and risk management. To further support consistent interpretation and implementation, the European Commission is issuing \textbf{official guidelines and codes of practice} \citep{eucomm:2025cop-gpaidraft3} on several aspects of the AI Act. Together, these instruments aim to balance regulatory coherence with innovation and improve legal certainty for stakeholders. However, their real-world impact on compliance and AI governance - both within and beyond the Union - remains to be seen.

Despite these efforts, significant \textbf{legal uncertainty remains}. The definition of an ``AI system'' under Article 3(1) is ambiguous and could encompass technologies not typically seen as AI and exclude others that should arguably fall within its scope, as noted by experts \citep{Wendehorst:24} and institutions such as the European Law Institute \citep{eli:2024guidelines-aisystem}. A similar ambiguity surrounds the terms ``General-Purpose AI'' and ``systemic risk''. Additionally, inconsistencies between the AI Act's legal terminology and the evolving vocabulary of standardization bodies risk complicating the practical implementation. These challenges are particularly acute for open-source developers, small and medium sized enterprises and research institutions, who often lack the resources to navigate complex compliance requirements. Fragmented enforcement between member states may further undermine the harmonization goals of the AI Act.

In summary, the AI Act represents \textbf{a bold regulatory step} toward aligning technological advancement with ethical principles and social values. It aims to balance progress with accountability while reaffirming the commitment of the EU to protect fundamental rights in the digital age. Its long-term success will depend on the practical interpretation of key terms, the proportional enforcement of obligations, and the framework's ability to accommodate different development modalities for AI models - including open-source and distributed approaches. 

\subsection{Conformity Assessment and Certification}

\begin{figure}[h]
    \centering
    \includegraphics[width=\linewidth, trim=1cm 1cm 1cm 1cm, clip]{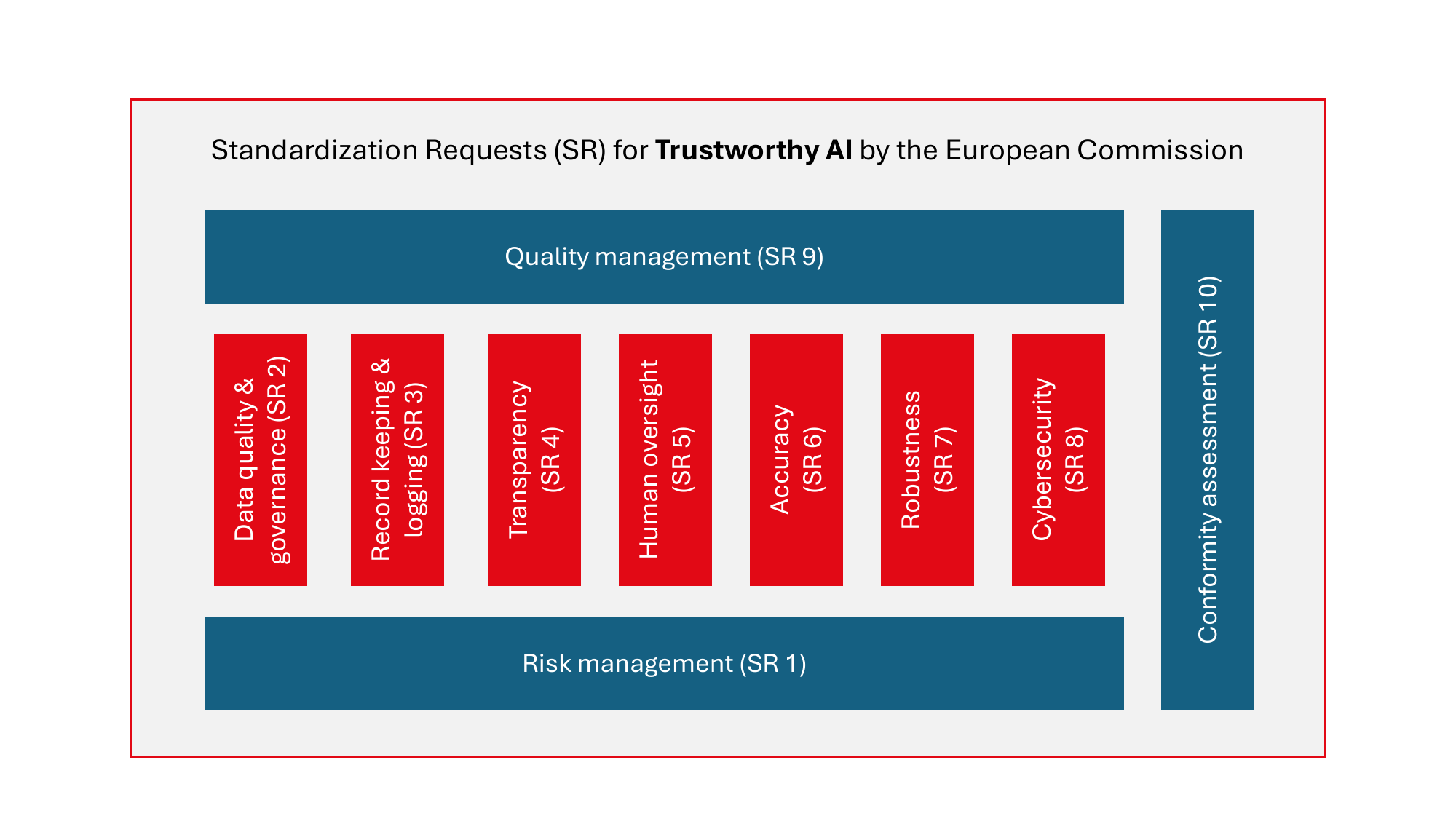}
    \caption{Standardization requests by the European Commission to CEN/CENELEC, also see \citet{eucomm:2023standardisationrequest}.}
    \label{fig:requests}
\end{figure}

Conformity assessments under the EU AI Act serve as an \textbf{overarching accountability framework for high-risk AI systems}. They play a pivotal role in the governance of AI in the EU, ensuring that AI systems meet the requirements of upcoming harmonized standards and regulations. Thus, the European Commission has defined several standardization requests for the CEN-CENELEC standardization committees, as shown in Figure \ref{fig:requests}. Concomitant assessments involve evaluating AI systems' compliance with performance, security, safety, transparency, fairness, and ethical considerations. The main objective is to ensure that AI systems operate as intended, without causing harm or bias. 

In order to classify the use of the term ``certification'' in the context of this paper, it is worthwhile to consider the taxonomy of the International Organization for Standardization (ISO) Committee for Conformity Assessment (CASCO). The term ``certificate'' is used to indicate that the AI assessment scheme presented in this paper can be used to attest that an audited AI system meets the specific requirements raised in the corresponding audit catalog.
This follows the functional conformity assessment approach defined in \citet{iso17000}. 
Our conformity assessment approach can be used by private organizations as a guideline for evaluating the performance of AI systems in the course of a first-party audit or by conformity assessment bodies to independently evaluate compliance with the requirements defined in our audit catalog. 

The objectivity and reliability of an assessment is determined by the independence of the conducting institution and by the qualification and objectivity of the assessors. However, the approach presented in this paper is not part of an accredited conformity assessment scheme or the notified conformity assessment required for compliance with the AI Act. Therefore, the term ``certification'' must be understood as a conformity statement based on a set of requirements and as a benchmark for the technical trustworthiness of AI systems. Nevertheless, we expect that this approach will meet the requirements of the upcoming standards.

Accredited certification schemes that existed at the time this paper was published focus more on management practices or individual aspects of functional correctness and performance, without addressing the issue of overarching AI trustworthiness on a technical level \citep{AIManagementSystem}. This is mainly due to the fact that these technical issues are still a matter of research and the state-of-the-art is evolving rapidly, which complicates the creation of harmonized standards and accreditation schemes. Standards under development aim at providing implementation guidance for the requirements on AI systems in high-risk applications in connection with the obligations under Articles 8 to 15 of the EU AI Act.

We expect the first accredited schemes to be published before the start of mandatory conformity assessments for high-risk AI systems according to the AI Act.
These will focus more on technical content and compliance with accreditation requirements.
This paper may be used as an additional basis for the definition of scientific requirements for standardization purposes and contribute to the ability of the Testing, Inspection and Certification (TIC) sector to build trust in the \textbf{safety, reliability and performance of AI systems}.

\clearpage

\blankpage

\section{The Audit Catalog for ML Models} \label{sec:description-audit-catalog}

In 2021, TÜV AUSTRIA presented one of the world's first audit catalogs for trustworthy AI applications \citep{winterTrustedArtificialIntelligence2021}.
The public perception and spread of use of AI applications has changed rapidly since November 2022, due to the release of OpenAI's \emph{ChatGPT}, followed by similarly capable models by other companies. 
Although ChatGPT was initially only based on text, within the last two years it has become possible to analyze images, search for the latest information on the web, and let the models provide reasoning on their answers.
Furthermore, systems have been introduced that can generate extremely detailed and realistic images or even videos.
Although less prominent in public perception, recent advances in AI systems for simulations are equally disruptive.
These systems can speed up simulations of industrial processes that used to take weeks to merely a few minutes, allowing much faster development.

To keep pace with the rapid technological advances in the field of AI, the \textit{TÜV AUSTRIA Trusted AI} audit catalog is continuously being developed. 
The requirements have advanced since the first release, on the one hand by incorporating the current state-of-the-art in AI assurance, and on the other hand by aligning it with the EU Commission's  legal requirements and standardization efforts. 

\begin{figure}[h]
    \centering
    \includegraphics[width=.9\textwidth]{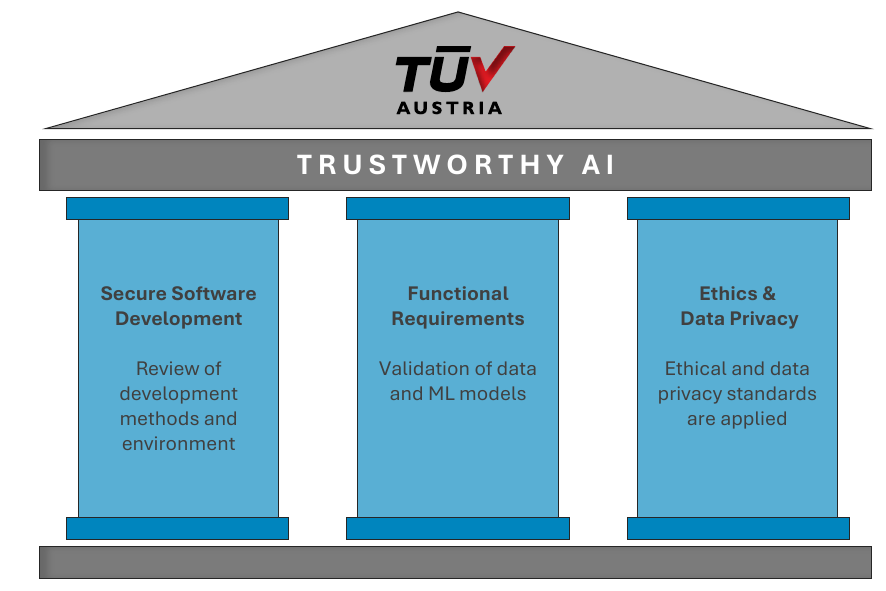}
    \caption{The three pillars of the \emph{TÜV AUSTRIA Trusted AI} audit catalog: Secure Software Development, Functional Requirements, and Ethics \& Data Privacy.}
    \label{fig:pillars}
\end{figure}

In the \textit{TÜV AUSTRIA Trusted AI} framework, the evaluation of trustworthy AI is based on three pillars, shown in Figure~\ref{fig:pillars}: Secure Software Development, Functional Requirements, and Ethics \& Data Privacy. 
The \textit{TÜV AUSTRIA Trusted AI} certificate ensures that the AI system developed by the auditee meets high development standards, performs well for its intended use, and is robust against various risks, including bias, security threats, and operational failures. 
Furthermore, our framework provides concrete technical requirements that operationalize the abstract general provisions laid out in the EU AI Act and emerging standards.

\subsection{Foundational Principle: Functional Trustworthiness}
\label{sec:functrust}
\begin{figure}[h]
    \centering
    \includegraphics[width=\linewidth, trim=0.5cm 0.2cm 0.5cm 0cm, clip]{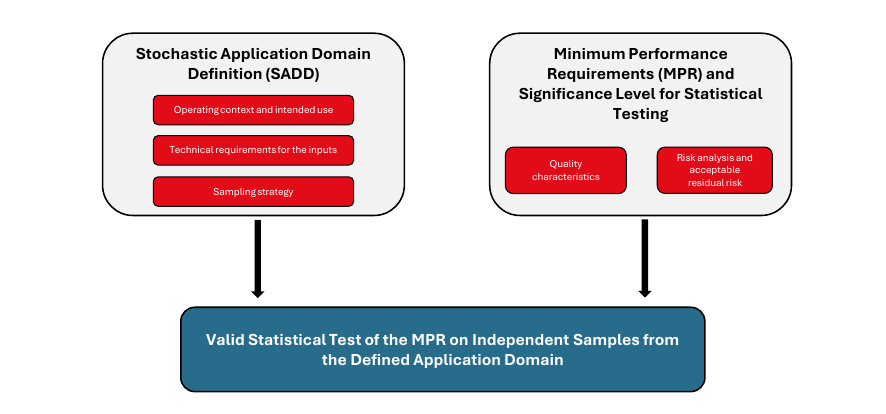}
    \caption{Ensuring the \emph{functional trustworthiness} of an AI system by valid statistical testing of the fulfilment of the minimum performance requirements on independently sampled test data from the application domain.}
    \label{fig:mpr}
\end{figure}

A central aim of the certificate is to provide \textbf{statistical guarantees of the expected performance of the system}. 
For this, we apply the \textbf{principle of functional trustworthiness} as proposed in \citet{Nessler:23}. 
A meaningful performance assessment for an AI system is based on three central components (see also Figure~\ref{fig:mpr}):

\begin{enumerate}
    \item \textbf{A precise definition of the (stochastic) application domain of the system.} The Stochastic Application Domain Definition (SADD) consists of three parts: (1) a description of the intended use and operating conditions of the AI system, (2) a detailed description of the technical requirements for the inputs to the AI system, and (3) a description of the sampling process that can be used to draw independent samples from the application domain for performance evaluation. 
    The description of the sampling process creates the reference distribution and enables a meaningful interpretation of the model performance measures. 
    Furthermore, third parties can reproduce the model evaluation by following the recipe outlined in the three parts of the Stochastic Application Domain Definition.

    \item \textbf{Minimum performance requirements (MPRs) that reflect the desired quality characteristics of the AI system and account for acceptable residual risks.} The minimum performance requirements define the performance that we expect from our system. 
    They can be based on desired output qualities, e.g., ``Our product is only trustworthy if it makes the correct decision in at least 99\% of the cases'', but also based on potential legal obligations and risks associated with the operation of the system. 
    This can, for example, include the risk of discrimination or the risk of adversarial attacks by third parties. 
    All of these considerations should be operationalized as measurable metrics that allow to test for fulfillment of the MPRs.
    
    \item \textbf{A statistically valid test of said performance requirements on independently sampled elements from the application domain.} In order to obtain a reliable estimate of the performance of the AI system in real life, it should be evaluated using appropriate statistical tests on a ``representative'' sample from the defined application domain. 
    The sample used to assess the minimum performance requirements should reflect the instruction set in the Stochastic Application Domain Definition.

\end{enumerate}

\subsection{The Heart of the Audit Catalog: Functional Requirements}

\begin{figure}[h!]
    \centering
\includegraphics[width=\textwidth, trim=0.5cm 0.7cm 0.cm 1cm, clip]{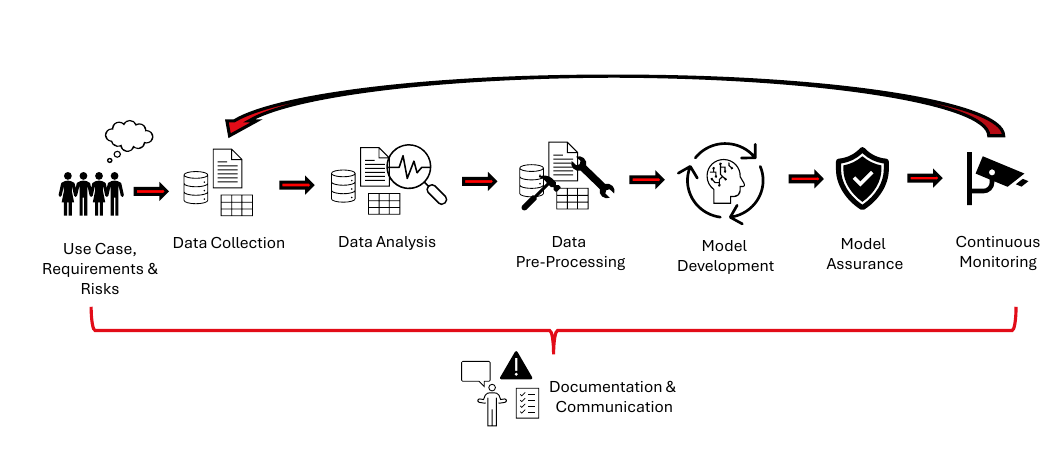}
    \caption{Lifecycle of AI system development and assurance}
    \label{fig:dev_cycle}
    \vspace{0.2cm}
\end{figure}

The functional requirements are the core of the \textit{TÜV AUSTRIA Trusted AI} certification and deal with the validation of the data and ML models. 
Most importantly, they check the AI systems' functionality for the given use case and application domain.
Furthermore, the functional requirements ensure that best practices were followed during the development of the AI system. All development phases must be thoroughly documented.
The respective sections of the audit catalog are oriented along the lifecycle of an AI system as illustrated in Figure~\ref{fig:dev_cycle}: Starting from specification of the use case and requirements for the AI system, through data collection, analysis, and preprocessing, the development of the model as well as the analysis and testing of the model on independent test data, and the deployment phase, including monitoring and fallback mechanisms. 

\subsection*{Structure and Overview of the Functional Requirements}

In the following, we present the 13 sections of the functional requirements.%
Furthermore, we provide a selection of typical mistakes that we have encountered within our certification audits conducted over the past years.
We hope these findings foster a deeper understanding of common pitfalls that arise in practice when developing AI systems and support better compliance with functional requirements in the market.

\subsubsection*{1. ML Use Case Definition}

    \begin{wrapfigure}{r}{0.12\textwidth} 
    \vspace{-1.2cm}
    \includegraphics[width=\linewidth]{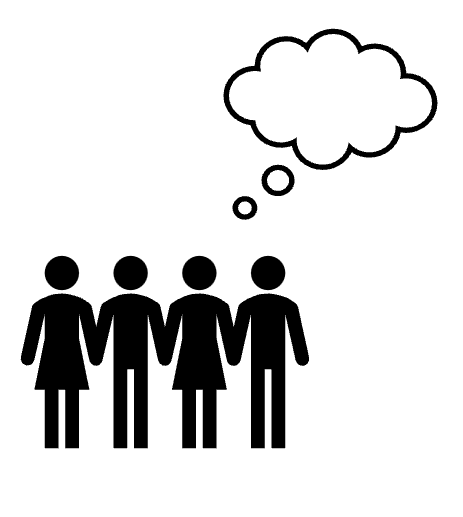}
    \end{wrapfigure}
    The business context and the ML task must be identified in detail. All relevant stakeholders and their roles in the development process have to be identified and documented.
\vspace{0.8cm}

\begin{tabular}{|p{15cm}|}
    \hline
    \rowcolor{gray!20}
    \textsf{{Typical Mistakes and Findings}} \\
    \hline
    \begin{itemize}[leftmargin=*, itemsep=3pt]
    \item The goal of the AI system is unrealistic and the business case behind it is unclear. 
    \item The ML task could be solved by a simple rule-based system based on expert input.
    \end{itemize} \\
    
    \hline
\end{tabular}

\subsubsection*{2. Quality Characteristics, Risk Assessment and Management}
    
    \begin{wrapfigure}{r}{0.12\textwidth} 
    \vspace{-0.7cm}
    \includegraphics[width=\linewidth]{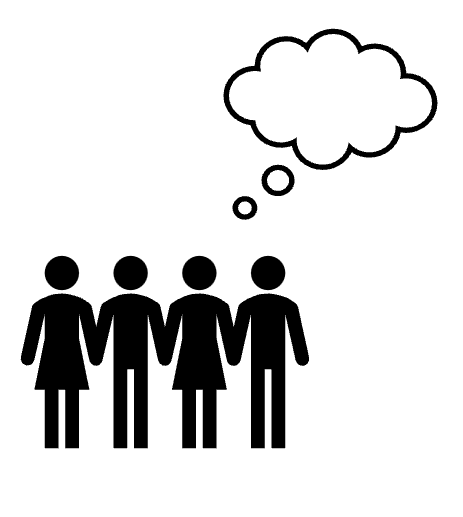}
    \end{wrapfigure}
    The requirements for the AI system should be specified in detail before the development process starts. The requirements are derived from the business case, from legal obligations such as the EU AI Act and the GDPR, and from a risk assessment for the application. The minimum performance requirements are derived from these considerations.
    
    A risk management system should be put in place to ensure that mitigation strategies, safety nets, and continuous improvement loops prevent damage to both image and finances.
   
    \vspace{.5cm}

    \begin{tabular}{|p{15cm}|}
    \hline
    \rowcolor{gray!20}
    \textsf{{Typical Mistakes and Findings}} \\
    \hline
    \begin{itemize}[leftmargin=*, itemsep=3pt]
    \item The risk of adversarial attacks by users that try to maliciously manipulate the system are not accounted for.
    \item The classification of the system under the AI Act was not considered, thus relevant requirements and regulations are overlooked. 
    \item The impact of incorrect model predictions is not taken into account properly. 
    \end{itemize} \\
    \hline
\end{tabular}

\subsubsection*{3. Application Domain Definition}

    \begin{wrapfigure}{r}{0.12\textwidth} 
    \vspace{-1cm}
    \includegraphics[width=\linewidth]{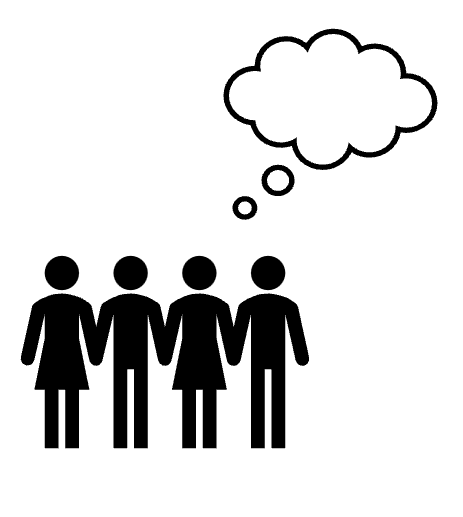}
    \end{wrapfigure} 
    Establishing a clear definition of the application domain of the AI system is crucial. This ensures that the intended use of the AI system is clearly specified and that a realistic estimate of the expected model performance can be obtained.
    
     \vspace{.5cm}

    \begin{tabular}{|p{15cm}|}
    \hline
    \rowcolor{gray!20}
    \textsf{{Typical Mistakes and Findings}} \\
    \hline
    \begin{itemize}[leftmargin=*, itemsep=3pt]
\item Geographical restrictions are not accounted for in the operational context, e.g. the AI system is only trained and tested on data from the DACH area, but is to be used Europe wide.
   \item The domain definition does not account for the technical details in the setup of the data recording. 
    \end{itemize} \\
    \hline
\end{tabular}

\subsubsection*{4. Data Collection}

    \begin{wrapfigure}{r}{0.12\textwidth} 
    \vspace{-0.5cm}
    \includegraphics[width=\linewidth]{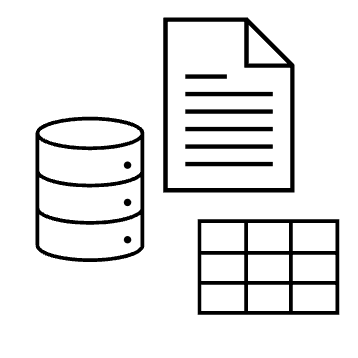}
    \end{wrapfigure}
    High-quality results start with high-quality data. This is why a detailed documentation of data sources, annotation procedures, and any third-party data integration is essential. 
    This reduces ambiguity and builds a solid foundation for trustworthy ML models by ensuring transparency from the outset.
   
   \vspace{.5cm}
   
   \begin{tabular}{|p{15cm}|}
    \hline
    \rowcolor{gray!20}
    \textsf{{Typical Mistakes and Findings}} \\
    \hline
    \begin{itemize}[leftmargin=*, itemsep=3pt]
    \item The collected data does not represent the defined application domain (underrepresented groups, etc.), e.g., instead of collecting images from certain operating conditions, these have been simulated through filters. The original conditions were thus never seen during training, and not considered when testing the model.
   \item Data is collected from different internet sources for different parts of the data domain, therefore possibly introducing data leakage.
    \end{itemize}
    \\
    \hline
\end{tabular}

    \subsubsection*{5. Data Analysis}  
    
    \begin{wrapfigure}{r}{0.12\textwidth} 
    \vspace{-0.5cm}
    \includegraphics[width=\linewidth]{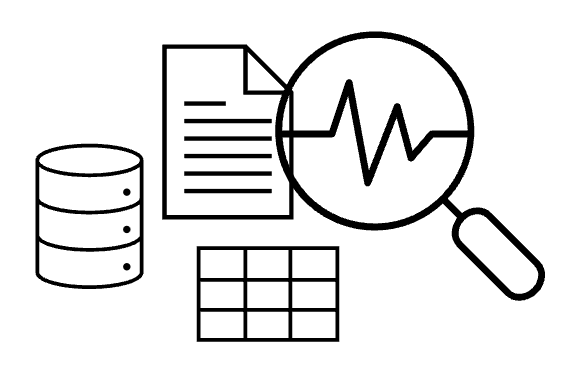}
    \end{wrapfigure}
    In-depth data analysis and documentation are required to verify the composition and consistency of the data. In particular, the data should cover the application domain as defined above. The requirements ensure that the data are suitable for the given ML task.
    
    \vspace{.5cm}
    
   \begin{tabular}{|p{15cm}|}
    \hline
    \rowcolor{gray!20}
    \textsf{{Typical Mistakes and Findings}} \\
    \hline
    \begin{itemize}[leftmargin=*, itemsep=3pt]
    \item The dataset contains a lot of duplicated values that need to be handled appropriately before proceeding with the model development. 
    \item Highly correlated features introduce artifacts and redundant information in the data.
    \end{itemize}

    \\
    \hline
\end{tabular}

    \subsubsection*{6. Data Pre-Processing} 

    \begin{wrapfigure}{r}{0.12\textwidth} 
    \vspace{-0.8cm}
    \includegraphics[width=\linewidth]{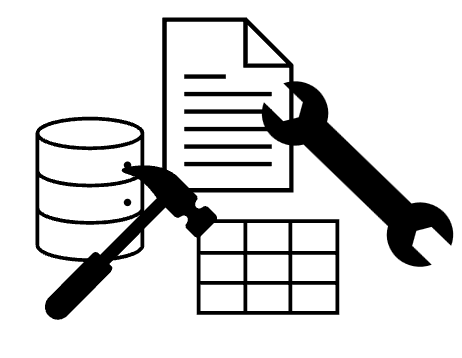}
    \end{wrapfigure}
    
    All steps in data pre-processing, such as data cleansing, feature normalization, and data augmentation, are reproducible, fully documented and follow ML best practices.

    \vspace{.5cm}
    
   \begin{tabular}{|p{15cm}|}
    \hline
    \rowcolor{gray!20}
    \textsf{{Typical Mistakes and Findings}} \\
    \hline
    \begin{itemize}[leftmargin=*, itemsep=3pt]   
    \item A dataset is first pre-processed, e.g., feature normalization, feature selection, outlier capping, ..., and then split into training and test sets, leading to data leakage.
    \item Images from the same video sequence are used partly in the training data and partly in the test data.
    \item Features are not processed in a way compatible with the chosen model class (e.g., nominal, categorical variables are mapped to numbers when using linear regression).
    \end{itemize}
    \\
    
    \hline
\end{tabular}

    \subsubsection*{7. Model Development}
    
    \begin{wrapfigure}{r}{0.12\textwidth} 
    \vspace{-1cm}
    \includegraphics[width=\linewidth]{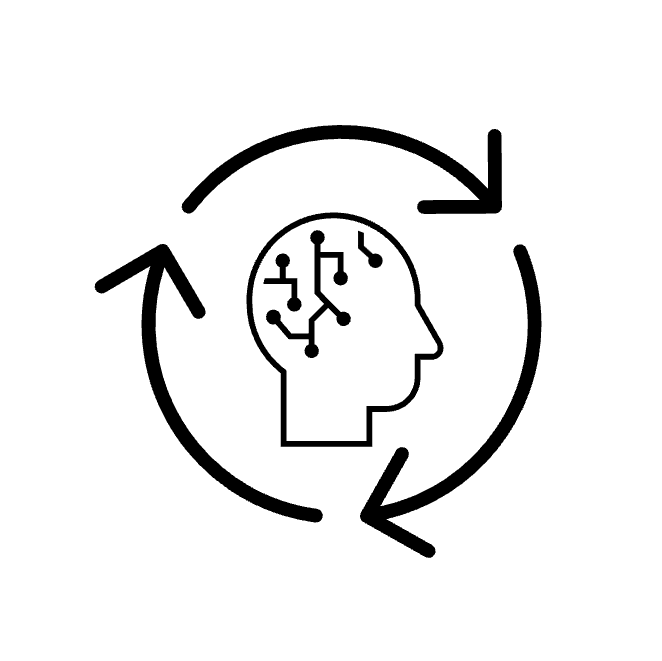}
    \end{wrapfigure}
    Established ML methodologies are utilized to select model architectures and evaluation metrics tailored to the task at hand. It is ensured that the entire training process can be tracked and is reproducible. The reasoning behind design decisions is clear and understandable.
  
    \vspace{.5cm}
    
   \begin{tabular}{|p{15cm}|}
    \hline
    \rowcolor{gray!20}
    \textsf{{Typical Mistakes and Findings}} \\
    \hline
    \begin{itemize}[leftmargin=*, itemsep=3pt]   
    \item No explanation as to why a certain methodology was used (why use a deep neural network if a simpler method would have been sufficient).
    \item Choice of an improper evaluation metric which does not consider possible risks (e.g., accuracy without evaluation of precision and recall).
    \item The size of the test set is not large enough to draw confident conclusions about the performance of the model.
    \end{itemize}
    \\
    \hline
\end{tabular}

\subsubsection*{8. Quantitative Model Inspection} 

    \begin{wrapfigure}{r}{0.12\textwidth} 
    \vspace{-0.8cm}
    \includegraphics[width=\linewidth]{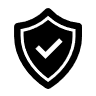}
    \end{wrapfigure}
    A quantitative assessment confirms that the model meets the predefined standards. 
    Appropriate metrics are measured on held-out test data to confirm the reliability of the model. If, for example, the model claims to detect defects with an accuracy of $99.5\%$, the claim should be verified with statistical methods to ensure trust in the results.
    \vspace{.5cm}
    
   \begin{tabular}{|p{15cm}|}
    \hline
    \rowcolor{gray!20}
    \textsf{{Typical Mistakes and Findings}} \\
    \hline
    \begin{itemize}[leftmargin=*, itemsep=3pt]   
    \item Statistical testing is neglected or is performed incorrectly, e.g., under wrong assumptions.
    \item The test data for the model inspection is not \textit{representative} of the actual application domain, or a distribution shift has occurred, rendering the performance measurement meaningless.
    \end{itemize}
    \\
    \hline
\end{tabular}

\subsubsection*{9. Qualitative Model Inspection} 

    \begin{wrapfigure}{r}{0.12\textwidth} 
    \vspace{-1cm}
    \includegraphics[width=\linewidth]{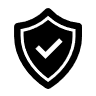}
    \end{wrapfigure}
    Model decisions should be inspected on a case-by-case basis. This deeper investigation ensures that the model is robust, stable, and prepared for real-world variations.
    
    \vspace{.5cm}
    
     \begin{tabular}{|p{15cm}|}
    \hline
    \rowcolor{gray!20}
    \textsf{{Typical Mistakes and Findings}} \\
    \hline
    \begin{itemize}[leftmargin=*, itemsep=3pt]   
     \item Focus on cases that meet expectations and overlook instances where the model performs poorly (confirmation bias).
     \item The edge-case analysis points to systematic shortcomings of the model (and the data) but is ignored in view of a seemingly sufficient validation performance.
    \end{itemize}
    \\
    \hline
\end{tabular}

    \subsubsection*{10. Explainability and Interpretability} 

    \begin{wrapfigure}{r}{0.12\textwidth} 
    \vspace{-1cm}
    \includegraphics[width=\linewidth]{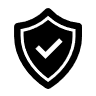}
    \end{wrapfigure}
    Interpretability and explainability methods should be applied during model development to understand and identify potential failure modes of the developed model. Furthermore, they increase the understanding of the data domain and its projection into the features of the model, and can help to detect data leakage.
  
  \vspace{0.5cm}
  
     \begin{tabular}{|p{15cm}|}
    \hline
    \rowcolor{gray!20}
    \textsf{{Typical Mistakes and Findings}} \\
    \hline
    \begin{itemize}[leftmargin=*, itemsep=3pt]   
    \item Overconfidence in the model's capabilities because a local explainability method indicates a focus on the center of the object in object recognition tasks. 
    \item Explainability methods are used as tools to indicate the degree of trustworthiness of the model prediction to the user, instead of correctly applying methods for uncertainty estimation.
    \end{itemize} \\
    \hline
\end{tabular}

 \subsubsection*{11. Operation}

    \begin{wrapfigure}{r}{0.12\textwidth} 
    \vspace{-1.2cm}
    \includegraphics[width=\linewidth]{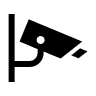}
    \end{wrapfigure}
    Once deployed, continuous monitoring of its performance in the field is required to ensure that it behaves as well as it did in testing.

    \vspace{0.5cm}
   
    \begin{tabular}{|p{15cm}|}
    \hline
    \rowcolor{gray!20}
    \textsf{{Typical Mistakes and Findings}} \\
    \hline
    \begin{itemize}[leftmargin=*, itemsep=3pt]       
    \item Failing to detect covariate shift (changes in the input data distribution) or concept drift (changes in the relationship between inputs and outputs).
    \item Focusing on technical performance while overlooking ethical concerns in deployment.
    \end{itemize}
    \\
    \hline
\end{tabular}

\subsubsection*{12. Failure Handling}

    \begin{wrapfigure}{r}{0.12\textwidth} 
    \vspace{-1cm}
    \includegraphics[width=\linewidth]{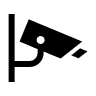}
    \end{wrapfigure}
    Even the best models can encounter unexpected scenarios. Robust fallback strategies in the event of unexpected inputs ensure that safety and quality standards are never compromised. 

    \vspace{.5cm}

    \begin{tabular}{|p{15cm}|}
    \hline
    \rowcolor{gray!20}
    \textsf{{Typical Mistakes and Findings}} \\
    \hline
    \begin{itemize}[leftmargin=*, itemsep=3pt]           
    \item No logging of fallback activations which prevents efficient error tracing and thus further improvement of the model.
    \item Parallel guardrails for input validation are not evaluated in isolation.
    \end{itemize}

    \\
    \hline
\end{tabular}

\subsubsection*{13. Transparency and Communication} 

    \begin{wrapfigure}{r}{0.12\textwidth} 
    \vspace{-0.8cm}
    \includegraphics[width=\linewidth]{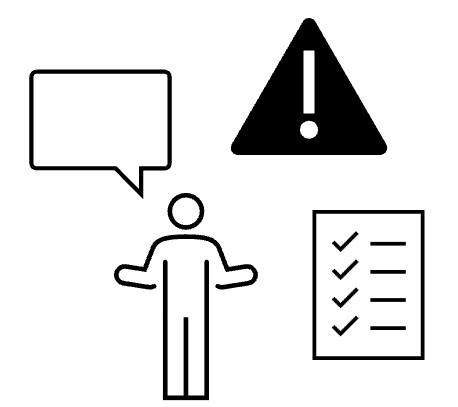}
    \hspace{.1cm}
    \end{wrapfigure}
    Lastly, well-structured technical documentation and user guides should be provided. These resources empower everyone (developers, maintenance teams and business stakeholders) to understand, trust and make the best use of the AI system over its entire lifecycle.
    
    \vspace{0.5cm}
    
\begin{tabular}{|p{15cm}|}
    \hline
    \rowcolor{gray!20}
    \textsf{{Typical Mistakes and Findings}} \\
    \hline
    \begin{itemize}[leftmargin=*, itemsep=3pt]  
    \item No harmonization of documentation approaches between different stakeholders in the development of the AI system
    \item No accountability for different aspects of the technical documentation
    \item No proper change management for the technical documentation, i.e., no availability of a change history for the document
    \end{itemize}

    \\
    \hline
\end{tabular}

\subsection{Secure Software Development}

Since machine learning applications are fundamentally software applications, the principles for secure engineering and the operation of software also apply to AI systems.
These principles cover areas such as regular training for developers, rigorous testing, effective patch management, proper container usage, and secure deployment practices. 

As with classical software security, it is essential to understand the assets involved in the AI system in order to understand the motives behind potential attacks.
Based on a threat analysis and a risk hierarchy based on the feasibility of the attack and the value of the linked assets, a defense strategy for the AI system must be designed based on best practices in all phases of the design process.

However, efforts to achieve AI security are not limited to the field of engineering practices and technical measures. 
Similar to classical security, the entire organizational process for the development and management of artifacts related to the AI system, including access control to relevant data, needs to be augmented by a dedicated security management approach.

The \textit{TÜV AUSTRIA Trusted AI} catalog thus covers a baseline of requirements for software security, from general awareness and development basics to testing, operation, and secure updating.
However, since the entire discipline of achieving secure software systems is quite extensive, the \textit{TÜV AUSTRIA Trusted AI} catalog does not try to cover all security requirements as defined in many established standards. 
AI systems with relevant risks require separate certification according to standards such as \citet{iso27001} or the IEC 62443 series.\footnote{\href{https://en.wikipedia.org/wiki/IEC_62443}{https://en.wikipedia.org/wiki/IEC\_62443}}

Nevertheless, it is necessary to emphasize the importance of including AI-specific threats in the risk analysis, which depend to a large extent on the type of model and the assets to be protected. 
Known examples of AI-specific attacks include inversion attacks to infer information from a model \citep{Fredrikson:15} or evasion attacks to manipulate classification outcomes \citep{Meyers:2023}.

Another factor that increases the level of security of AI systems is regular monitoring of known AI-specific attacks so that countermeasures can be implemented. 
The latest information on this topic can be found in online databases such as OWASP,\footnote{\href{https://owasp.org/www-project-machine-learning-security-top-10}{https://owasp.org/www-project-machine-learning-security-top-10}} which has several projects focusing on ML risks, or MITRE Atlas ML,\footnote{\href{https://atlas.mitre.org/matrices/ATLAS}{https://atlas.mitre.org/matrices/ATLAS}} which focuses on a classification of attack paths.

\subsection{Ethics \& Data Privacy}

AI systems have the potential to have an enormous impact on society, in both a positive and a negative way. 
The first question here is whether or not the realization of a specific AI system is ethically justifiable in the first place. 
In particular, systems with potentially far-reaching consequences for individuals or society warrant a thorough analysis and discussion of ethical implications.

Although functional requirements already address the identification and mitigation of unwanted biases and aspects of fairness, this chapter of the audit catalog further focuses on ensuring important aspects of ethics such as accessibility and discrimination in a more holistic way.

Whenever an AI system involves personal data, all aspects of data protection as laid down in the General Data Protection Regulation (GDPR) must be taken into account. 
Similar to security aspects, GDPR compliance requires considerable effort and goes beyond purely AI-related concerns.  
This is why the \textit{TÜV AUSTRIA Trusted AI} catalog focuses on the core requirements of data protection. 

A frequent topic that arises in certification audits is the question of where data is processed and to what extent an AI system allows the identification and tracking of individuals. 
As a general guideline it can be said that in every case where personal data is processed, an assessment of the impact on rights of the data subject must be carried out.
This includes the documentation of required protection measures in the AI system.

\newpage

\section{Key Aspects for Certification of AI Systems} \label{sec:details}

Since the  \textit{TÜV AUSTRIA Trusted AI} audit catalog for AI applications was released in 2021, we have been continuously working to improve the catalog. 
In addition to the ideas set out in \citet{winterTrustedArtificialIntelligence2021}, we identified further key aspects for AI certification, which we present in the following section. 
One major consideration is Article 15 of the EU AI Act, which relates to the \textbf{Accuracy, Robustness and Cybersecurity} of AI systems and requires the providers and deployers of high-risk AI systems to ensure that they can be operated safely and without harm to the health and safety of human beings. 
The audit catalog concretizes those legal requirements by providing explicit and detailed criteria for auditing AI systems. 

Our auditing scheme focuses on the \textbf{assessment and assurance of the AI system's performance under the expected operating conditions} (i.e., accuracy).
The validity of these performance evaluations must be ensured by examining and ruling out possible data leakage scenarios.
Assuming the accuracy of the system is sufficient, we further assess various aspects of the robustness of AI systems. 
This includes auditing the methodology used to detect out-of-distribution samples and distribution shifts during the operation of the AI system.
Moreover, we recommend the use of uncertainty estimation to deal with samples that are difficult to predict, as well as for edge cases and samples at the boundaries of the application domain, which rarely occur during training.
In addition, an important aspect is the implementation of appropriate defenses against malicious adversarial attacks. 
Explainability and interpretability methods should be used to identify potential failure modes in the AI system.
Finally, by assessing potential biases as well as algorithmic fairness, we aim to ensure that the AI system does not promote discrimination or violate fundamental human rights.

\subsection{AI Assurance Through Functional Trustworthiness} \label{sec:functional-trustworthiness}

\subsubsection{Defining the (Stochastic) Application Domain of the AI System} \label{sec:add}
The definition of the application domain of the AI system is a crucial part of AI assurance. It should consist of three parts, which we refer to as \textit{Stochastic Application Domain Definition} (SADD) in the following.

\begin{itemize}
\item[\textbf{(1)}] A description of the data-generating process and the potential operating context of the AI system. The procedure describes how the data samples on which the system operates are generated in practice.
\item[\textbf{(2)}] A description of the technically controllable requirements for acquiring input data for the AI system. The description pays special attention to the relevant categories and metadata for each data point that is recorded, system configurations (such as camera settings when capturing an image), and potential limiting factors.
\end{itemize}
These two descriptions ensure that an informed person can decide whether the AI system developed is applicable in principle to a certain sample or not (e.g., whether the input image has the required light exposure conditions and resolution and was recorded in the right setting).
In order to understand whether the performance of the AI system is sufficient in a given use case, we also need to define a reference distribution for the performance evaluation. This is particularly important to enable third parties to replicate the evaluation. Thus, the third component of the Stochastic Application Domain Definition is:
\begin{itemize}
\item[\textbf{(3)}]
A detailed description of the sampling strategy that, given access to the full application domain, can be used to obtain independent samples from the application domain. The description should enable a third party to interpret the reported performance metrics and reproduce the performance assessment(s) for the AI system. Thus, the description specifies the reference distribution $p(x)$ for the application domain.
\end{itemize}

In order to illustrate the concept, consider the following example: An AI system was developed to automatically detect scratches in the paint of metal parts manufactured by machine A, which is used in different locations in Austria. In order to determine whether the AI system is applicable to the machine in company B, which falls within the operating context defined in (1), one first needs to study all the technical requirements for the machine and the data collection process as outlined in part (2); for example, which sizes and colors of the metal parts are acceptable, what resolution is required for the photos taken during production, etc.
If the technical requirements are met, part (3) of the Stochastic Application Domain Definition helps us to decide if the performance of the AI system is sufficient in the given use case and the specific operating context of the machine. We thus need to understand the reference distribution, which is described by the sampling strategy.

\begin{figure}
    \centering
    \includegraphics[width=\linewidth]{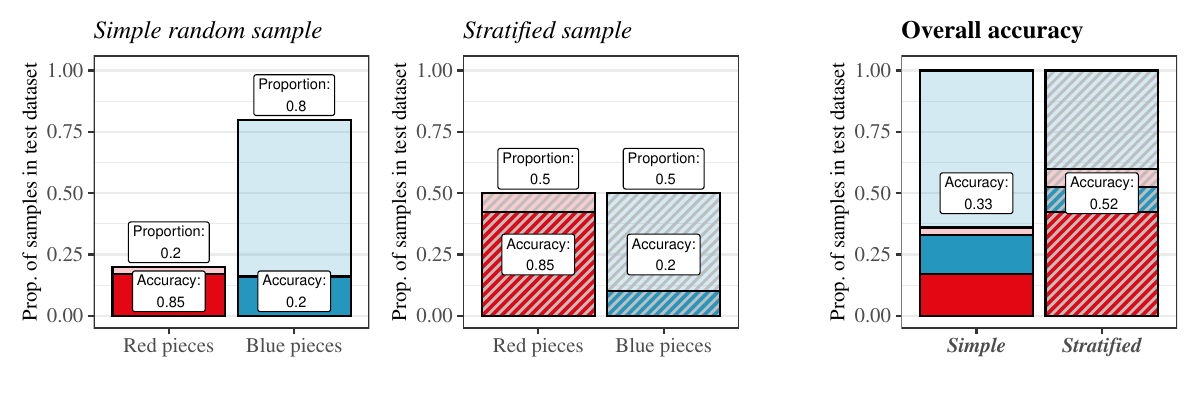}
    \includegraphics[width=\linewidth]{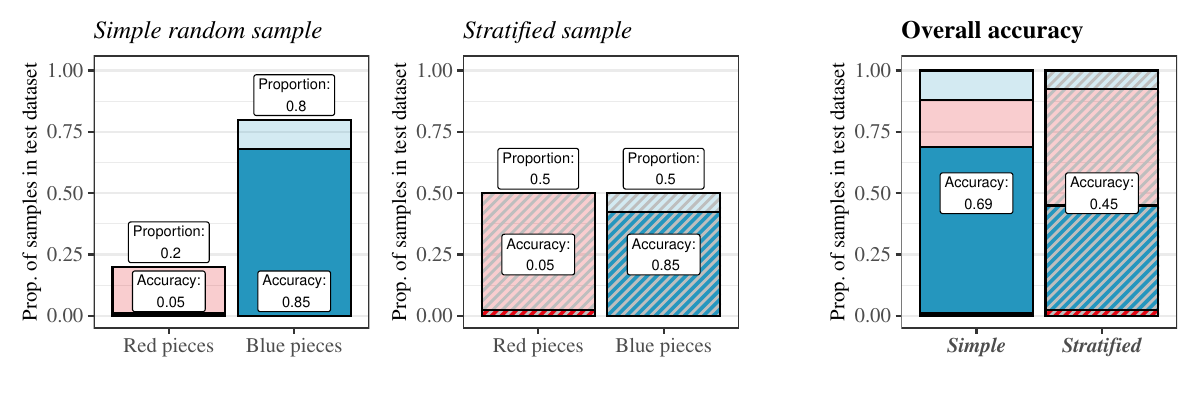}
    \includegraphics[width=\linewidth]{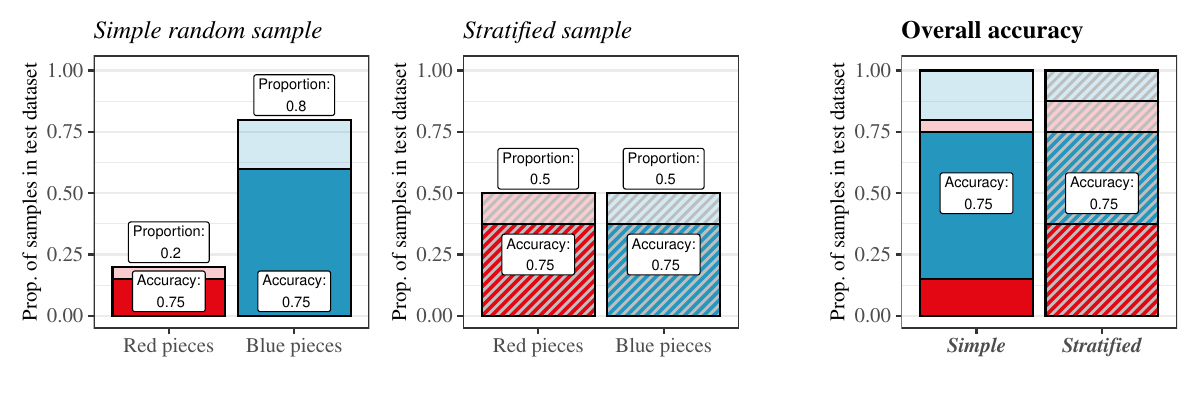}
    \caption{Comparison of two sampling strategies with different resulting distributions. The performance estimate can change drastically depending on the sampling strategy and the underlying ground truth proportions.}
    \label{fig:sampling_strategy_comparison}
\end{figure}

For this example, we assume that the metal parts are produced in two different colors: red and blue. The accuracy of the system should be measured as the classification accuracy of pieces as faulty or intact. One could imagine two different sampling strategies for (3), which result in the distributions shown in Figure~\ref{fig:sampling_strategy_comparison}.

\textbf{Strategy A (``Simple random sample''):} Select $n$ parts at random from all parts produced. 
Since 80\% of the parts produced are blue, we can expect a ratio of 80-20 for the blue and red parts in the sample.
In this case, an overall accuracy of 80\% can be interpreted as the expectation that a randomly selected sample from production is correctly classified as defective or intact. However, this accuracy can be achieved if the algorithm correctly classifies all blue parts and not on a single red part. 

\textbf{Strategy B (``Sampling stratified by color''):} 
If we want to ensure that the system works equally well for both colors, samples from each color must be selected in equal proportions. Thus, the final sample would contain the same number of red and blue parts. In this case, an accuracy of 80\% would imply that even if the algorithm works perfectly on the blue parts, we still need at least 60\% accuracy for the red parts to reach the average accuracy of 80\%. However, if the red parts are actually very rare in production, or if there are many classes, it may not be desirable to assign them a high weight when evaluating the AI system. 

The impact and possible consequences of the sampling strategy on the performance measurement of the AI system are shown in Figure~\ref{fig:sampling_strategy_comparison}.
\textbf{In conclusion, the choice of sampling strategy depends to a large extent on the desired performance measurement. Without knowing the sampling strategy, it is not possible to interpret the performance measure and draw conclusions about the applicability of the AI system to a specific use case.} In particular, the performance evaluation cannot be replicated without knowledge of the precise sampling strategy.  More details on the practical sampling of test data are presented in Section~\ref{sec:sampling-test-data}. Popular sampling strategies that have been well researched in literature include \textit{simple random sampling}, \textit{cluster sampling}, \textit{stratified sampling} as well as \textit{multistage sampling}. A detailed introduction to statistical sampling theory is given, for example, in \citet{Lohr:21}.

\subsubsection{Guideline on Defining Quality Characteristics and MPRs}
\label{sec:mpr-definition}

Defining quality characteristics for the AI system is highly specific to the use case and is a delicate task. Before starting, or in the early stages of developing the AI system, one should clearly define the desired properties of the AI system. 
The requirements for the AI system need to be specified in a structured way.

It is possible to differentiate between \textbf{qualitative} and \textbf{quantitative} requirements.
Quantitative requirements are requirements that enable certain thresholds to be set that must be met. However, certain requirements cannot be quantified (such as the user controllability of an AI system) and these therefore require a qualitative evaluation. 

The initial thresholds for the identified quantitative quality characteristics are derived from the business case and potential legal obligations.
In the next steps, they must be refined through a risk analysis of the AI system. 
By combining the output of the quality requirements and the risk analysis, thresholds can be defined that specify the minimum performance requirements (MPRs) for any requirement that can be quantitatively assessed. The MPRs should be translated into statistical measures or performance metrics that can be evaluated for the AI system (e.g., MSE, RMSE, Accuracy, etc.) on independent test data. Statistical tests must confirm that all MPRs are fulfilled at predefined family-wise significance level $\alpha$. The corresponding significance level can be set even higher for particularly critical elements of the MPRs. This decision should be based on the residual risk that the developers of the AI system are willing to bear.

Helpful guidance on defining quality characteristics for AI systems is provided in \citet{iso25058} and \citet{iso25059}. Examples of such characteristics include
\begin{itemize}[noitemsep]
\item \textbf{Functional correctness:} (quantitative) Degree to which a product or system provides the correct results.
\item \textbf{Robustness and fault tolerance:} (quantitative) Maintaining the performance level of AI systems under changing environmental conditions, without re-training or varying the initial configuration of the system.
\item \textbf{Societal and ethical responsibility:} (quantitative / qualitative) Degree to which an AI system is developed and used in accordance with societal values and ethical principles, ensuring fairness, non-discrimination, and respect for human rights.
\item \textbf{User controllability:} (qualitative)  Property of an AI system that enables a human or another external agent to intervene in its functioning in a timely manner. 
\item \textbf{Transparency:} (qualitative) Degree to which appropriate information about the AI system is communicated to stakeholders.
\end{itemize}

\subsubsection{Sampling Test Data} \label{sec:sampling-test-data}

ML models learn patterns, relationships, and correlations from data. Learning complex relationships typically requires a large amount of data during the model training phase. However, the ultimate goal of an ML model is to perform well on data it has never seen before. This is where the concept of generalization comes into play. An ML model should be able to generalize from its training data to unseen data and make accurate predictions or classifications.

The purpose of test data is to simulate this real-world scenario with unknown data. The test data is a separate dataset that is withheld from the model during training and is used solely to evaluate how well the model can generalize to new data. It serves as a proxy for real applications and provides information on how the AI system will behave in practice.

The effectiveness of the test data depends on two aspects. 
Firstly, the test data should reflect the true operating conditions of the AI system. In other words, it should be \textit{representative} of the application domain of the system with respect to the defined minimum performance requirements. The application domain corresponds to what is termed \textit{population} in statistical sampling literature.
Secondly, its independence from the training data. The test data must not overlap with or be influenced by the training data. An ML model should never have seen the test data during the training phase, and even the development team should not have the possibility to inspect the test data in any way.

In particular, the performance of an AI system in real life is estimated on the basis of a random sample from the application domain. The goal of this procedure is to obtain an estimate of the performance that is \textit{unbiased} for the parameter of interest, i.e., the performance of the system. In statistical literature, unbiased means that the estimator reflects the true value across the application domain if it is evaluated on a large enough sample and that there are no systematic deviations. However, every estimator is accompanied by an uncertainty that is quantified by its \textit{variance}.  
It quantifies the accuracy of the estimated value and depends on the sample size, but also on the sampling strategy employed. The unbiasedness of the estimator for the population value can be compromised when it is not estimated from a purely random sample of the population. In this case, the estimate can be misinterpreted. In the context of the evaluation of AI systems, this can lead to severe over- or underestimation of the performance. 
When conducting a statistical test to evaluate the performance of an AI system, it is particularly important to ensure that the (estimated) variance of the estimator is accounted for properly. The results of the test may otherwise become meaningless.

As outlined in Section~\ref{sec:add}, the sampling strategy can have a great influence on the value and interpretation of the reported performance estimate. 
The underlying distribution for the evaluation is defined by the Stochastic Application Domain Definition. If the wrong sampling strategy is employed, this can lead to performance estimates that do not reflect the context of the AI system envisioned by its developers.

\subsubsection{``Representativeness'' of Test Data}

\begin{figure}[h]
    \centering
    \begin{subfigure}[b]{0.42\textwidth}
        \includegraphics[width=\textwidth, trim={4cm 4cm 4cm 4cm}, clip=true]{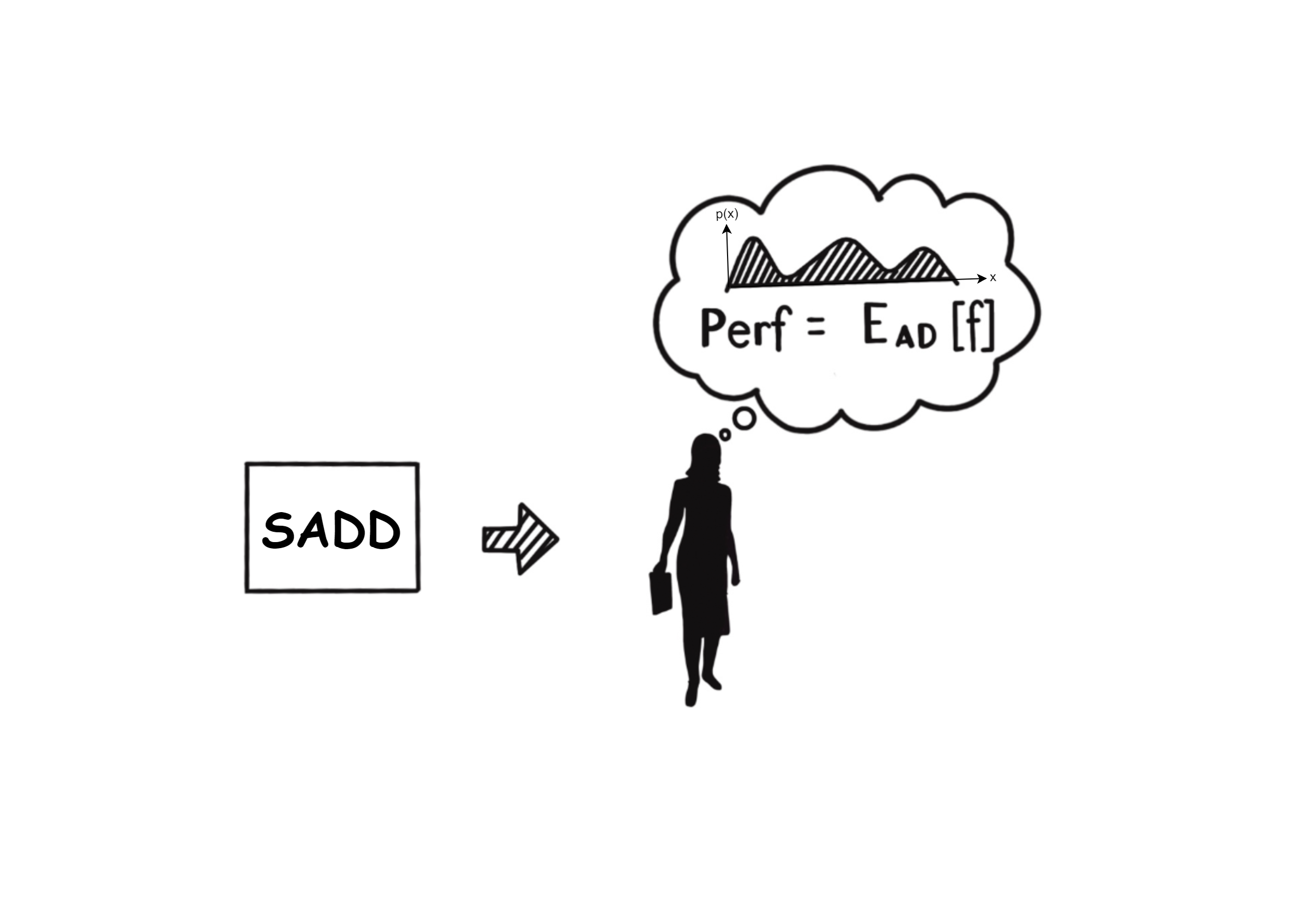}
        \caption{Single user's interpretation}
        \label{fig:single_user}
    \end{subfigure}
    \hfill %
    \begin{subfigure}[b]{0.57\textwidth}
        \includegraphics[width=\textwidth, trim={2cm 3cm 2cm 3cm}, clip=true]{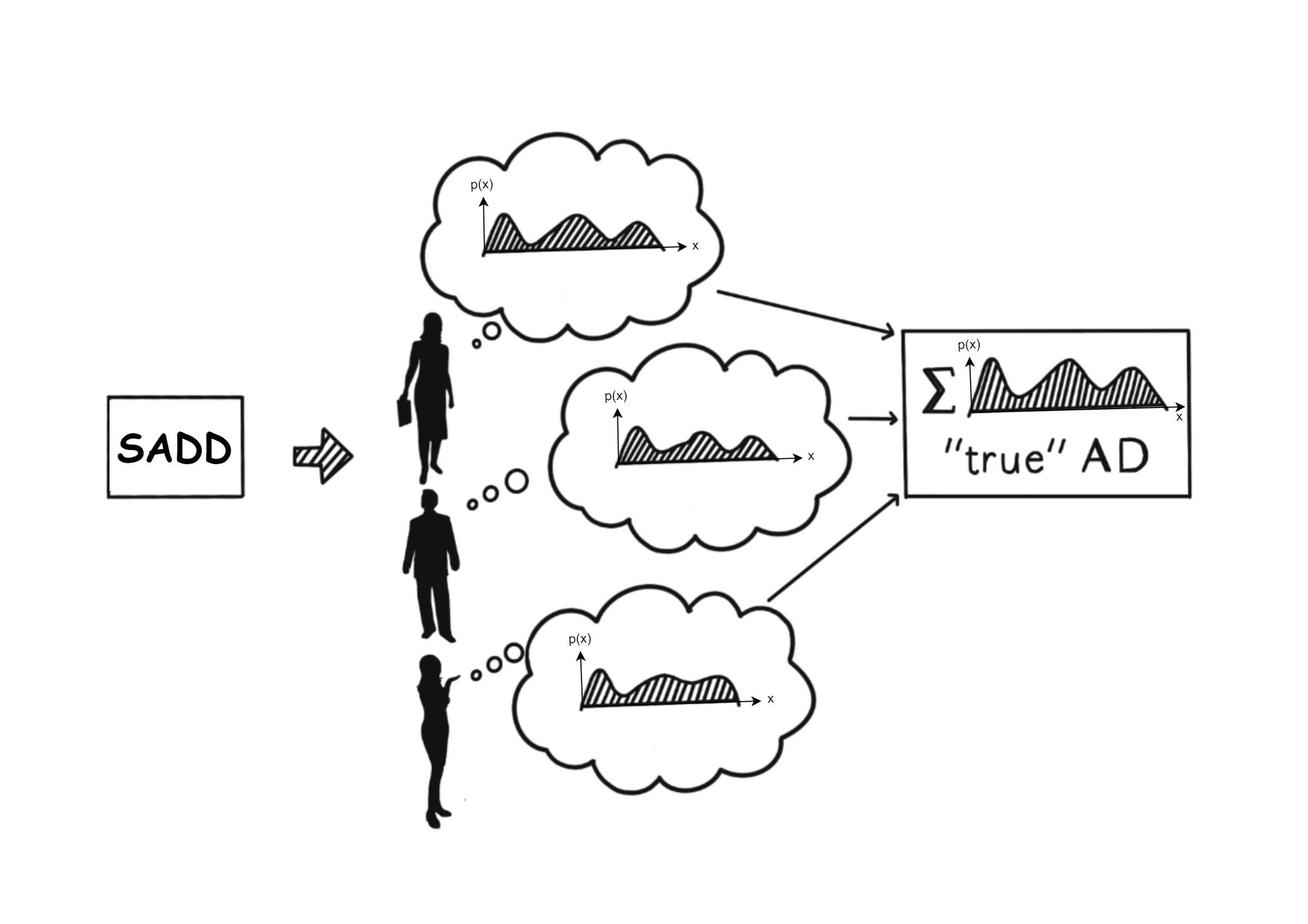}
        \caption{Collective interpretation}
        \label{fig:average_user}
    \end{subfigure}
    \caption{Interpreting the Stochastic Application Domain Definition (SADD). (a) A single informed user’s interpretation of the SADD enables understanding of ML model performance in the application domain (AD). (b) The collective interpretation, based on the average informed user's understanding, defines the ``true'' AD. Adapted from \citet{add_airov}.}
    \label{fig:user}
\end{figure}

One question that naturally arises in the context of certification is how to assess whether the test dataset can be considered \textit{representative} for the defined application domain with respect to the minimum performance requirements. 
This is the case if the samples are drawn independently and randomly following the descriptions in the Stochastic Application Domain Definition (SADD) described in parts (1), (2), and (3) defined in Section~\ref{sec:add}. 
However, the SADD is usually provided as a text and cannot be fully formalized in a mathematical sense. Instead, we need to rely on human interpretation (see Figure~\ref{fig:single_user}) of the textual descriptions. Even if these are worded carefully, their interpretation will differ slightly from human to human. In order to determine whether a sample is representative with respect to the defined application domain, we therefore need to rely on the concept of the ``average informed user'', commonly used in the legal context. In domains where individual opinions may vary, the concept of an average informed user covers a range of subjective opinions. This approach is similar to the legal standards used in product liability cases, where the expectations and reactions of a hypothetical ``average consumer'' are used to evaluate the accuracy and completeness of product information, such as labels, instructions, and advertising claims. In the context of this concept, the true application domain can be seen as the aggregated understanding of the SADD based on the individual interpretations of reasonably informed users. 
This concept is illustrated in Figure~\ref{fig:average_user}.
For a more detailed introduction to the concept of the SADD, refer to \citet{add_airov}.

The representativeness of training, validation, and test data is addressed in Article 10 (3) of the EU AI Act.
However, there is as yet no exact definition of what representativeness entails in the specific context of high-dimensional datasets for machine learning.
We believe that our approach of determining the SADD and sampling independently at random according to this definition is a technically sound way of interpreting the term ``representativeness'' of the test data.

\newpage

\subsubsection{Statistical Testing for Performance Evaluation}

Statistical testing when evaluating machine learning models ensures that observed performance metrics such as accuracy, precision, recall, and mean squared error (MSE) are reliable indicators of the model's performance on unseen data.
Since the test data is typically a random subsample of a larger population (the application domain of the AI system), the observed performance metrics can vary due to sampling variability. Statistical testing helps to account for this randomness by determining whether the observed performance is statistically significant and can be generalized to the larger population.
Specifically, when evaluating the performance of a machine learning model, we are interested in checking whether the model performance significantly exceeds the thresholds set by the MPRs as described in Section~\ref{sec:mpr-definition}. The following hypotheses are typically formulated for the statistical test:
The \textbf{null hypothesis $H_0$}, which states that the performance of the model is less than or equal to a given MPR, i.e., the performance is not sufficient, and the \textbf{alternative hypothesis $H_1$}, which states that the performance of the model meets or exceeds the MPR, implying that it performs well enough for the task.
Rejecting the null hypothesis implies that there is enough statistical evidence to conclude that the model performance meets the required standard. The MPR is not just exceeded by chance.

\begin{figure}[h]
    \centering
    \includegraphics[width=\linewidth, trim=0 0 0 -0.3cm]{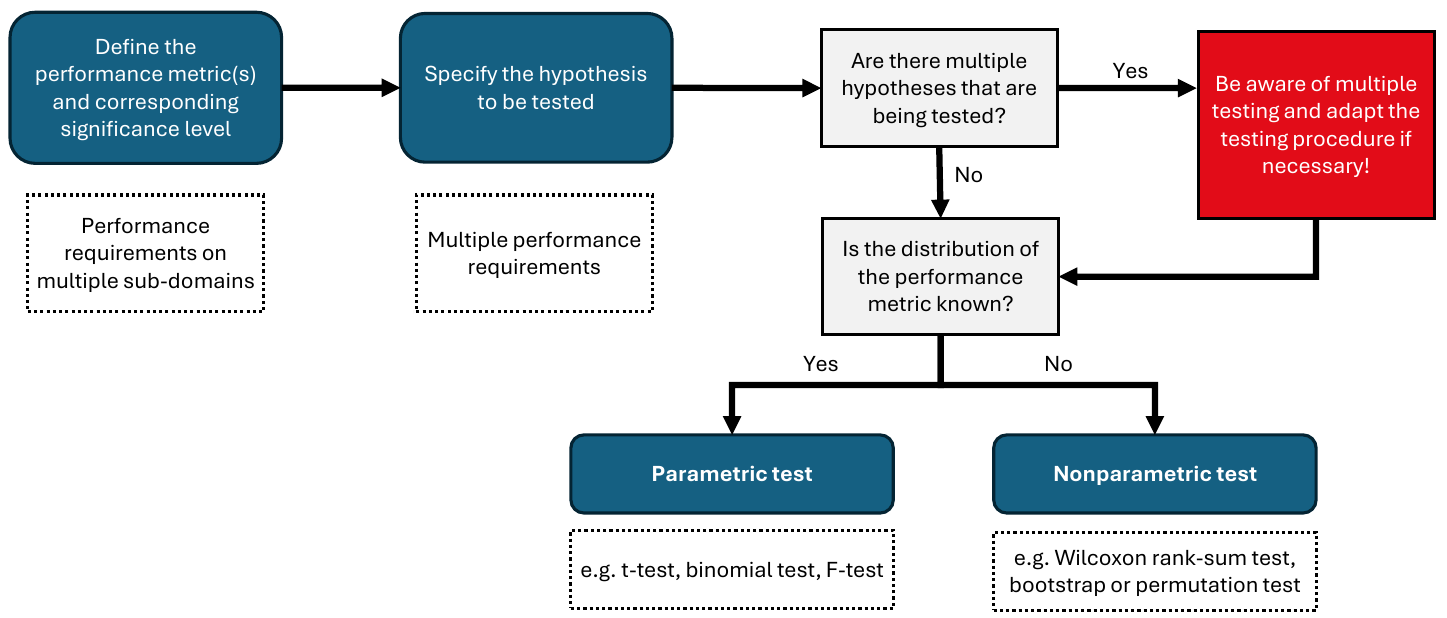}
    \caption{Decision flowchart for statistical testing of performance metrics: The procedure starts by defining the performance metrics and corresponding significance level for the tests and the specification of the hypotheses that are being tested. Depending on the theoretical distribution of the performance metric, we can either carry out a parametric test (see e.g.\ \citet{casella2024statistical}) or a non-parametric test (see e.g.\ \citet{wasserman2006all}).\vspace{0.3cm}}
    \label{fig:decision_flowchart}
    \label{fig:placeholder}
\end{figure}

Decisions in statistical testing are taken by calculating a so-called \textit{test statistic} from the data. Its value is then compared to a reference value to reach a conclusion. However, since the test dataset is randomly drawn from the overall population, there is always a certain probability of error. There are two types of errors:
A \textbf{Type I error ($\alpha$)} occurs if the null hypothesis is incorrectly rejected when it is actually true. The significance level $\alpha$ represents the probability of making this error. $\alpha$ is typically set to 0.05, reflecting a $5\%$ chance of incorrectly rejecting $H_0$.
The \textbf{Type II error ($\beta$)} occurs when the null hypothesis is not rejected even though it is false. In this case, the conclusion would be that the model does not meet the performance requirement, even though it actually does.

It is not possible to control both errors at the same time: lowering $\alpha$ to avoid type I errors (i.e., to be more certain when rejecting the null hypothesis) increases the probability of commiting a type II error where a false null hypothesis is not rejected. Thus, only the rejection of the null hypothesis leads to a meaningful decision ~\citep{barnard1949statistical}.

Multiple factors must be taken into account when selecting an appropriate statistical test for the performance of the AI system. These include (knowledge of) the theoretical distribution of the performance metrics of interest, the number of relevant (sub-)domains with separate performance criteria, and whether multiple performance metrics are being tested on the same dataset. The flow chart in Figure \ref{fig:decision_flowchart} illustrates the decision flow.

\textbf{Example: Binomial test to validate performance metrics that are measured as proportions}

For a classification system whose performance is measured on $n$ samples based on accuracy, defined as 
\begin{align}
    \widehat{\text{acc}} = \frac{\#\text{correctly classified samples}}{\#\text{all samples}} \ ,
\end{align}
we assume the MPR for accuracy is 0.9. We further assume the observed accuracy is $\widehat{\text{acc}}=0.94$. We aim to determine whether the observed accuracy exceeds the MPR at significance level 0.05. Thus, we aim to test the hypotheses
\begin{align}
    H_0:~\widehat{\text{acc}}\leq 0.9 \quad \text{vs.} \quad H_1:~\widehat{\text{acc}} > 0.9 \ .
\end{align}
Since $\widehat{\text{acc}}$ is a proportion, we can apply the binomial test (see, e.g., Ch.\ 9 in \citet{degroot2012probability}).
 Using the R implementation \citep{R:23}, the test yields the following results, assuming the test data consists of $n=100$ and $n=200$ samples, respectively:
\begin{center}
\begin{tabular}{ccccc}
    $n$ & correct classifications & $\widehat{\text{acc}}$ & confidence interval & p-value \\ \hline
    100 & 94 & 0.94 & [0.885, 1.000] & 0.117 \\
    200 & 188 & 0.94 & [0.904, 1.000] & 0.032 \\
\end{tabular}
\end{center}
The p-value for the test with $n=100$ is 0.117, thus larger than the significance level. The evidence from the test data is not large enough to reject the null hypothesis. 
However, if we estimate the same accuracy based on $n=200$ samples, the test yields a p-value of 0.032, lower than 0.05. Observing an accuracy of 0.94 with 200 samples is sufficient to reject the null hypothesis. In this case, the accuracy of the system \textbf{significantly} exceeds the MPR.

\subsubsection{Testing Multiple Hypotheses: The Multiple Comparison Problem}
\label{sec:multiple_testing}

\begin{figure}[b!]
    \centering
    \includegraphics[width=\linewidth]{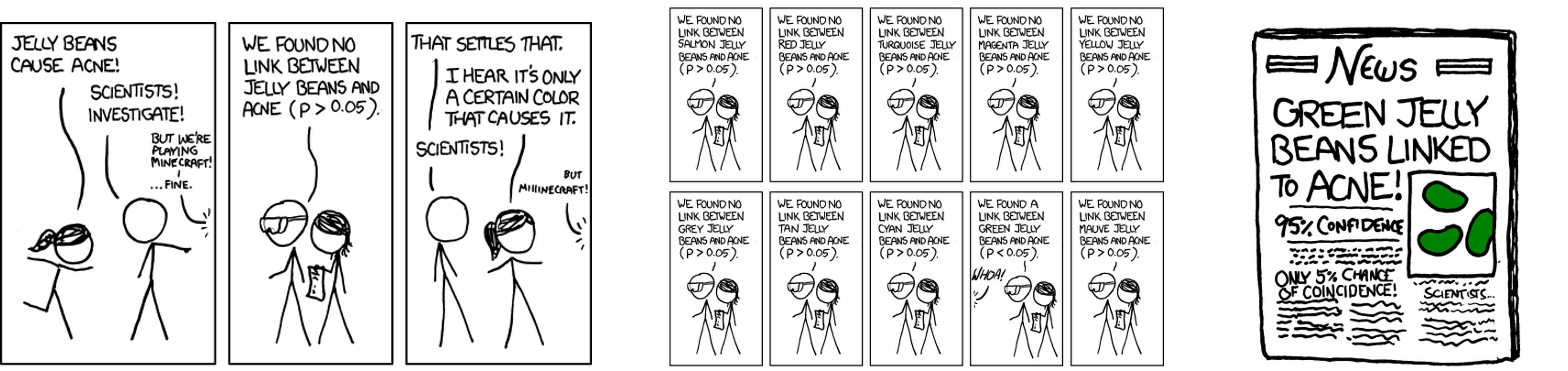}
\caption{Multiple testing increases the probability of falsely rejecting the null hypothesis and arriving at wrong conclusions. Comic taken from \href{https://xkcd.com/882/}{https://xkcd.com/882/}.}      
    \label{fig:mulitple-testing-comic}
\end{figure}

Often, multiple performance requirements need to be tested on the same dataset. This leads to a so-called \textit{multiple comparison problem}.
In fact, the multiple comparison problem can arise in many settings during the certification and assurance of AI systems, including monitoring and the assurance of AI systems that learn and are continuously updated. 
As every statistical test that is performed has a type I error probability of $\alpha$, each new test increases the probability of making at least one mistake, even though $\alpha$ is controlled for the individual test. This problem is illustrated in Figure~\ref{fig:mulitple-testing-comic}. The more tests are carried out, the more likely it is that a false conclusion is drawn.
Since we perform a test at level $\alpha$, the probability of not falsely rejecting the null hypothesis is $1-\alpha$. If we now perform $n$ comparisons on the same dataset, the probability of not falsely rejecting at least one of the null hypotheses is $(1-\alpha)^n$, which can very quickly become very low. 
The probability of making at least one error in a series of tests is referred to as the Family-Wise Error Rate (FWER).
The FWER increases as each additional test compounds the chance of error in the family of tests conducted. This leads to an overall increase in the probability that at least one of these tests will produce a false positive result\citep{tukey1953problem}.
There are different ways to control the FWER. One well-known and effective correction is the \textit{Bonferroni correction}~\citep{bonferroni1936teoria}. Each of the $n$ tests is hereby performed at the level $\alpha/n$, leading to an expected error rate of
\begin{align}
    \left(1-\frac{\alpha}{n}\right)^n \ \geq \ (1-\alpha) \ .
\end{align} 
However, if $n$ is large, this can lead to tests that are over-conservative. Other, more sophisticated methods are available to conserve a higher power of the statistical tests (see also Section~\ref{sec:continuous_learning}).

\subsection{Data Leakage} 

\begin{figure}[h]
    \centering
\includegraphics[width=\textwidth, trim=0 0 0 -0.3cm, clip]{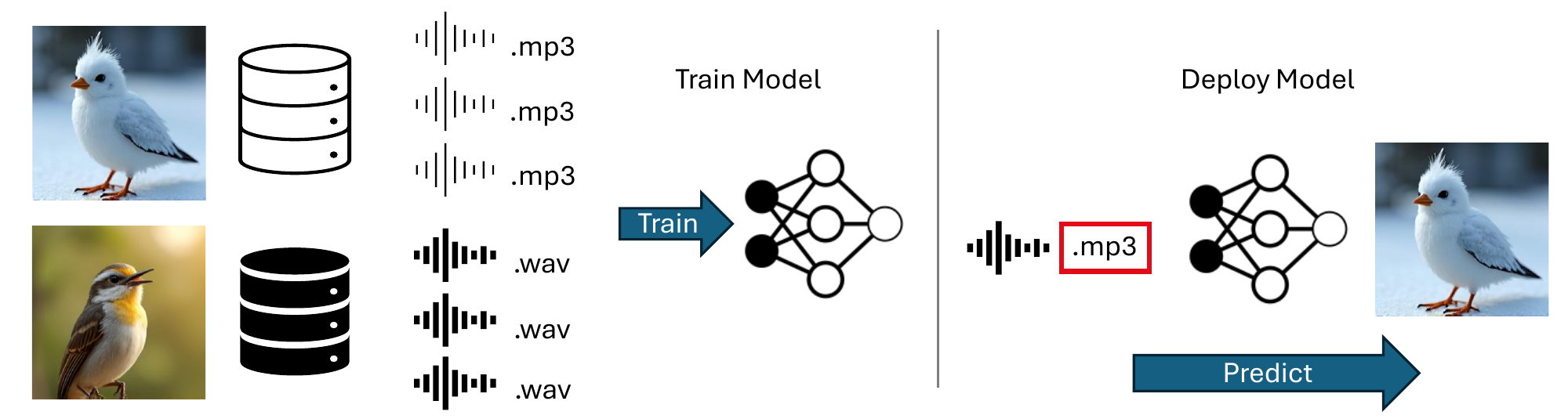}
    \caption{Example of data leakage: Consider a model designed to classify the audio of bird calls. Assume the training data for the classes (bird species) comes from different sources that use different audio encoding methods, such as lossy versus lossless encoding. If no pre-processing is performed to unify the file formats, the model could inadvertently learn to differentiate between the encoding formats rather than the actual audio content representing the bird calls. Consequently, if the model encounters new data during operation, all encoded using a single method, it may fail to generalize correctly and provide inaccurate predictions.
}
    \label{fig:DataLeakage}
\end{figure}

Data leakage is a widespread problem in machine learning that can tacitly undermine the reliability and generalizability of otherwise carefully crafted models. It occurs when information from the target variable inadvertently ``leaks'' into the training data, allowing the model to exploit this information to make predictions. This leakage can take many forms, from small errors during data collection, over seemingly innocuous data pre-processing steps, right through to subtle issues with feature engineering. An example of a data leakage scenario is shown in Figure~\ref{fig:DataLeakage}.

We identify the possible occurrence of data leakage at two different points of model development, leakage in data collection, and leakage in pre-processing \citep{kapoor2023leakage}.

\subsubsection{Data Leakage in Data Collection}

We identified three major causes of data leakage during data collection: duplicates, privileged information that is not available during prediction, and group information.

\paragraph{Duplicates in datasets:} 
When duplicates are present, the model essentially sees the same data point multiple times, once in the training set and again in the test set. This can cause the model to overfit the duplicates, learning to memorize the specific data points rather than learning generalizable patterns. As a result, the model's performance on the test set will be artificially inflated, leading to overly optimistic estimates of its true performance. This can be particularly problematic if the duplicates are not representative of the broader population, as the model may learn to recognize and exploit the duplicates rather than learning to generalize to new, unseen data. 

\paragraph{Target leakage:} 
Can occur when the target variable or a proxy of the target variable is inadvertently included as a feature in the training dataset. This can lead to data leakage, as the model essentially gains access to information that it wouldn't have at the time of prediction, allowing it to cheat and achieve an artificially high performance. For example, if a feature is a post-event indicator that only makes sense in the context of the target variable, the model is able to exploit this extra information to make more accurate predictions on the training set. However, when the model is deployed to make predictions on new, unseen data, it will not have access to this additional information, and its performance is likely to deteriorate significantly. 

\paragraph{Group leakage:} 
This occurs when information about the group or subset of data to which an instance belongs is inadvertently included in the training data. This can happen when the group membership is not explicitly included as a feature, but other features are correlated with the group membership. As a result, the model may learn to rely on these correlated features to make predictions, effectively leaking information about the test data. One famous example of group leakage was the first version of \citet{rajpurkar2017chexnet}, where a model was presented for vision-based pneumonia detection. The model was trained on the ChestX-ray14 dataset \citep{wang2017chestx}, which contains 112,120 frontal view X-ray images of 30,805 unique patients. For training purposes, the dataset was randomly split into 80\% training and 20\% validation data. This created a group leakage scenario, since the same patients were present in both the training and the validation data, and the model memorized patient-specific features to predict the target during validation. Later revisions took this grouping into account for the split, which resulted in a poorer performance.

\subsubsection{Data Leakage in Pre-processing}

Furthermore, we identified several conditions that can lead to data leakage during pre-processing.
These include feature selection or engineering without explicitly excluding the test data, temporal leakage, and generally privileged information during model development.

\paragraph{Feature selection:} 
This is a common step in machine learning pipelines, where a subset of the most informative features is chosen for use in model training. However, if feature selection is performed on the entire dataset, including the test set, it can lead to data leakage. As a result, the selected features may be overly optimistic and biased towards the test set, leading to an overestimation of the model's performance. 
In order to avoid data leakage, it is essential that feature selection is only performed on the training set, using techniques such as cross-validation to evaluate the feature selection process.

\paragraph{Feature engineering:} 
Data leakage occurs when information from the test set or future data is inadvertently incorporated into the training process, resulting in overly optimistic model performance estimates. 
Feature engineering can lead to data leakage if features are created using aggregated statistics or transformations that rely on the entire dataset, including the test set. 
For example, scaling or normalizing features using statistics computed from the entire dataset can leak information from the test set into the training set. 
To avoid data leakage, it is essential that feature engineering is performed exclusively on training data and that all aggregations or transformations for training and test set are computed separately.

\paragraph{Temporal leakage:} 
This occurs when a machine learning model is trained on data that includes information from the future, which is then used to make predictions on past data. This is normally the case when dealing with sequential data. A common error here is the use of random splits during dataset-splitting. When dealing with sequential data, a random split will inevitably lead to a training set that contains data which lies chronologically after data that belongs in the test set. The model is essentially being given access to information that would not have been available at the time of testing. As a result, the model performance may be inflated. For example, if the training data in a stock market prediction model includes stock prices after the prediction date, the model may learn to use these future prices to inform its predictions, rather than relying on only the information available prior to the prediction date.
Therefore, temporal splitting is usually performed to avoid this issue, where the most recent data is used for testing.

\paragraph{Privileged information during development:}
The dangers of privileged information during development are twofold.
Firstly, if the model developer has access to or uses knowledge about the test dataset, it can lead to overfitting to that specific test set. 
This invalidates the purpose of using test data for an unbiased evaluation. Model design should rely only on the training (and possibly validation) data.
Secondly, models are sometimes trained with features (privileged information) that will not be available in real-world use,
for example, knowing the number of people in a household when predicting energy consumption. 
Even if this feature is present in both the training and test sets, relying on it can lead to a significant drop in performance when the model is deployed without it.

There are various ways to detect and investigate potential data leakage. 
When using tabular data, a thorough data exploration can reveal certain features that exhibit unusually strong correlations with the target. 
Similarly, training a model using each feature individually can reveal data leakage if a model achieves very high accuracy on only one feature.
Explainability methods are also a possible way to detect leakage, see Section~\ref{sec:explainability}.
Data leakage can be quite subtle and extremely difficult to detect, and it is crucial to consider its risks throughout the development lifecycle, beginning with data collection. 
A thorough understanding of the data and a healthy skepticism, especially when the results appear too good to be true, are essential to avoid the dangers of data leakage.

\subsection{Robustness} \label{subsec:robustness}

\begin{figure}[h]
    \centering
    \includegraphics[width=\linewidth]{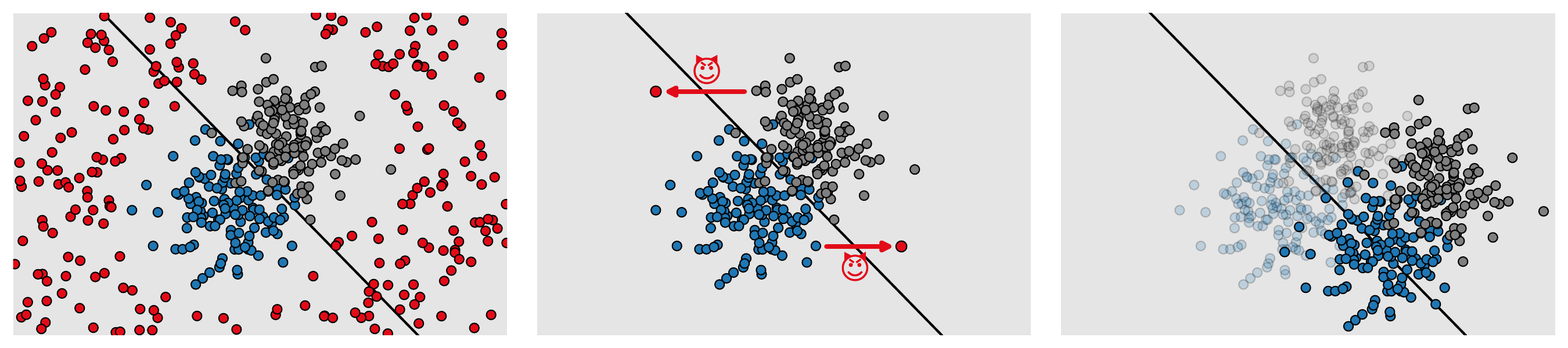}
    \caption{Different aspects of robustness. The application domain yields datapoints that are either class 1 (blue) or class 0 (gray). A model is trained to classify new samples, the decision boundary is shown as a black line. Left: Out-of-distribution samples (red points). Middle: A malicious attacker creates adversarial examples (red points). Right: A distribution shift occurs, after which the model performance drops.}
    \label{fig:robustness}
\end{figure}

One of the key advantages of machine learning models over traditionally programmed functions is their ability to generalize across a wide range of scenarios. 
In practice, the input space for any given application is often so large that it becomes impractical to enumerate and test every possible case exhaustively. 
Moreover, machine learning models are frequently used in dynamic, open environments rather than strictly controlled settings such as specific industrial machines. 
As a result, they may encounter situations that were not present in the training data but still require correct handling, ideally through a failsafe mechanism. 
Therefore, ensuring the robustness of these models is critical.
We define robustness as the retention of a certain level of performance under deviations from the application domain.
This section will explore several key aspects of assessing and improving robustness (see Figure~\ref{fig:robustness}), including out-of-distribution (OOD) detection, adversarial example detection, uncertainty quantification, and domain shift detection.

\subsubsection{Out-Of-Distribution (OOD) Detection} \label{sec:ood}

The use of high-risk ML applications often suffers from the possibility that anomalous or out-of-distribution (OOD) inputs to the application can occur. For example, in autonomous driving, it is important that the system warns the driver and transfers control when it encounters unfamiliar situations or objects that it has not been trained to process and for which it cannot make a safe decision. 
High-risk ML applications thus need a way to properly detect and handle inputs on which they are not designed to operate. 
For this reason, OOD detection plays a crucial role in ensuring the reliability and safety of machine learning systems.
Following \citet{Ruff:21}, a new input $x$ is OOD, if it is within the set
\begin{equation}
    \mathcal{O} = \left\{ x \in \mathcal{X} \mid p(x) \leq \epsilon \right\}, \quad \epsilon \geq 0 \ .
\end{equation}
Most current machine learning models are trained under the closed-world assumption~\citep{Krizhevsky:12,He:15}, which assumes that future data encountered by the model are independently and identically distributed (i.i.d.) with respect to the training data (also known as the \emph{in-distribution} (ID) assumption).
The ID assumption implies that both the training data and the test or future data are i.i.d.\ samples drawn from the reference distribution $p(x)$, which is defined within the context of our Stochastic Application Domain Definition, as detailed in Section~\ref{sec:functional-trustworthiness}.

However, when these models are used in an open-world setting~\citep{Drummond:06},  samples may come from out-of-distribution (OOD), which requires careful consideration.
Related tasks, such as anomaly detection, novelty detection, open set recognition, and outlier detection, have a similar motivation and methodologies to OOD detection~\citep{yang2022openood}. All of these tasks define a specific in-distribution and share the common objective of identifying out-of-distribution samples within the context of an open-world assumption.
They can be unified under the notion of ``generalized OOD'' detection~\citep{Yang:24} by using the general concept of distribution shifts. The individual tasks are special cases of OOD detection depending on the specific shift scenario, e.g., covariate shift (shift in $p(x)$ while $p(y \mid x)$ remains the same) or prior shift (also called label shift; shift in $p(y)$ while $p(x\mid y)$ remains the same).

\begin{figure}[h]
    \centering
    \includegraphics[width=\textwidth]{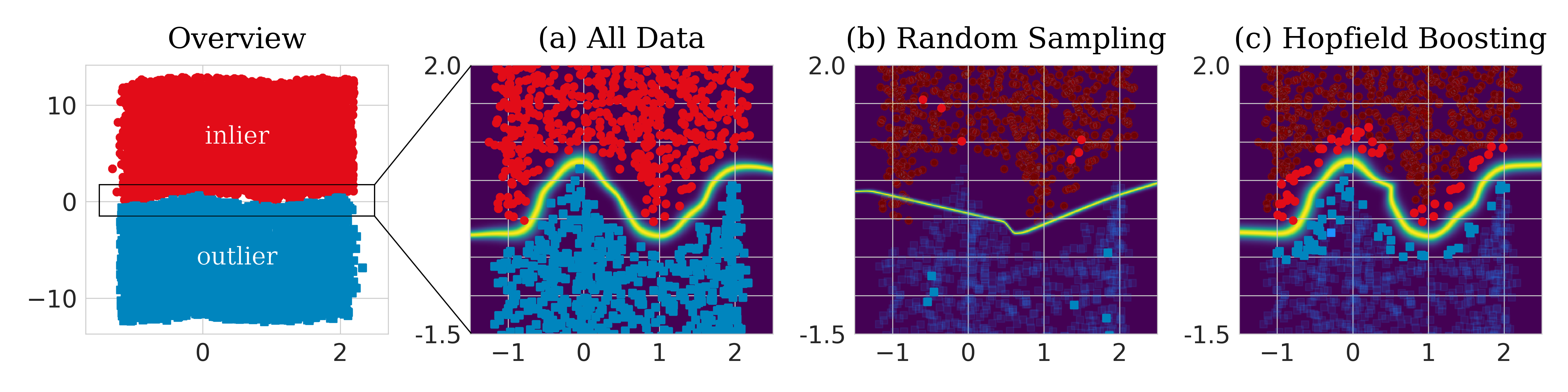}
    \caption{Hopfield Boosting creates a strong learner by sampling and combining a set of weak learners close to the decision boundary of inlier and outlier samples. Adapted from \citet{Hofmann:23}.}
    \label{fig:patchcore}
\end{figure}

The main lines of work in OOD and closely related anomaly detection can be categorized into
\begin{enumerate}[noitemsep]
    \item classification-based methods,
    \item density-based methods,
    \item distance-based methods,
    \item and reconstruction-based methods.
\end{enumerate}
\textit{Classification-based methods} train explicit classifiers \citep{hendrycks2018deep} and use their output, such as softmax scores, to distinguish between ID and OOD samples.
\textit{Density-based methods} work by modeling the in-distribution using probabilistic models and identifying data points in low-density~\citep{Zong:18} %
areas or with high uncertainty predictions~\citep{hendrycks2016baseline} as OOD data points.
Similarly, density ratios can be used for inlier-based outlier detection~\citep{sugiyama2012density} where a sample is likely to be an outlier or OOD if the estimated density ratio is close to zero. We introduced algorithms~\citep{Gruber:24,gruberimproved} that improve a large class of density ratio estimation methods for problems with higher regularity and where hyperparameter selection is difficult.
The fundamental concept behind \textit{distance-based methods} is that out-of-distribution test samples should be located at a relatively greater distance from the centroids or prototypes of in-distribution classes~\citep{Lee:17,Hofmann:23}, thus exploiting the similarity of inputs to the training data~\citep{Roth22}. One example of a distance-based approach that uses Hopfield boosting is shown in Figure~\ref{fig:patchcore}.
Finally, \textit{reconstruction-based methods} rely on the assumption that an encoder-decoder framework trained on in-distribution data typically produces distinct results for ID and OOD samples. This performance disparity can serve as a signal for OOD detection. For example, reconstruction models trained solely on ID data struggle to accurately recover OOD data~\citep{Denouden:18}, making it possible to identify OOD samples.
Furthermore, well-established OOD detection methods based on classical ML algorithms such as SVMs and Random Forest~\citep{scholkopf2000support, liu2008isolation} that are well suited for structured datasets, such as tabular data.

\textbf{In the context of certification}, we require auditees to explicitly define OOD / anomalous scenarios, 
in manner similar to the Stochastic Application Domain Definition described above.
Test data for the OOD detection mechanism should be sampled from each scenario, yielding respective datasets that can be used to empirically validate the OOD / anomaly detection approach.
A statistically valid test requires the selection of a threshold for OOD samples using the validation set and the calculation of performance in detecting OOD or anomalous samples using the respective test dataset. 

Our requirements focus on the correct selection of the anomaly / OOD detection method for the application domain of the AI system, the correct collection of the relevant datasets, the correct empirical evaluation and the fallback procedure if samples are detected to be OOD or anomalous during operation of the AI system.

\subsubsection{Adversarial Examples}

\begin{figure}[h]
    \centering
    \includegraphics[width=\linewidth]{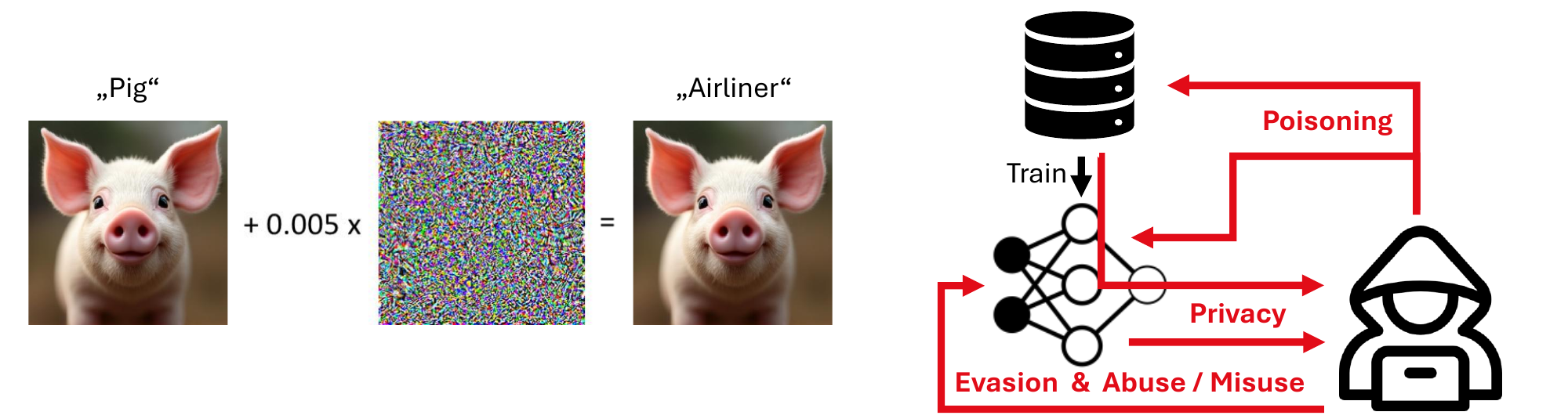}
    \caption{Instructional illustration for an adversarial example and broader picture of adversarial machine learning. The left image (recreated after \href{https://gradientscience.org/intro_adversarial/}{https://gradientscience.org/intro\_adversarial/}) shows a classical instance of an adversarial example, fooling a classifier to predict that the pig is an airliner. The right image (based upon \href{https://github.com/Trusted-AI/adversarial-robustness-toolbox}{https://github.com/Trusted-AI/adversarial-robustness-toolbox}) depicts an overview of adversarial machine learning scenarios.}
    \label{fig:adversarial_examples}
\end{figure}

Adversarial examples can be compared to OOD samples in that they are samples where machine learning system will not achieve good results.
However, they are usually intentionally created from in-distribution samples to fool the network into making a wrong prediction. This is called evasion attack in the adversarial/robust machine learning literature.
Formally, the attacker solves some variant of the following constrained optimization problem:
\begin{equation} \label{eq:ae}
    \max_{\delta} L(f(x + \delta), y) \quad \text{s.t.} \quad \|\delta\|_{p} \leq \epsilon \ ,
\end{equation}
where $\delta$ is the adversarial perturbation on the input and $\|\cdot\|_{p}$ is a suitable Lp norm\footnote{\href{https://en.wikipedia.org/wiki/Lp_space}{https://en.wikipedia.org/wiki/Lp\_space}} -- widely considered are $p = 1,2 \ \text{or} \ \infty$.
Intuitively, the aim is to find a perturbation $\delta$ to the input $x$ that maximizes the loss while satisfying the boundary condition for the perturbations norm, i.e., the perturbation is not easily noticeable (see Figure~\ref{fig:adversarial_examples} left).
For images, $\epsilon$ is often set to small positive values such as $8/255$, motivated by the sensitivity of human perception.
Note that Eq.~\eqref{eq:ae} is a very generic formulation of an adversarial attack for illustrative purposes. 
Individual attack methods represent different optimization problems, and white-box or targeted attack settings, for example, would need additional specification.

The currently prevalent understanding of why adversarial examples can be constructed is the local linearity of neural networks \citep{Goodfellow:15}.
Although this is not a problem exclusive to neural networks, the term was coined for this type of machine learning model, which is why we are focusing on it here.
The perhaps most surprising fact about adversarial examples is their high transferability between individual models for the same data \citep{Szegedy:14}.
This also makes models vulnerable in black box settings, where the model is unknown if an approximation can be constructed by the attacker \citep{Athalye:18}.
In white box settings, where direct access to the model architecture is provided, defense mechanisms continue to struggle against attacks that are specifically designed against the defense mechanism \citep{Carlini:17, Carlini:17b}.
The exact mechanisms of how to obtain adversarial examples also depend on the input domain, i.e. if there are spatial relationships between features and whether features are discrete or continuous.
However, adversarial examples are also a valuable tool for understanding the robustness of machine learning models.
They allow the quantification of minimal input perturbation under which a reasonable performance can still be expected.

We have adopted a three-fold approach to assess robustness against adversarial examples.
Firstly, we require an assessment of potential attack vectors.
This generally requires defining a meaningful distance measure in the input space that can be used to quantify the severity of an adversarial perturbation.
Secondly, we require an assessment of possible defense mechanisms and their feasibility.
Thirdly, we require an assessment of the robustness of the machine learning model by state-of-the-art adversarial attack algorithms suitable for the application domain and the model used.
Furthermore, if a specific defense mechanism is implemented, the additional robustness arising from this defense has to be assessed.

Future work will focus on covering other aspects of adversarial machine learning, such as data poisoning, privacy attacks, and abusion / misuse of machine learning models (see Figure~\ref{fig:adversarial_examples} right).
Data poisoning refers to placing a backdoor within the training data to trigger a certain behavior in the deployed machine learning model.
For example \citet{Carlini:24} showed that one can easily and cheaply buy domains and replace images for web-scale datasets, widely used nowadays to pre-train large machine learning models.
Privacy attacks are particularly relevant for generative models.
\citet{Carlini:23}, for example, showed that it is possible to reconstruct individual training samples under certain conditions.
However, more subtle privacy attacks, such as membership inference, can also pose a problem for applying machine learning models in practice.
Abuse / misuse attacks are particularly relevant for generative language models.
For example, \citet{Zou:23} showed that there are universal and transferable attacks on aligned language models, which trigger actionable answers to questions such as how to steal from a charity and the like.
Most importantly, they showed that attack suffixes obtained from smaller open-source models can be transferred to large-scale closed-source models provided via a public API.

\subsubsection{Domain and Distribution Shifts} \label{sec:dshift}

A recent Nature review on AI~\citep{Wang:23} identifies out-of-distribution (OOD) generalization as a critical and unresolved challenge at the forefront of AI research. Neural networks trained on data from a specific domain may capture patterns that do not generalize when applied to data from a shifted distribution. In fact, standard ML models frequently struggle when faced with input data that exhibit statistical properties different from those present in the training set. This limitation is evident in various applications, including the following:
\begin{itemize}[noitemsep]
    \item Medical diagnostics, where AI systems provide incorrect predictions when they encounter previously unseen variations in human physiology~\citep{guan2021domain}
    \item Industrial quality inspection, where models fail when deployed under modified system configurations~\citep{zellinger2020multi}
    \item Chemical measurements systems require recalibration following changes in experimental setups~\citep{nikzad2018domain}
    \item Computer vision benchmarks, where the performance of state-of-the-art models declines on ImageNet test sets derived from more recent internet datasets, despite adhering to the original data collection protocol~\citep{recht2019imagenet}
\end{itemize}

Distribution shifts in machine learning refer to the scenario in which the distribution of data on which a model is trained and tested differs from the distribution of data it encounters in the real world. 
Such discrepancies can be caused if the system's training and test data do not reflect the operating conditions of the AI application, i.e., if the application domain is not defined properly and not represented by the system's training and test data, but also if the distribution in the real world changes over time.
Distribution shifts can lead to a range of adverse outcomes, including poor generalization of the AI application, inaccurate or invalid risk estimation, and arbitrarily severe performance degradation. These effects are not just theoretical concerns, but have real-world implications, especially in high-stakes applications such as healthcare~\citep{Habib:21}, self-driving cars and the legal domain~\citep{Hao:19}. 
More precisely:
\begin{itemize}[noitemsep]
    \item Poor generalization due to distribution shifts can occur if the model's training data no longer represent the environment in which the model operates. Thus, the assumptions imposed by the model are no longer valid.
    \item Inaccurate or invalid risk estimation can be caused by violations of the i.i.d.\ assumptions of the test set and unknown future data. This can lead to arbitrarily severe performance degradation~\citep{Rabanser:19,Ovadia:19}.
    \item Distribution shifts can also exacerbate the biases present in the training data or introduce new biases. This can lead to unequal service quality and high economic costs, as well as violations of legal regulations, ethical principles, and guidelines.
\end{itemize}

A wide range of methods for measuring the similarity between datasets has been proposed in the literature~\citep{stolte2024methods}. These methods vary in complexity, including simple summary statistics, comparisons of densities and cumulative distributions, discrepancy measures for distributions, and techniques based on kernel embeddings.

A simple method for assessing the similarity between two distributions, given finite samples, is to compare fundamental summary statistics such as mean and variance. However, it is well known that distinct distributions can exhibit highly similar summary statistics, leading to potential misinterpretations. Moreover, techniques exist that are designed to generate datasets with visually diverse structures while preserving identical summary statistics, further complicating reliance on these measures for distributional comparison~\citep{matejka2017same}.

\begin{figure}[b!]
    \centering
    \includegraphics[width=\textwidth]{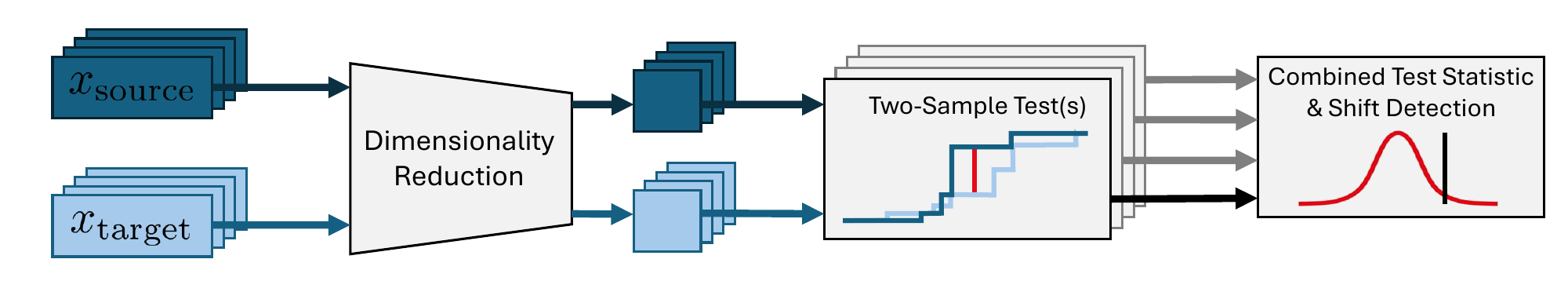}
    \caption{Dimensionality reduction and two-sample tests for distribution shift detection, recreated after \citet{Rabanser:19}.}
    \label{fig:2sample}
\end{figure}

For univariate distributions, the detection of distribution shifts using statistical two-sample tests such as the Kolmogorov-Smirnov test (KS) or the chi-square test (CS) is a well-established method. However, extending these methods to high-dimensional data, which are prevalent in ML applications, poses significant challenges and is by no means a simple matter.
Discrepancy measures between distributions, such as f-divergences, are a potential alternative; however, they have been shown to be vulnerable to adversarial attacks. Similarly, while standard kernel-based multivariate two-sample tests are attractive, they suffer from poor scalability as the dataset size increases, and their statistical power (the probability of correctly detecting a shift) deteriorates significantly in high-dimensional settings~\citep{ramdas2015decreasing}.

A common strategy to address these issues involves a two-stage approach, as shown in Figure~\ref{fig:2sample}. High-dimensional features are initially encoded using a deep learning-based dimensionality reduction technique. Two-sample tests are then applied to these extracted features to facilitate shift detection. In this context, univariate methods such as the KS or CS tests, when used in conjunction with the Bonferroni correction, provide a viable solution. Alternatively, multivariate approaches based on kernel embeddings are frequently used for distribution shift detection~\citep{Rabanser:19}.

We approach the problem of distribution shift detection by leveraging a neural network-based encoding of potentially high-dimensional data. This transformation enables the application of statistical hypothesis testing to assess distributional changes. Specifically, one approach involves conducting multiple univariate two-sample tests--one per feature dimension--while employing Bonferroni correction to control the family-wise error rate. Alternatively, a kernel-based embedding approach can be used for scenarios requiring a multivariate perspective, followed by a permutation test on the resulting kernel matrix. This methodology offers the dual advantage of maintaining stringent control over the probability of false positive shift detections while simultaneously enhancing the sensitivity to detect genuine distributional shifts.

We also advocate the adoption of advanced kernel embedding strategies, in which feature encoders are optimized to maximize the statistical power of the test, as proposed by~\citet{liu2020simple}. In addition, successor methodologies that improve kernel adaptability by using a collection of kernels~\citep{schrab2023mmd}—while concurrently improving both test power and computational efficiency~\citep{biggs2024mmd}—can be similarly integrated to improve performance.

We also consider the situation where distribution shifts are not harmful e.g., the probability of different subgroups in data changes where each respective subgroup still fulfills the MPRs. Therefore, we advocate for a combination of distribution shift tests with performance metrics.
Detecting distribution shifts is not a one-time task, but a continuous necessity. Since models are used in dynamic real-world environments, continuous monitoring and adaption are critical to maintaining performance. This ongoing process requires sophisticated systems capable of detecting shifts in real-time and either adjusting the model accordingly or alerting human operators to the need for intervention. 
Furthermore, the classification of distribution shifts as malignant and benign must be investigated by combining the detection methodology with monitoring of the model performance in cases where labeled data is readily available. 
One strategy to mitigate the consequences of distribution shift is continuous retraining, which will be discussed in Section~\ref{sec:maintaining}.

\subsubsection{Uncertainty Estimation}
\label{sec:uncert}

The risk of actionable prediction often has to be assessed with the help of predictive uncertainty. 
This is particularly important in high-risk ML applications, such as medical diagnosis or drug development, where human lives or extensive investments are at risk.
The prediction risk of ML methods is most commonly computed by averaging over a set of data points. 
Nevertheless, it is crucial for many applications to obtain
a risk assessment for each prediction.
Methods for quantifying predictive uncertainty provide such an estimate. 
Furthermore, LLMs become more widespread, yet struggle with hallucinated outputs.
It has been established that a particular type of hallucination -- confabulations -- is caused by the predictive uncertainty of the LLM \citep{Farquhar:24, Yadkori:24}.
Therefore, uncertainty estimation methods are becoming increasingly popular for hallucination detection \citep{Malinin:21, Kuhn:23, Aichberger:25, Farquhar:24}.

\begin{figure}[b!]
    \centering
    \begin{subfigure}[b]{0.48\linewidth}
        \centering
        \includegraphics[width=0.8\textwidth, trim=0mm 3mm 0mm 3mm, clip]{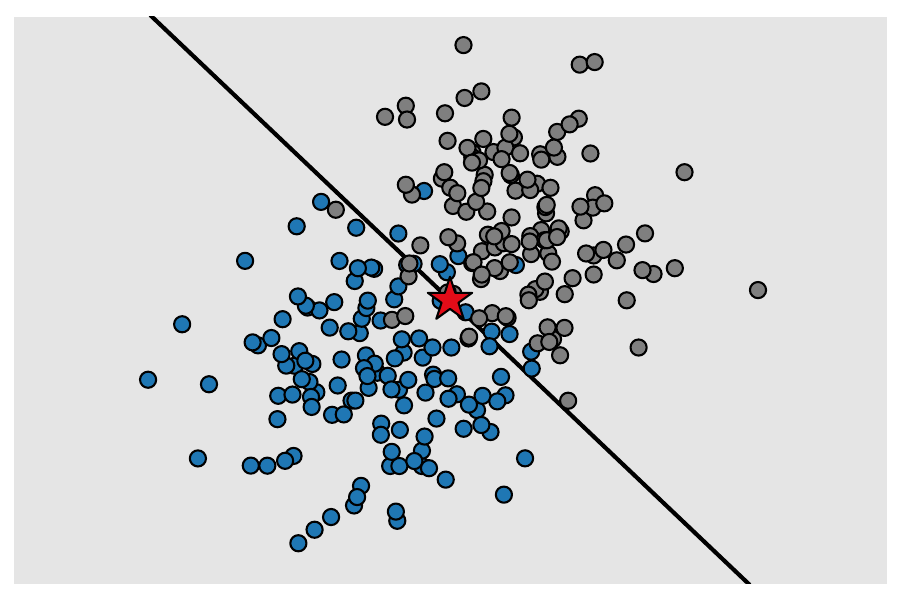}
        \subcaption{High aleatoric uncertainty}
        \label{fig:aleatoric}
    \end{subfigure}
    \begin{subfigure}[b]{0.48\linewidth}
        \centering
        \includegraphics[width=0.8\textwidth, trim=0mm 3mm 0mm 3mm, clip]{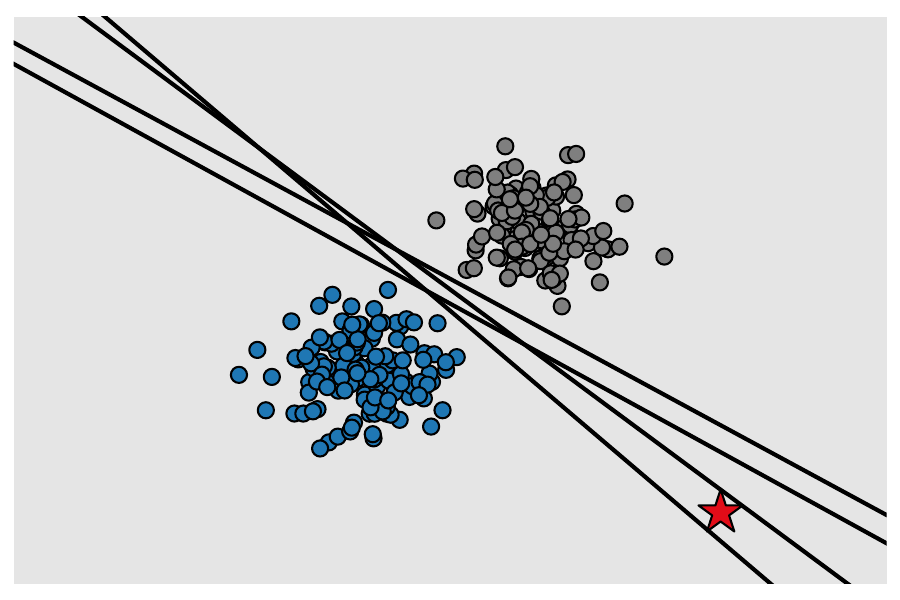}
        \subcaption{High epistemic uncertainty}
        \label{fig:epistemic}
    \end{subfigure}
    \caption{Example considering a binary classification dataset, represented by blue and gray dots. Machine learning models are optimized on the dataset, the decision threshold is shown by the black line. We are interested in the uncertainty in the prediction at a new test point, represented by the red star. (a) If the aleatoric uncertainty is high, it is inherently unclear to which class the new test point belongs to. (b) If the epistemic uncertainty is high, there is a lack of knowledge about the true model, many are possible and are consistent with the dataset, but differ in their prediction for the new test point.}
    \label{fig:uncert}
\end{figure}

The correct measurement of predictive uncertainty is an active area of research.
Although numerous approaches exist to measure various kinds of uncertainty for ML applications, the most widely distinguished notions of uncertainty are aleatoric uncertainty (inherent stochasticity) and epistemic uncertainty (lack of knowledge) as shown in Figure~\ref{fig:uncert}. 
The most commonly considered definition of a measure of predictive uncertainty is the Shannon entropy of the posterior predictive distribution \citep{Gal:16thesis, Kendall:17} given by
\begin{align} \label{eq:uncertainty_1}
    \underbrace{\mathrm{H}(p(y \mid x, \mathcal{D}))}_{\text{total}} \ = \ \underbrace{\mathrm{E}_{p(w\mid \mathcal{D})}\left[ \mathrm{H}(p(y \mid x, w)) \right]}_{\text{aleatoric}} \ + \ \underbrace{\mathrm{I}(p(y, w \mid x, \mathcal{D}))}_{\text{epistemic}} \ .
\end{align}
A Bayesian approach is thus used to represent uncertainty over possible models by using the posterior distribution $p(w \mid \mathcal{D}) = p(\mathcal{D} \mid w) p(w) / p(\mathcal{D})$ under a given dataset $\mathcal{D}$.
In addition, $p(y \mid x, w)$ is the predictive distribution for a single model with parameters $w$. 
For example, in classification, $p(y \mid x, w)$ is a categorical distribution represented by the softmax output of the model. 
Furthermore, $p(y \mid x, \mathcal{D}) = \mathrm{E}_{p(w\mid \mathcal{D})} \left[ p(y \mid x, w) \right]$ is the posterior predictive distribution.
Intuitively, the posterior predictive distribution is the average of the predictive distributions of models, weighted according to their posterior probability.
These uncertainty measures are defined by means of information-theoretic quantities; $\mathrm{H}(\cdot)$ denotes the Shannon entropy and $\mathrm{I}(\cdot)$ denotes the mutual information, %
see \citet{Cover:06} for an introduction.

It should be noted that there are ongoing debates regarding the behavior of these measures in different settings \citep{Malinin:21, Wimmer:23, Schweighofer:23b}.
Alternatives to the way posterior expectations are performed in Eq.~\eqref{eq:uncertainty_1} have recently been suggested \citep{Malinin:21, Schweighofer:23a, Schweighofer:23b, Hofmann:23, Kotelevskii:24, Schweighofer:24}, which are more suitable than Eq.~\eqref{eq:uncertainty_1} in certain practical settings.
Furthermore, there are alternatives to these types of measures, e.g., variance-based measures \citep{Depeweg:18, Sale:23b, Sale:24b} or distance-based measures \citep{Sale:24}.

Calculating uncertainty measures according to Eq.~\eqref{eq:uncertainty_1} requires approximating posterior expectations, generally through Monte Carlo sampling.
We identified well-established methods \citep{gal2016dropout, lakshminarayanan2017simple, liu2020simple, daxberger2021laplace, Sensoy:18} and introduced a state-of-the-art method \citep{Schweighofer:23a}.
Depending on the available computing budget both during the training stage and per prediction, different methods should be favored.
The most efficient, yet potentially suboptimal estimates can be provided by the Laplace approximation \citep{daxberger2021laplace} or evidential models \citep{Sensoy:18}.
Efficient during training, yet more demanding during inference are dropout ensembles \citep{gal2016dropout}.
Deep ensembles \citep{lakshminarayanan2017simple} are more computationally demanding during training and inference, but empirically very well performing.
The QUAM method \citep{Schweighofer:23a} improves upon deep ensembles without any extra effort compared to a single model during training, but incurs a higher effort during inference.

Most methods for \textbf{hallucination detection in LLMs} focus on accurately measuring the aleatoric uncertainty, as the model sizes are too large to reasonably quantify the variability over the model parameters.
State-of-the-art approaches attempt to estimate the semantic uncertainty of the LLM \citep{Kuhn:23, Farquhar:24}.
However, they rely on chance to sample possible output sequences and do not explicitly search for likely outputs of the model with different semantics.
The recently published SDLG method \citep{Aichberger:25} improves on the previously best methods by targeted sampling of alternative outputs with different semantics.

We approach uncertainty quantification in the context of certification by deriving requirements that focus on the correct choice of predictive uncertainty quantification methods for the data domain of the application.
Furthermore, we derived requirements for an empirical evaluation of the selected uncertainty method.
Finally, we derived requirements for the use of the uncertainty information in the context of the application of the product, such as the implementation of a fallback procedure in the case of a high uncertainty for a given prediction.

It is worth noting that empirical evaluation is a critical aspect of uncertainty estimation, as there is usually no ground-truth uncertainty as a reference.
Therefore, the quality of uncertainty estimates is generally evaluated for different tasks.
One possibility is selective prediction, where uncertainty estimates are used to select a set of samples to predict for, thus its ability to select a good set of such samples is evaluated.
Another is misclassification detection, which evaluates the correlation of the uncertainty score with the correctness of the model predictions.
Finally, out-of-distribution detection is a widely considered task, where the correlation between the uncertainty score and whether the input is from within or outside the distribution is evaluated.
Similar principles hold for LLMs, where hallucination detection tasks are akin to misclassification detection.

\subsection{Biases}

The implementation of any predictive system naturally comes with a variety of different biases \citep{Mehrabi:19}.
This is especially true for machine learning systems that learn from data. 
These systems often have implicit biases that are difficult to identify for developers, let alone users.
Bias can broadly be defined as a systematic deviation from a true or desired value.
Values can be measurement values, i.e., an input or a target for a given input, or they can be values in a social sense, i.e., a social norm such as equal payment irrespective of the gender.
We also distinguish between biases of a technical and societal nature.

\begin{figure}[h!]
    \centering
    \begin{subfigure}[b]{0.49\textwidth}
        \centering
        \includegraphics[width=0.8\textwidth, trim=0mm 0mm 0mm 0mm, clip]{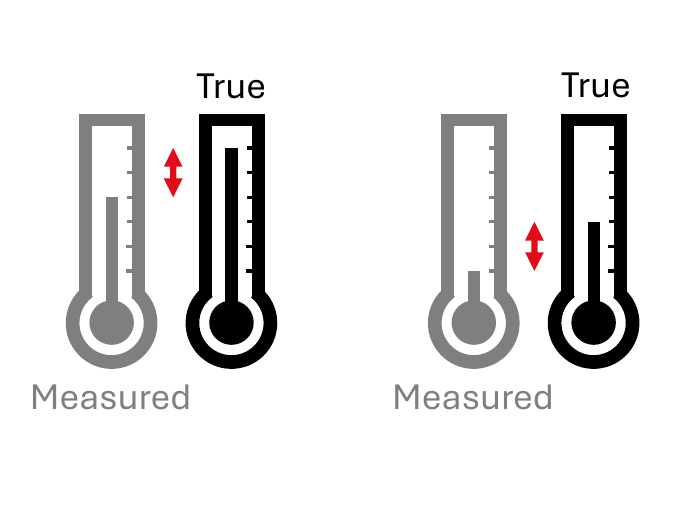}
        \subcaption{Technical bias}
        \label{fig:technical_bias}
    \end{subfigure}
    \begin{subfigure}[b]{0.49\textwidth}
        \centering
        \includegraphics[width=0.8\textwidth, trim=0mm 0mm 0mm 0mm, clip]{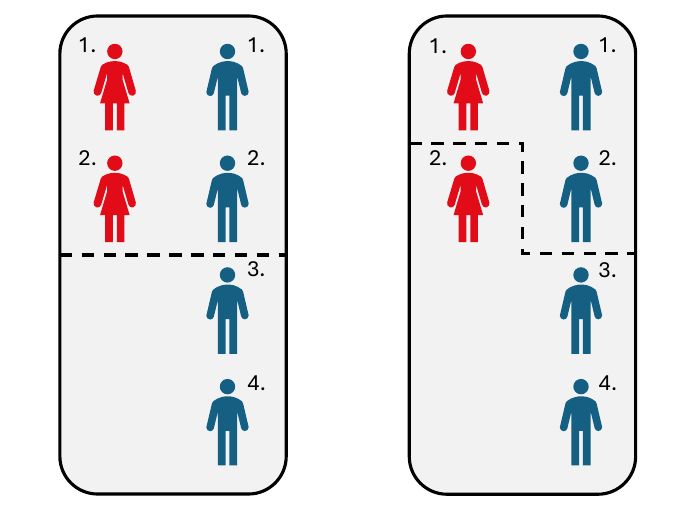}
        \subcaption{Societal bias}
        \label{fig:societal_bias}
    \end{subfigure}
    \caption{Technical and societal bias. Technical bias corresponds to a systematic deviation from a true value, exemplified by the measured values being lower than the true value by the same amount for each measurement. Societal bias refers to systematic deviations from social norms. For example picking the best job applicants. One norm (left) could be an equal number of applicants per group, another (right) could be equal rates, e.g. 50\%, per group.}
\end{figure}

\textbf{Technical biases} arise from the design, development, and implementation processes of machine learning models. 
These biases often stem from issues such as incomplete or skewed datasets, algorithmic limitations, or choices in model architecture.
An example of a technical bias would be a thermometer that systematically measures two degrees lower than the actual temperature (cf. Figure~\ref{fig:technical_bias}).

\textbf{Societal biases}, on the other hand, are rooted in the social, cultural, and historical contexts from which data and models emerge (cf. Figure~\ref{fig:societal_bias}).
These biases reflect existing inequalities, stereotypes, and power dynamics in society.
For example, a model trained on job application data may learn to perpetuate gender or racial biases present in historical hiring practices.
Gender and race are two examples of protected attributes, which according to anti-discrimination law are not eligible as grounds for decision making.
However, simply excluding protected attributes as features from the model may be too simplistic an approach as it is, not suited to alleviate such biases. 
This is because protected attributes are often highly correlated with other features that a model could simply learn as a proxy in a less obvious but equally harmful manner. 

Our approach to assessing potential biases, whether they are of a technical or societal nature, is as follows.
Firstly, there are explicit requirements aimed at detecting common biases in the data collection
and labeling process. 
These are so general that they apply to the vast majority of possible machine learning applications.
Secondly, we curate a list of biases that are commonly encountered throughout the machine learning lifecycle and across model types.
A qualitative and, if possible, quantitative investigation of at least our listed biases, if they are applicable to the machine learning approach under test, has to be conducted and is evaluated.
The list of biases is continuously being updated to address new biases arising from the rapid advancements of machine learning technology.
For example, we identified multiple biases that are unique for large language model (LLM) chatbot systems, such as their bias as to how they are prompted (asked) to generate an answer.

\subsection{Algorithmic Fairness} \label{subsec:fairness}

Algorithmic fairness is also concerned with societal biases, though more through the lens of the predictions of a machine learning system \citep{Mehrabi:19, Caton:20}.
Implicit or sometimes even explicit biases due to the state of society and the currently prevalent norms are investigated in this field.
Furthermore, the regulatory environment, such as anti-discrimination law, must be considered and adhered to \citep{Weerts:23}.
The field of algorithmic fairness does not generally aim to define what it means to make fair decisions, which is the domain of philosophy, jurisdiction, and politics.
The goal of algorithmic fairness is to operationalize different notions of fairness and to assess machine learning systems with respect to their impact \citep{Barocas:23}.
Effectively, this can mean measuring disparities from the desired notion of fairness, but various methods have also been developed to alleviate these disparities.
The central consideration for any given problem is always the associated ``cost'' to an individual of being subject to the prediction of a machine learning system.
These may include direct financial costs, such as higher loan interest rates, as well as opportunity costs, such as being rejected from a job.

Commonly used definitions of fairness in machine learning studies are the notions of statistical parity \citep{Dwork:12, Kamishima:12} and equality of opportunity (or odds) \citep{Hardt:16}.
Those notions have originated from societal and legislative considerations in the United States and have since been broadly adopted within the machine learning community.
Formally, statistical parity is defined as
\begin{align} \label{eq:statistical_parity}
    p(\hat{y} = 1 \mid a = 1) \ = \ p(\hat{y} = 1 \mid a = 0) \ ,
\end{align}
where $\hat{y} = f(x)$ is the target predicted by a model $f$ and $a$ is a protected group attribute such as age, gender, or race.
Note that $p(\hat{y}=1 \mid a) = \int_{\mathcal{X}} \mathbbm{1}\{f(x) = 1\} p(x \mid a) \ \mathrm{d}x$, in other words, the fraction of predictions with positive outcomes in the whole application domain conditioned on the group attribute.
In practice, this is approximated through sampling, i.e. with a test dataset of finite size.
The definition of equalized odds and equal opportunity is similar, yet additionally conditioned on the true target $y$, resulting in
\begin{align} \label{eq:equal_odds_opp}
    p(\hat{y} = 1 \mid a = 1, y = \upsilon) \ = \ p(\hat{y} = 1 \mid a = 0, y = \upsilon) \ ,
\end{align}
where $\upsilon = 1$ for equal opportunity and $\forall \upsilon \in \{0, 1\}$ for equalized odds.
Note that these definitions follow the widespread convention that $y=1$ is a positive or desired outcome and $a=1$ is the advantaged majority group.
Importantly, even the basic definitions of fairness in Eq.~\eqref{eq:statistical_parity} and Eq.~\eqref{eq:equal_odds_opp} have been shown to be generally incompatible with each other \citep{Chouldechova:17, Kleinberg:16}.
Although a fairness assessment should never be reduced to a single metric, this shows that there are inherent trade-offs and decisions to be made on a societal and legislative level.

\begin{figure}[b!]
    \centering
    \begin{subfigure}[b]{0.49\textwidth}
        \centering
        \includegraphics[width=0.8\textwidth, trim=0mm 3mm 0mm 3mm, clip]{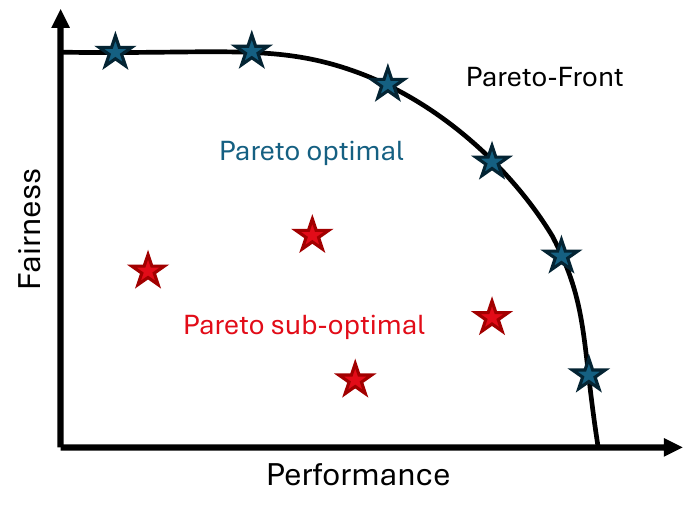}
        \subcaption{Pareto optimal and sub-optimal models}
        \label{fig:pareto_front}
    \end{subfigure}
    \begin{subfigure}[b]{0.49\textwidth}
        \centering
        \includegraphics[width=0.8\textwidth, trim=0mm 3mm 0mm 3mm, clip]{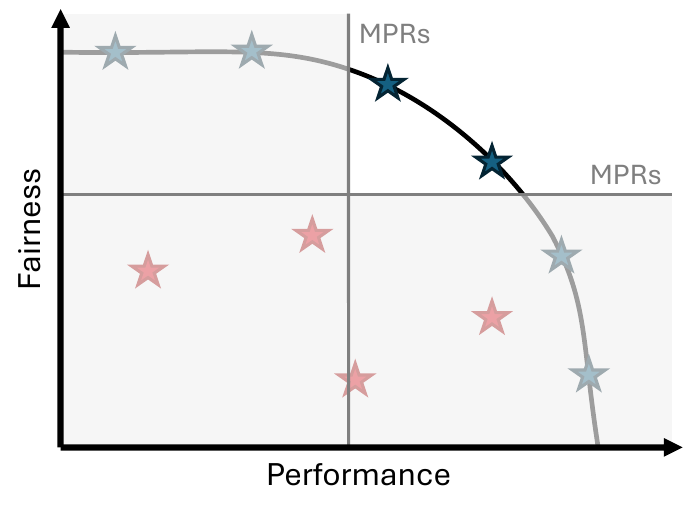}
        \subcaption{Eligible pareto-optimal models under MPRs}
        \label{fig:pareto_front_mprs}
    \end{subfigure}
    \caption{Pareto-front of a machine learning system. Pareto-optimal models are on the pareto-front, pareto sub-optimal models are within the pareto-front. MPRs define eligible sections on the pareto-front, where the system can be certified.}
\end{figure}

Central to the field of algorithmic fairness is the \textbf{Performance-Fairness trade-off}, which can be thought of as the \emph{Pareto-Front}.
Being on the Pareto-Front means that the machine learning system cannot achieve higher performance without leading to a higher violation of a fairness measure and vice versa can not decrease the violation of the fairness measure without impeding the performance.
Machine learning systems should strive to operate at some point on this front, as shown in Figure~\ref{fig:pareto_front}.
The Pareto front is not the same for every model. For example, ensembles are usually Pareto dominant in relation to their individual models. \citep{schweighofer:24b}.
A risk assessment is needed to determine where on the Pareto front it is permissible for the machine learning system to be.

Our approach to assess algorithmic fairness is threefold.
Firstly, we have to define appropriate notions of fairness and corresponding measures that are suitable for the application domain of the machine learning system.
Furthermore, desirable minimum (or maximum) values for these measures have to be defined a priori, considering the ``cost'' of the prediction and other ethical and legal considerations as outlined above.
This is analogous to the definition of performance measures and minimum performance requirements. See Figure~\ref{fig:pareto_front_mprs} for a representation.
Attaining the desired values of a fairness measure generally requires fairness interventions.
Fairness interventions are changes to the ML training pipeline on either the data (pre-processing intervention), the training process (in-processing intervention) or how the trained model predicts (post-processing intervention) with the aim of increasing the fairness of the model's predictions.
This means that secondly, we require to justify the choice of fairness intervention and assess the impact of the fairness intervention.
Thirdly, we require to document the findings rrelated to the fairness implications of the machine learning system within the documentation required by Annex IV of the EU AI Act, especially due to the requirements regarding ``potentially discriminatory impacts'' in clause 2 (g) of that Annex.

\subsection{Explainability and Interpretability} \label{sec:explainability}

Explainability and interpretability of ML models aim to provide human-understandable descriptions of (1) why a model made a specific prediction and (2) how the model works internally.
The need for such approaches arises from the opaque nature of modern machine learning methods.
With access to the model, i.e. its functional form and parameters, it is possible to understand exactly why a specific prediction was made and how the model works internally. 
However, it is generally incomprehensible for most humans.
This is similar to understanding a circuit board in a computer.
A domain expert can understand it perfectly without any additional information. Other users, however, need additional descriptions that are potentially more abstract (e.g., that a certain area of the board handles the power supply without describing all the components).
Similar to larger circuit boards, it becomes increasingly challenging to describe very large ML models in a way that is both comprehensible and accurate.

Explainability / interpretability methods are developed along three main considerations.
\begin{itemize}[noitemsep]
    \item First, whether the method is \textit{model-specific or agnostic}.
    \item Second, whether the method tackles\textit{ local or global interpretability}, i.e., whether a single prediction or the overall model behavior is to be explained.
Examples of local, single prediction, methods are the widely known LIME~\citep{ribeiro2016whyitrustyou} and SHAP~\citep{lundberg2017unifiedapproachinterpretingmodel} methods.
Examples of global, general model behavior methods are feature importance methods or variants specific to neural networks such as integrated gradients \citep{Sundararajan:17}.
\item 
Third, \textit{post-hoc vs. intrinsic methods}, i.e., whether the method is applied to analyze models after training or whether the mechanism that provides explainability / interpretability is an intrinsic component of the model.
\end{itemize}
However, it is not entirely clear how useful the current explainability / interpretability methods are in real-world scenarios \citep{Colin:22}, i.e., if there is an added benefit for a user in understanding the ML models and their individual predictions.

Our approach to the use of explainability and interpretability is as follows.
Firstly, it is important to note that we do not see these explainability and interpretability methods as a way to directly increase trust in the predictions of an ML model. 
We see them as a way to detect failure modes.
To our knowledge, it is impossible to ascertain whether a model uses the ``correct'' features in the data to make its predictions with current methods. 
However, current methods can potentially help to determine whether the model uses ``incorrect'' features, such as watermarks or other spurious features.

Another fruitful application of explainability / interpretability methods could be as an additional tool to understand the dataset or to detect distribution shifts.
This is orthogonal to the majority of interpretations of the results of explainability / interpretability methods, where the focus is on the model and its predictions.
An example of such a method is VisDiff \citep{Dunlap:24}, which uses language models to create a description of the differences between two image datasets.
Furthermore, this can be used to understand failure modes of the model, by describing the differences between samples where the model predicts correctly and those where it predicts incorrectly on some dataset.
Similarly, \citet{riccio2025imageset2text} developed a methodology to generate rich textual descriptions of image datasets. 
Descriptions of the training, validation, and test dataset are an important requirement as part of the technical documentation according to the EU AI Act Article 11, detailed in Annex IV 2(d).
Thus, such methods can help automate compliance with regulatory requirements while enhancing the interpretability and traceability of AI systems.

\newpage

\section{Post-Certification Requirements} \label{sec:maintaining}

In this section, we discuss topics that are inherent to the safe operation of an AI system after certification. 
Once the AI system is deployed, a comprehensive monitoring system needs to be set up to oversee the system. 
The monitoring system should trigger an alarm if the model performance decreases. 
This can be the case if external circumstances of the model operation change, e.g., in the case of a distribution shift (also see Section~\ref{sec:dshift}).

If the ongoing monitoring of the AI system indicates that the prerequisites for the certificate are no longer valid, an update or retraining of the ML models underlying the AI system may be necessary. 
Possible reasons for this could be a distribution shift in the real world, or unforeseen complications or risks related to using the AI system. 
However, in these cases, appropriate safety measures must be taken to ensure that the updated system continues to meet the certification requirements.

In the following, we outline a possible approach to monitoring and the basic statistical assurance procedures in the case of model retraining. 
These findings are the subject of ongoing research and will be examined in greater depth in future work.

\subsection{Monitoring Deployed AI Systems}

\begin{figure}[b!]
    \centering
    \includegraphics[width=\linewidth]{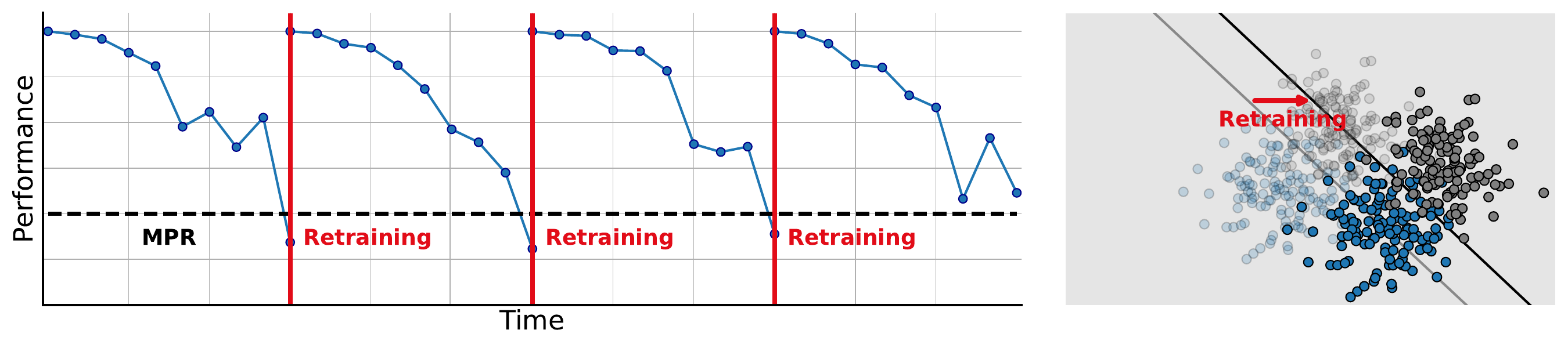}
    \caption{Left: performance degradation over time. Right: Retraining after a distribution shift.}
    \label{fig:monitoring}
\end{figure}

Monitoring ML models poses distinct challenges arising from their inherently different nature and architecture compared to conventional software systems. Unlike traditional systems, which are typically deterministic and rule-based, ML models are data-driven and require monitoring approaches based on statistical analysis~\citep{Zhang:20}. 

Before an ML model is deployed online, it is tested offline on a test dataset that is sampled according to the Stochastic Application Domain Definition (see Section~\ref{sec:functional-trustworthiness}). 
The test dataset should be representative of the expected operating conditions of the system. 
The performance estimate obtained from those tests serves as an indicator of the performance that we can expect from the AI system after deployment. 
However, an ML model trained on data from a particular domain may identify patterns that do not generalize to a different setting where the underlying data distribution has changed~\citep{Wang:23}. 
In particular, changes in the operating environment of the AI system can lead to new challenges for the model and cause it to operate in situations that were not anticipated during model training.
Since the offline test set cannot fully represent future data~\citep{Werpachowski:19}, online testing and comprehensive monitoring of the deployed AI system become an integral component to maintain the certificate and ensure the safe operation of the system~\citep{Zhang:20,Schroder:22}.

Ideally, the performance, e.g., accuracy, of the AI system should be monitored to detect any deterioration in the performance during deployment in a timely manner. 
However, measuring performance often requires the true value of the target variable, that is, labeled test data. 
This may be difficult to obtain in practice since the labeling process can be expensive and time-intensive, for example, conducting a wet lab experiment during drug development, and replacing this process is often the reason why the AI system is developed in the first place. 
However, an indicator for the model performance can be obtained by monitoring the \textbf{input data} to the system to detect potential domain shifts in a timely manner (see also Section~\ref{sec:dshift}).

The connection between domain shifts and learning performance can be understood from the following schoolbook problem~\citep{vapnik2013nature} of learning an unknown dependency $h(x)$ that assigns a probability to an input $x$, e.g., the probability that the image $x$ contains a cat.
We assume that we have successfully found a learning model $f(x)$ that has a low error $\mathrm{err}_p(f)=\mathrm{E}_{x\sim p}[|h(x)-f(x)|]$ in our application domain $p(x)$. 
We then know that the performance decrease $|\mathrm{err}_p(f)-\mathrm{err}_{q}(f)|$ caused by a shift from $p$ to some domain $q$ satisfies
\begin{align}
    \label{eq:domain_shift_and_model_performance}
    \max_{h:\mathbb{R}^n\to [0,1]} |\mathrm{err}_p(f)-\mathrm{err}_{q}(f)|=d(p,q) \ ,
\end{align}
where $d$ is the (total-variation) distance between $p$ and $q$~\citep[Theorem~2.1]{zellinger2020moment}. 
Eq.~\eqref{eq:domain_shift_and_model_performance} proves that the distance $d(p,q)$ between a domain $p$ and its shifted version $q$ determines the performance decrease (left-hand side) of a machine learning model. 
\textbf{In other words, if the domain shift is large, then the performance decrease can also be large.}
Eq.~\eqref{eq:domain_shift_and_model_performance} also implies that, for any dependency $h$, 
\begin{align}
    \label{eq:domain_shift_and_model_performance2}
     |\mathrm{err}_p(f)-\mathrm{err}_{q}(f)|\leq d(p,q) \ .
\end{align}
If direct monitoring of the model performance is infeasible, the relationship above implies that a viable approach can be obtained by monitoring the system for distribution shifts of the input data.
This can be achieved, e.g., using the techniques presented in Section~\ref{sec:dshift}. 
Of course, not every distribution shift has to be malignant~\citep{Rabanser:19}. 
However, once a distribution shift is detected, it can be classified as malignant or benign by point checking the performance of the AI system on a labeled, up-to-date test dataset.  
If the performance of the model is negatively affected by the detected shift, the possibility of adapting the model needs to be evaluated (see Figure~\ref{fig:monitoring} for an illustration of retraining the model after detecting a performance drop). 
However, before actually deploying the retrained AI system, suitable measures need to be taken to ensure its continued functional trustworthiness, as we will explain in the next section.

\subsection{Countering Deteriorating Performance: Continuous Re-training} \label{sec:continuous_learning}

The easiest way to counteract domain shifts is to update the model with new data. 
There are several factors that have to be taken into account when updating a model. 
When adapting a model to counter domain shifts, one significant challenge is \textit{catastrophic forgetting}--the tendency for neural networks to lose or overwrite previously learned knowledge when trained on new data. 
This issue arises because, as the model adjusts to the nuances of the new domain, it often ``forgets'' patterns and representations that are crucial for the domains it previously mastered. 
This loss of prior knowledge can be especially detrimental in settings where the model is expected to handle multiple domains or where the domain shift may later revert, requiring the model to retain versatility. 
Incorporating state-of-the-art methods to prevent catastrophic forgetting is thus a prerequisite when updating the model with new data.

\textbf{However, a more fundamental problem is modifying a certified model in the first place.} 
Modifying the model is problematic, as it invalidates the guarantees given by the original assessment, specifically by the statistical performance tests, as they hold only for the set of parameters that they were performed on, i.e., the unmodified model. 
As described in Section~\ref{sec:functional-trustworthiness}, statistical tests ensure that the model's performance is above the defined minimum requirements with some predefined error probability. 
Those test results cannot be transferred to an updated model to ensure its quality.  

\begin{figure}[h]
\centering
\includegraphics[width=\linewidth, trim=0 0.8cm 0 -0.8cm, clip]{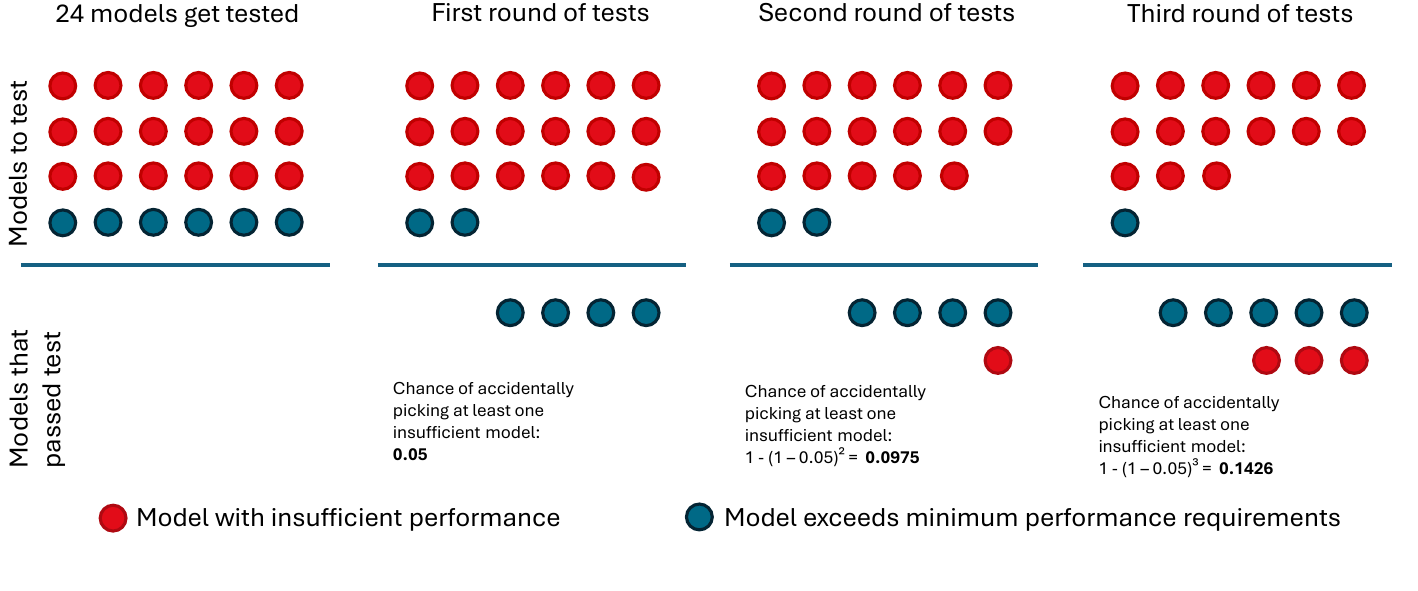}
\caption{Illustration of the multiple testing problem. Each statistical test increases the the likelihood of incorrectly approving a model with insufficient performance.\vspace{0.5cm}}
\label{fig:multiple_testing2}
\end{figure}

The straightforward solution would be to simply repeat the statistical tests with the updated model. 
However, this introduces the multiple comparison problem as described in Section~\ref{sec:multiple_testing} (also see Figure~\ref{fig:multiple_testing2}).

This can be particularly problematic when certifying a model for high-stakes or regulated applications where guarantees on performance are critical. 
Thus, merely re-testing an updated model is not viable, as each iteration would risk undermining the statistical validity of the model performance assessment. 
It is therefore crucial to control the Family-Wise Error Rate (FWER) using suitable methods. 
The proposed methods for this are \textit{Fixed Sequence Testing} and the \textit{Fallback Procedure}, which are sequential approaches that allow flexible and incremental testing while maintaining control over the FWER \citep{maurer1995multiple,westfall2001optimally,wiens2003fixed}. 
These methods are particularly suited for continuously re-trained models, as they do not necessitate the simultaneous execution of all tests as do other methods (e.g., the Bonferroni correction), thus allowing for iterative improvements without inflating the risk of false positives.

The Fixed Sequence Testing procedure is based on a predefined order of hypotheses $H_1,\ldots,H_n$. In the context of continually trained models, this predefined order naturally arises from the temporal ordering of the model updates. Each hypothesis is tested conditional on the previous ones, with a significance level that is fixed beforehand.
Testing therefore starts with $H_1$, and each subsequent test is performed without adjusting the selected significance level $\alpha$ to correct for multiple tests. 
Any hypothesis $H_i$, $i=1,\ldots,n$, can only be rejected if all previous hypotheses were also rejected, thus all p-values:
\begin{equation}
    p_j \leq \alpha, \ j=1,\ldots,i.
\end{equation}
The sequence terminates as soon as a hypothesis cannot be rejected, as all subsequent hypotheses will also be non-rejectable.
This limitation is addressed by the Fallback procedure. 
Similar to the Fixed Sequence Testing procedure, hypotheses $H_1,\ldots,H_n$ are temporally ordered, but each hypothesis is now assigned a weight, $w_1,\ldots,w_n$, which specifies the proportion of the overall significance level allocated to this hypothesis by $\alpha w_i, i=1,\ldots,n$. 
Note that the weights $w_i$ must add up to 1, i.e., $\sum_{i=1}^{n} w_i = 1$, to ensure that the total assigned significance level is equal to $\alpha$. 
The Fallback procedure is then as follows:

\begin{itemize}
    \item $H_1$ is tested at $\alpha_1 = \alpha \ w_1$. If $p_1\leq\alpha_1$, $H_1$ can be rejected.

    \item For $i=2,\ldots,n$: $H_i$ is tested at $\alpha_i=\alpha_{i-1}+\alpha \ w_i$ if $H_{i-1}$ was rejected. Otherwise, $H_i$ is tested at $\alpha_i = \alpha \ w_i$, which implies that the previously allocated $\alpha_{i-1}$ is ``lost''.
\end{itemize}

It is worth noting that the Fixed Sequence Testing procedure is a special case of the Fallback procedure with $w_1 = 1$ and $w_2 = \ldots = w_n = 0$. 
\citet{wiensFallbackProcedureEvaluating2005} proved that the Fallback procedure (and therefore the Fixed Sequence Testing procedure) to be a closed testing procedure, meaning it controls the FWER in the strong sense. 
A procedure controls the FWER in the strong sense if the control is guaranteed for any configuration of true and non-true null hypotheses, regardless of whether the global null hypothesis is true or not \citep{dmitrienko2009multiple}.

In summary, although unwarranted modification of the certified model will generally cause a loss of the certificate, a careful refinement of the model should be possible if it undergoes the same assessments as its predecessor.
However, the statistical implications of carrying out multiple tests ``until one of the models finally produces significant results'' as visualized in the comic in Figure~\ref{fig:mulitple-testing-comic} (p.~\pageref{fig:mulitple-testing-comic}) have to be countered by applying an appropriate correction for the significance level as described in this section. 

It is understood that a great amount of future work is necessary, both in research and in standardization, to explore the implications of particular changes to AI systems and the possibilities of preserving certain traits or qualities of the models.

\newpage

\section{Summary and Outlook} \label{sec:conclusions}

This white paper outlines the scientific basis of the \textit{TÜV AUSTRIA Trusted AI} audit catalog for conformity assessments of AI applications. It provides a comprehensive framework for auditing and ensuring the quality of AI systems with a specific intended use.
This takes into account the specific purpose and the inherent risks of the particular application. 

\textbf{Our auditing approach:} \quad
At the heart of our auditing approach (see Section~\ref{sec:description-audit-catalog}) is the statistical assessment of functional trustworthiness. 
In this context, we have presented the Stochastic Application Domain Definition (SADD), a unique approach to connect the technical sampling process of the test dataset to the legal requirements of responsibility and accountability.
In order to benchmark the performance of the AI system, a set of minimum performance requirements must be established based on a specific risk assessment and the desired quality characteristics of the intended functionality of the AI system. 
The careful sampling of independent test data allows a meaningful evaluation of the relevant operating conditions and an assessment of the generalization capabilities of the model.
Statistical tests are then applied to confirm that the observed performance significantly exceeds the defined minimum performance requirements. 
However, it is an illusion to assume that performance tests alone are sufficient to conclusively judge the risks and qualities of an AI system with all its complexities. 
The validity of the performance tests and the quality of the overall composition of the AI system have to be established by a thorough investigation of all design choices and implementation details of the AI system.
This entails assessing all of the steps, from the design of the Stochastic Application Domain Definition, the data collection and the model training process, right through to the user interface design and the subsequent deployment and monitoring of the system.
Beyond the technical level, a proper quality management framework, e.g., as in \citet{AIManagementSystem}, is necessary to establish overall trust in the AI system. 
Our catalog covers these aspects, and while it is intended to serve as a valuable resource for the ongoing international standardization effort, it will ultimately be aligned with upcoming standards and regulations in that area once they are adopted.

\textbf{Contributions beyond the auditing approach:} \quad
In Section~\ref{sec:background} of this white paper,  we provide an overview of the current state-of-the-art techniques for developing ML models. We also explain widely used buzzwords from Deep Learning to Agentic AI and Reasoning LLM systems in a technical context that makes them understandable.
We assume that these principles and the basic know-how are common to and understood jointly by both the AI community and the legal and standardization experts involved.
Similarly, we provide an introduction to the legal and standardization landscape for technical experts in Section~\ref{sec:regulatory-background}.
In Section~\ref{sec:details}, we have focused on providing a deeper understanding of those ML methods and phenomena that are crucial to the creation, analysis and auditing process with respect to quality and risks of the AI system. 
We would like to emphasize the sections on robustness (\ref{subsec:robustness}) and algorithmic fairness (\ref{subsec:fairness}).
We aim to contribute the scientific results presented therein to current discussions within the context of ongoing standardization initiatives.
Furthermore, we discuss uncertainty estimation (\ref{sec:uncert}), which is still often neglected but is an imperative prerequisite in the context of high-risk applications with a high degree of stochasticity of the intended output.
Finally, in Section~\ref{sec:maintaining} we outline monitoring approaches for deployed AI systems and address challenges arising from re-training.
These aim to ensure the safe operation of an AI system after an initial assessment.

\textbf{Future work:} \quad
Despite the establishment of a comprehensive audit catalog, the pursuit of trustworthy artificial intelligence remains an ongoing and evolving research endeavor.
On the one hand, we will of course further refine the alignment of our requirements and proofs to the currently emerging ISO and CEN standards (also see Figure~\ref{fig:requests} on p.~\pageref{fig:requests}) once they are available. 
More fundamentally, however, there are some very challenging technical topics ahead that will go beyond our current assessment scheme.

Domain shift and certification of the corresponding automated compensation procedures are one of those highly challenging topics. 
There are a number of difficult questions to answer: 
\textit{Did the application domain itself shift or was my original perception of the application domain just a subdomain of the full permissible set?}
\textit{Is the apparent shift significant, and does it induce a relevant drop in performance that makes a re-certification necessary?}
The next cluster of similar questions arises around the fine-tuning of models on the part of the deployer or user of the AI systems: 
\textit{How can we preserve certified qualities if certified models have to be adapted on the client side to specific use cases?}
\textit{How are legal responsibilities distributed in these cases?}
It is apparent that these two topics already cover the obvious pressing questions about long-term monitoring necessities and secure automated re-certification following model refinement.

AI systems are increasingly being deployed in domain-specific applications, such as healthcare, finance, and autonomous driving. 
Each domain presents unique challenges and requirements for AI assessments. 
Future research will also have to cover the development of customized assessment methodologies that address the specific needs and constraints of different data domains. 

Another consideration is the security of the AI system, especially in safety-critical scenarios.
This requires a systematic evaluation of both the source code and the model behavior. 
Traditional software analysis techniques, such as static and dynamic code analysis or formal verification, have long been used to assess correctness, maintainability, and security. 
However, these methods face limitations when it comes to handling the complexity and evolving nature of modern AI-based software, which involves dynamic dependencies, non-deterministic behavior, and interactions with uncertain data sources.
At the same time, AI systems themselves are vulnerable to security threats such as adversarial attacks, model stealing, and privacy breaches. 
Understanding these vulnerabilities is crucial for determining whether a model can operate safely in real-world environments.

Assessing risks and quality measures of foundation models without a specific intended use is another area of future work that is already receiving significant attention \citep{eucomm:2025cop-gpaidraft3} and will be pivotal in the near future. 
The ability to generate meaningful, high-quality output that meets user expectations and abstract functional requirements remains a major challenge. 
We assume that contributions analysis methods and specialized uncertainty estimation methods will play a significant role when it comes to assigning real-world decisions to agents driven by large language models.

\textbf{Conclusion:} \quad
The second edition of the \textit{TÜV AUSTRIA Trusted AI} audit catalog transfers science-based rigor to the reality of safely deploying AI systems. 
This marks a major improvement to the audit framework, incorporating the findings of the ongoing collaboration between TÜV AUSTRIA and its scientific partners.
Our approach operationalizes the abstract notion of trustworthiness into specific, reproducible requirements for real-world performance. 
These technical safeguards are based on three mutually reinforcing pillars: Secure Software Development, Functional Requirements, and Ethics \& Data Privacy. 
However, the rapidly evolving AI landscape remains an open research field and continuous adjustments to emerging technologies will be necessary.

For practitioners, the audit catalog serves as a step-by-step manual. 
It guides teams through the complete lifecycle of the AI system: From use case definition to data collection, model development, quantitative and qualitative inspection, operation, and post-market monitoring. 
Furthermore, our framework is anchored in the current regulatory efforts and the EU AI Act, concretizing the abstract requirements laid out therein.
In this way, we aim to bridge the translation gap between the regulators who define \emph{what} must be achieved and the engineers who need to know \emph{how} to achieve and demonstrate it.

\newpage

\bibliographystyle{abbrvnat}

\renewcommand{\refname}{Bibliography\vspace{0.5cm}}
\addcontentsline{toc}{section}{Bibliography}
\begingroup
    \setlength{\bibsep}{5pt}
    \setstretch{0.8}
    \footnotesize{
        \bibliography{literature}
    }
\endgroup

\blankpage

\includepdf[pages=1,fitpaper=true]{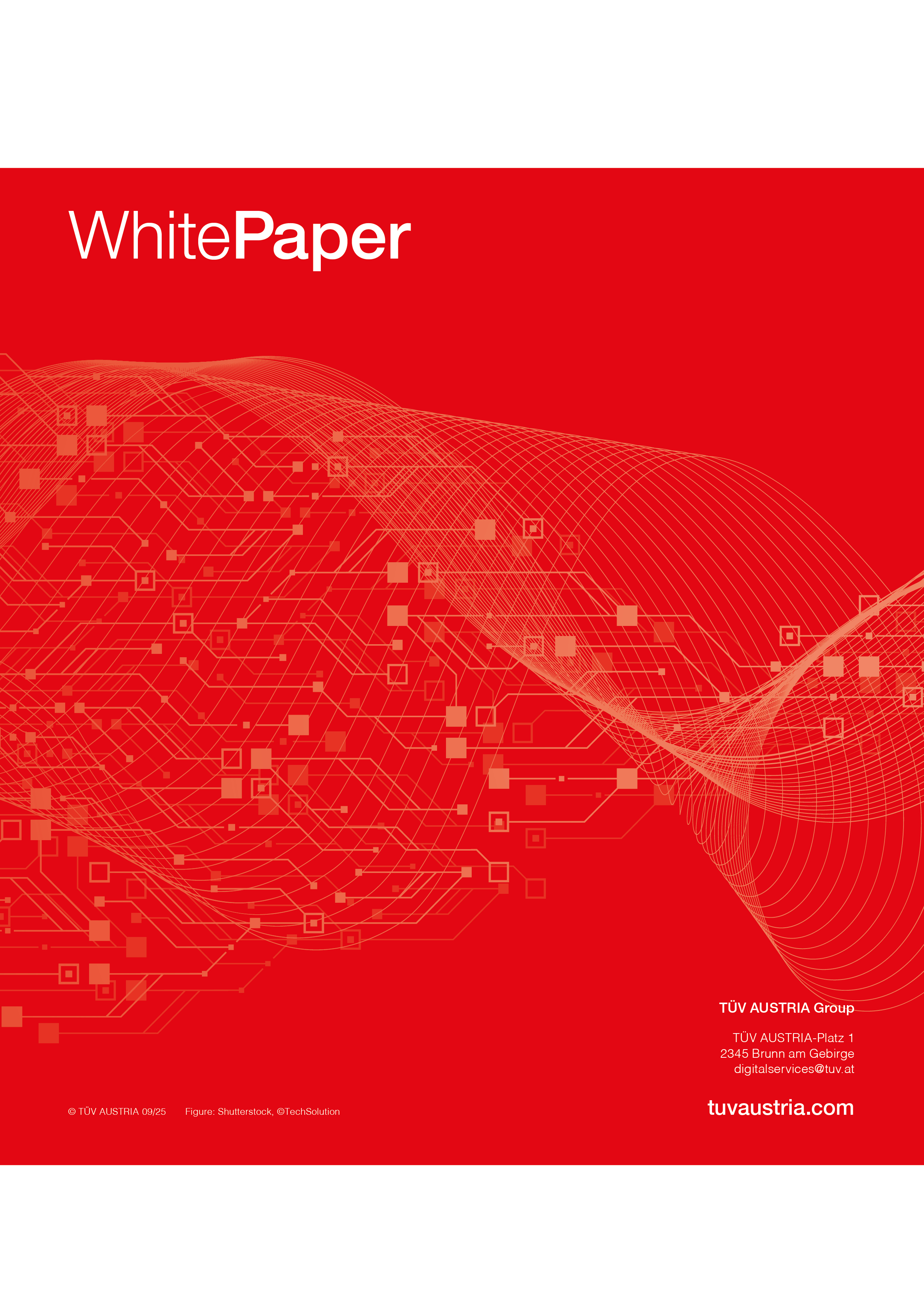}

\end{document}